\documentclass[pdflatex,sn-mathphys-num]{sn-jnl}% Math and Physical
\usepackage{graphicx}%
\usepackage{multirow}%
\usepackage{amsmath,amssymb,amsfonts}%
\usepackage{amsthm}%
\usepackage{mathrsfs}%
\usepackage[title]{appendix}%
\usepackage{xcolor}%
\usepackage{textcomp}%
\usepackage{manyfoot}%
\usepackage{booktabs}%
\usepackage{algorithm}%
\usepackage{algorithmicx}%
\usepackage{algpseudocode}%
\usepackage{listings}%
\usepackage[normalem]{ulem}
\newcommand{\ket}[1]{|#1\rangle}
\newcommand{\bra}[1]{\langle #1|}
\newcommand{\braket}[2]{\langle #1 | #2 \rangle}
\newcommand{\editor}[2]{%
  \expandafter\newcommand\csname #1note\endcsname[1]{%
    \textcolor{#2}{(\textbf{#1:} \textit{##1})}}%
  \expandafter\newcommand\csname #1\endcsname[1]{%
    \textcolor{#2}{##1}}%
  \expandafter\newcommand\csname #1cancel\endcsname[1]{%
    \textcolor{#2}{\sout{##1}}}%
  \expandafter\newcommand\csname #1change\endcsname[2]{%
    \textcolor{#2}{\sout{##1} ##2}}%
  \newenvironment{#1text}{\color{#2}}{\color{black}}
}
\definecolor{byzantine}{rgb}{0.74, 0.2, 0.64}
\editor{MQ}{red}
\definecolor{byzantin2}{rgb}{0.2, 0.2, 1.00}
\editor{MC}{byzantin2}
\definecolor{byzantin3}{rgb}{0.24, 1.0, 0.24}
\editor{NM}{byzantin3}
\AtBeginDocument{\hypersetup{%
  pdftitle={Spectral signatures of spin textures and the emergence of a pseudogap},
  pdfauthor={M. Quinzi, A. Pintus, T. Chiarotti, M. Capone, N. Marzari}}}
\begin{document}
%
% ---- TITLE -----
\title[spectral-signatures]
{Spectral signatures of spin textures and the emergence of a pseudogap}

%%=============================================================%%
%% GivenName	-> fnm{Matteo}
%% FamilyName	-> \sur{Quinzi}
%% \author*[1,2]{\fnm{Matteo} \sur{Quinzi} 
%% \email{iauthor@gmail.com}
%%=============================================================%%
\author[1]{\fnm{Matteo} \sur{Quinzi}}%\email{matteo.quinzi@epfl.ch}
\author[1]{\fnm{Andrea}\sur{Pintus}}
\author[2]{\fnm{Tommaso}\sur{Chiarotti}}
\author[3,4]{\fnm{Massimo} \sur{Capone}}
\author*[1,5,6]{\fnm{Nicola} \sur{Marzari}}\email{nm901@cam.ac.uk}

\affil[1]{\orgdiv{Theory and Simulation of Materials (THEOS), and National Centre for Computational Design and Discovery of Novel Materials (MARVEL)}, \orgname{\'Ecole Polytechnique F\'ed\'erale de Lausanne}, \orgaddress{ \city{Lausanne}, \postcode{1015}, \country{Switzerland}}}

\affil[2]{\orgdiv{Department of Applied Physics and Materials Science}, \orgname{California Institute of Technology}, \orgaddress{\city{Pasadena}, \postcode{91125}, \country{California, USA}}}

\affil[3]{\orgdiv{SISSA}, \orgname{Scuola Internazionale Superiore di Studi Avanzati}, \orgaddress{ \city{Trieste}, \postcode{34136}, \country{Italy}}}

\affil[4]{\orgdiv{CNR-IOM}, \orgname{Istituto Officina dei Materiali}, \orgaddress{ \city{Trieste}, \postcode{34136}, \country{Italy}}}

\affil[5]{\orgdiv{PSI Center for Scientific Computing, Theory and Data}, \orgname{Paul Scherrer Institute}, \orgaddress{\city{Villigen}, \postcode{5232}, \country{Switzerland}}}

\affil[6]{\orgdiv{Theory of Condensed Matter, Cavendish Laboratory}, \orgname{University of Cambridge}, \orgaddress{\city{Cambridge}, \postcode{CB3 0US}, \country{United Kingdom}}}

\abstract{
Strongly correlated solids exhibit phenomena -- from Mott localization to high-temperature superconductivity --
that escape conventional band theory.
Efforts to address this unconventional physics have been typically cast in the language of Green's function theories, with the associated self-energies capturing strong correlations beyond the reach of mean-field theories.
Here, we underscore that frequency- and momentum-dependent self-energies emerge also from mean-field solutions that break translational symmetry, and that these self-energies can display the characteristic fingerprints of strong correlations for certain, physically meaningful, spatial patterns.
In the paradigmatic case of the 2D Hubbard model, we find that already at the mean-field level the doped antiferromagnet develops, in addition to long-range ordered stripes, many quasi-degenerate self-consistent solutions with antiferromagnetic textures organized into antiphase domains separated by hole-rich interfaces of convoluted shape, reminiscent of the phase-separation scenario.
Remarkably, these solutions map into spectral functions that display very distinctively the phenomenology of the pseudogap, including the nodal-antinodal dichotomy, Fermi arcs, and a complementary Luttinger surface.
The result is a conceptually simple description of such a complex phenomenon, where a pseudogap emerges as the momentum-space spectral portrait of the real-space interplay between antiferromagnetic domains and their metallic interfaces.
}

\maketitle

%%%%%%%%%%%%%%%%%%%%%%%%%%%%%%%%%%%
%   INTRODUCTION
%%%%%%%%%%%%%%%%%%%%%%%%%%%%%%%%%%%

%\section{Introduction}
\label{sec:intro}
Interacting electrons display some of the most spectacular phenomena in condensed-matter physics, from the emergence of insulating gaps or satellites beyond the conventional band picture to strong quasiparticle renormalization, momentum-selective suppression of spectral weight (pseudogap), unconventional superconductivity, and bad metallic behaviour~\cite{Mott1968Metal-Insulator, Georges1996Dynamical, Keimer2015From}.
These phenomena, and many others, are  naturally described in terms of frequency- and momentum-dependent quantities that can be directly detected via, e.g., photoemission experiments, optical conductivity, and Raman scattering.

From a theoretical point of view, the effects of interactions on single-particle properties, like photoemission spectra, are captured by the electronic self-energy, a momentum- and frequency-dependent quantity which reshapes the independent-particle picture, broadening the quasiparticle excitations by finite lifetimes and transferring their spectral weight to additional satellites and incoherent structures. Many-body approaches naturally yield a frequency dependence: within perturbation theory (e.g., GW)~\cite{Hedin1965New} it
arises from the dynamical screening of the Coulomb interactions, while nonpertubative approaches like  dynamical mean-field theory (DMFT)~\cite{Georges1996Dynamical} can introduce a variety of collective phenomena, resulting in a rich frequency dependence.
The necessity of a frequency-dependent self-energy to describe strongly correlated quantum states is usually linked to the inherently many-body nature of these states. The paradigmatic example is the  Mott insulator, a system which displays insulating properties despite an odd number of electrons per primitive cell, in sharp contrast with the band theory of solids. This suggests that mean-field theories such as Hartree-Fock or Kohn-Sham density-functional theory (DFT)~\cite{Kohn1965Self-consistent}, which reduce the many-body problem to a single-particle theory, are bound to fail in the description of these phenomena. DMFT captures instead Mott physics via a momentum-independent local self-energy with a frequency dependence characterized by divergences at the Fermi level. The description of doped Mott insulators including, e.g, the celebrated pseudogap requires extensions of DMFT including non-local correlations~\cite{Schafer2021Tracking}, that can require a substantial computational effort.

At the same time, an interaction-driven insulating state can also be obtained by breaking spin- or translation symmetry, as in the antiferromagnetic phase characterizing many correlated materials and models. Here, the primitive cell is doubled and the single band is split in two in the reduced Brillouin zone, leading to an effective band picture which can be captured by mean-field theories.
These are, however, reliable only at weak couplings, and can not describe accurately correlated systems in primitive cells for fundamental reasons: one-particle excitations remain sharp and coherent, there are no intrinsic scattering rates or finite quasiparticle lifetimes, and no incoherent spectral features~\cite{Sangiovanni2006Static}.
 In short, a symmetry-breaking solution can open a spectral gap, but it does not capture the full spectrum of a Mott insulator. 
This expectation limits also the accuracy of mean-field calculations to study doped Mott insulators, starting from its simplest realization, the two-dimensional Hubbard model~\cite{Arovas2022Hubbard, Qin2022Computational}, which is widely considered central to understanding high-temperature superconductors. Nonetheless, mean-field calculations have revealed a rich phase diagram, displaying a variety of phases with long-range order depending on doping and hopping parameters~\cite{Scholle2023Comprehensive}.

In this work we provide a novel perspective on the physical content of mean-field solutions of the single-band Hubbard model, with significant implications for the interpretation of the signatures of electronic correlations. 
We first show that, in very general terms, a  solution that breaks translational and spin symmetry in a supercell -- even at the mean-field level (e.g., Hartree--Fock or Kohn--Sham DFT) -- can be recast as the solution of a translationally invariant Dyson equation in the primitive cell, with a momentum- and frequency-dependent self-energy arising as an embedding potential.
When translational symmetry is broken, each momentum $\mathbf{k}$ in the primitive Brillouin zone (PBZ) can be coupled to all the folded $\mathbf{k+q}$ of the supercell, where $\mathbf{q}$ is a reciprocal lattice vector for the supercell.
Eliminating these folded momenta by a Schur complement produces a dynamical self-energy $\Sigma(\mathbf{k},\omega)$.
As detailed in Methods, for a single-particle potential $V_\mathbf{q}$ breaking the translational symmetry of the primitive cell, this embedding self-energy reads
\begin{equation}
\label{eq:embedding-self-energy}
    \Sigma(\mathbf{k},\omega)=\bra{\mathbf{k}} V_\mathbf{q}\left(\omega-H_\mathbf{k+q}\right)^{-1}V^\dagger_{\mathbf{q}}\ket{\mathbf{k}},
\end{equation}
and it reproduces the $k$-diagonal component of the supercell Green's function $G(\mathbf{k},\omega)$ via a translationally invariant Dyson equation. 
For interacting electrons treated at the mean-field level a spontaneous symmetry breaking in the charge, spin, or pairing channels provides a direct physical interpretation of the resulting self-energy in terms of electrons scattered by the respective density modulations.
This picture shares also similarities with the concept of hidden fermions~\cite{Sakai2016}, derived from the cellular DMFT description of the pseudogap.
Of course, a single ordering vector merely splits the bands, while a dense distribution of ordering vectors leads to a very different scenario, as we unveil in the following.

\begin{figure}
    \centering
    \includegraphics[width=0.95\linewidth]{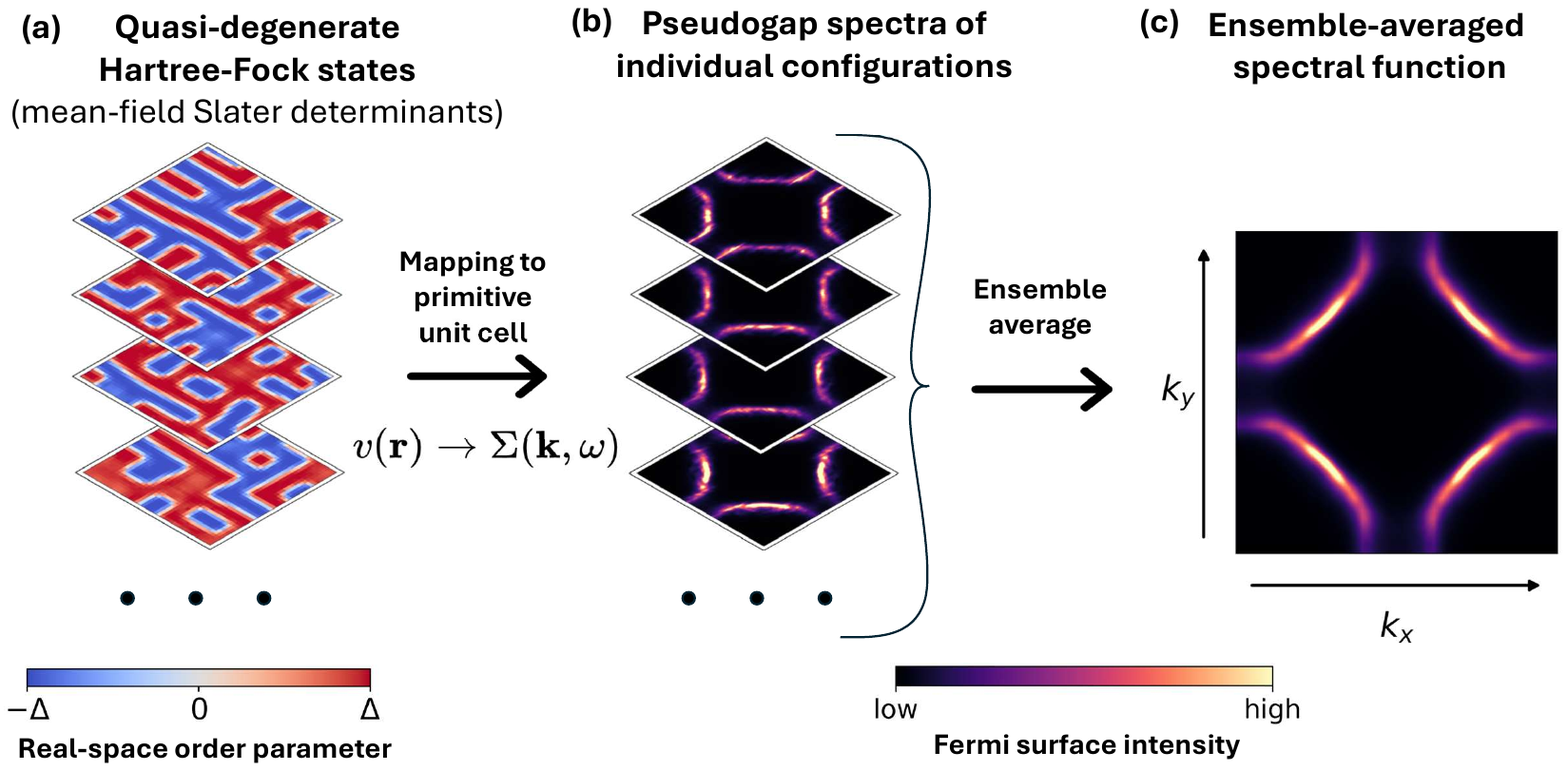}
    \caption{
    \textbf{Spectral signatures of spin textures},
    in the hole-doped two-dimensional Hubbard model ($p=0.10$ holes per primitive cell).
    \textbf{(a)} Each quasi-degenerate real-space spin texture represents a single Slater determinant and it is obtained as a stationary Hartree-Fock solution of the Hubbard model in a large supercell.
    \textbf{(b)} For each stationary state, the mean-field potential is mapped, tracing out the folded momenta generated by the supercell, to a frequency- and momentum-dependent self-energy $\Sigma(\mathbf{k},\omega)$ in the primitive cell, and hence to an interacting Green’s function $G(\mathbf{k},\omega)$;
    despite their different spatial patterns, each of these quasi-degenerate spin textures produces a closely similar spectrum.
    \textbf{(c)} An ensemble average over these removes residual configuration dependence, yielding a smooth spectral function approaching the thermodynamic limit, and highlighting the distinctive emergence of a pseudogap.
    }
    \label{fig:intro}
\end{figure}

Our main results can be conveyed using the hole-doped repulsive Hubbard model with a finite next-nearest-neighbor hopping $t'$.
 Already in the early days~\cite{Zaanen1989Charged, Poilblanc1989,
 Machida1989Magnetism, Kato1990Soliton, Schulz1990Incommensurate, Verges1991} it was shown that the antiferromagnetic (AFM) solution becomes rapidly unstable with doping to the formation of domain walls.
This led to the concept of stripes~\cite{Kivelson1998Electronic, Emery1999Stripe}, which have been recently confirmed with state-of-the-art numerics~\cite{Xu2024,Wietek2021Stripes}.
Here, using random initial configurations in the mean-field iterative implementation, we identify  a variety of stationary solutions (i.e., local minima) characterized by antiphase antiferromagnetic domains separated by hole-rich, weakly magnetic interfaces whose length increases with doping (Fig.~\ref{fig:intro}, panel (\textbf{a})).
These solutions are distinct from long-range ordered stripes, since the hole-rich regions and the magnetic domains display convoluted textures and a large degree of local randomness.
In the following we will refer to these solutions as spin textures (ST).
The crucial observation is that, when each of these real-space STs is mapped into the primitive cell, it leads to dynamical self-energies that display the spectral signatures of the pseudogap: metallic single-particle excitations first emerge along the Brillouin-Zone diagonals (nodes), while the regions along the coordinate axes (antinodes) remain gapped.
We interpret this surprising result recalling the embedding construction of the self-energy as the result of a large number of lattice momenta being coupled in the supercell.
In particular, in the thermodynamic limit a rich real-space pattern leads to a continuum of momenta, which in turn can lead to singularities in the embedding potential. Averaging over the many equivalent solutions helps reaching this limit, leading to even smoother distributions of relevant momenta; we illustrate these statements in Fig.~\ref{fig:intro} (panels (\textbf{b}) and (\textbf{c})).
Importantly, the STs appear as a manifold of a large number of quasi-degenerate stationary solutions, differing in their microscopic
real-space arrangement but representing essentially the same physical state.
We underscore that these are not the lowest-energy solutions, but they have energies comparable to the most stable stripes (see Fig.~\ref{fig:energy-diagram} later).
 The STs are therefore likely to have a much larger entropy than stripes, owing to the large number of quasi-equivalent solutions; even when a stripe or another ordered phase is more stable at low temperature, it is reasonable to expect the STs to become thermodynamically favored at some nearby intermediate temperatures.
 It is tempting to interpret the quasi-degenerate textures as the Slater determinants of a multiconfigurational correlated wavefunction: each determinant breaks the symmetries of the Hamiltonian, while their coherent superposition restores most of the symmetry, allowing however for momentum-space differentiation.

%%%%%%%%%%%%%%%%%%%%%%%%%%%%%%%%%%%%%%
%  PSEUDOGAP
%%%%%%%%%%%%%%%%%%%%%%%%%%%%%%%%%%%%%%

\section*{Phenomenology of the spin textures}
\label{sec:ST}
We focus on the 2D hole-doped repulsive Hubbard model with a finite next-nearest-neighbor hopping, where a wide range of accurate nonperturbative many-body methods has established the presence of a highly unconventional metallic state at (relatively) high temperature~\cite{Schafer2021Tracking, Simkovic2024Origin}.
This state is characterized by a momentum-selective suppression of the spectral weight at the Fermi level -- the pseudogap -- and it is typically addressed by momentum-dependent self-energies, for instance in nonlocal extensions of DMFT, such as cluster DMFT~\cite{
%Kotliar2001Cellular,
Maier2005} or diagrammatic extensions like the dynamical vertex approximation ($\mathrm{D\Gamma A}$) or the dual-fermion approach~\cite{Rohringer2018Diagrammatic}.
While many studies agree on short-range AFM fluctuations as the mechanism underlying the formation of the pseudogap at least at weak coupling and higher temperature
~\cite{
Gunnarsson2015Fluctuation,
Krien2020Fluctuation, 
%Kyung2004Pseudogap,
Kyung2006Pseudogap,
Civelli2008,
Macridin2006Pseudogap,
Senechal2004Hot,
Simkovic2024Origin,
%Wu2017Controlling, 
Wu2017Pseudogap,
%Scheurer2018Topological,
Schafer2021Tracking,
Gull2015,
Chubukov1998,
Schmalian1999}, the low-temperature regime is instead characterized by a diversity of differently ordered phases~\cite{Xu2024}, with mean-field solutions characterized by spiral order and stripes~\cite{Scholle2023Comprehensive}.
Here, we perform collinear unrestricted Hartree-Fock calculations, allowing the charge and magnetization to relax to self-consistency.
Each ST is obtained by initializing the mean-field iterative procedure with a different random guess for the spin-resolved charge density; this yields a set of different STs, each representing a stationary point of the mean-field total energy surface.
The details of the Hubbard Hamiltonian and of how these textures are selected and converged are reported in the Methods and Supplementary Information.
Our focus is on the spectral signatures of these solutions, even if they are not the ground state; for a comprehensive mean-field phase diagram, including noncollinear spiral order, we refer to Ref.~\cite{Scholle2023Comprehensive}.

\begin{figure}[h]
    \centering
    \includegraphics[width=0.95\linewidth]{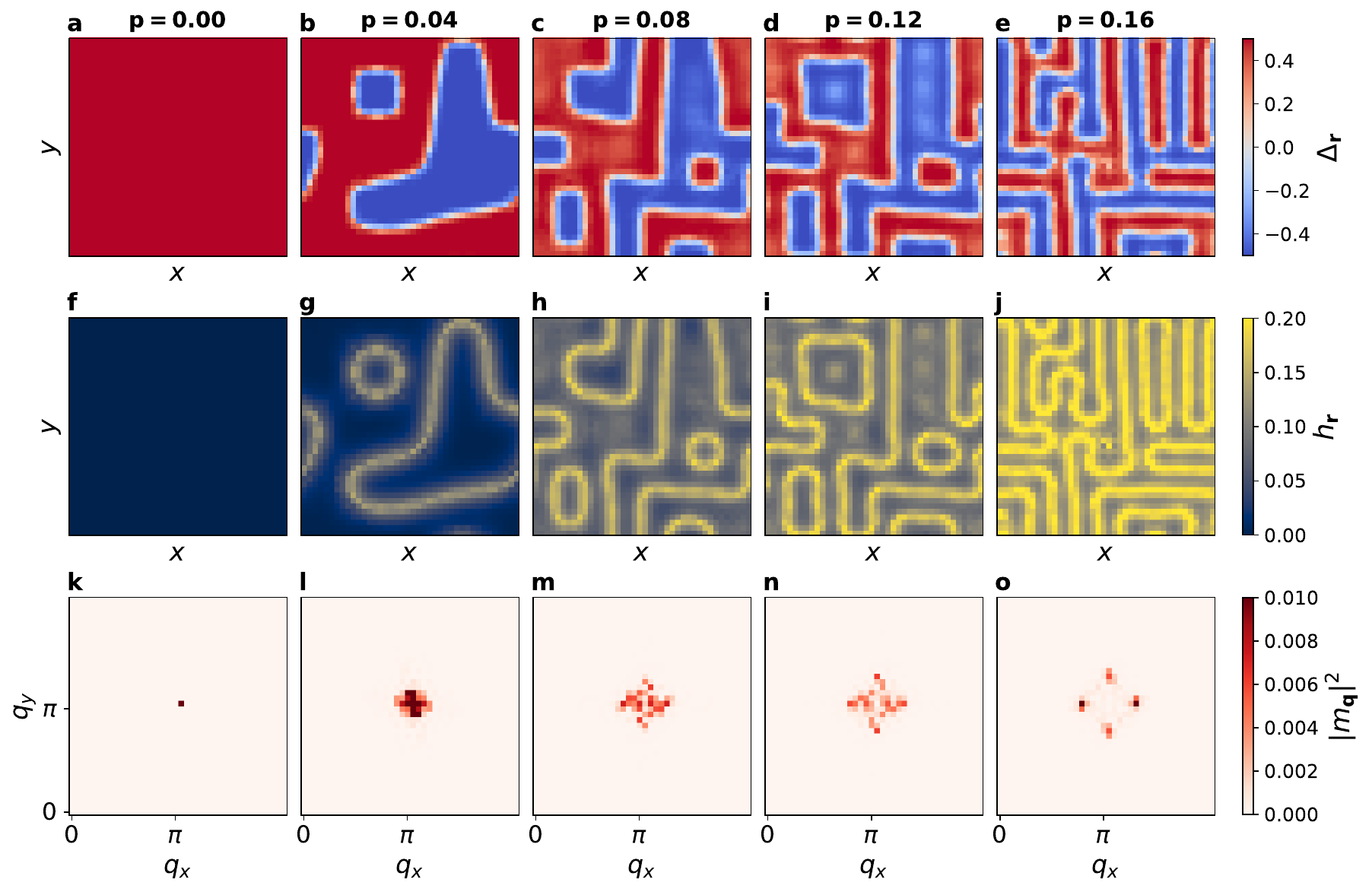}
    \caption{
    \textbf{Evolution of spin textures with hole doping.}
    Spin textures appear as Hartree--Fock self-consistent solutions in the two-dimensional Hubbard model.
    Here, calculations are performed in a $40\times 40$ supercell with model parameters $U=3t$ and $t'=-0.3t$.
    Initializing the profile of the charge density with the same guess, we track the evolution of a single solution at different values of hole doping $p$.
    (\textbf{a-e})~The real-space Néel parameter $\Delta_{\mathbf{r}}$ evolves from the half-filled antiferromagnetic state into a separation of antiphase antiferromagnetic domains bounded by weakly magnetic interfaces, whose spatial extent increases with doping.
    (\textbf{f-j})~The real-space hole density $h_{\mathbf{r}}$ shows that doped holes accumulate preferentially along these interfaces.
    They remain strongly localized to the domain walls at low doping, but progressively spill into the antiferromagnetic regions as the doping increases.
    (\textbf{k-o})~The Fourier components of $|m_{\mathbf{q}}|^2$ evolve first from the single antiferromagnetic peak at $(\pi,\pi)$ to a broadened distribution associated with short-range antiferromagnetic order, and finally to sharp and incommensurate peaks signalling the emergence of stripe-like correlations.
    }
    \label{fig:neel-charge}
\end{figure}
In Fig.~\ref{fig:neel-charge} we show the evolution with hole doping of one of these STs, displaying the real-space Néel parameter, the hole density and the squared absolute value of the Fourier components of the magnetization.
The latter is related to the magnetic structure factor that can be measured experimentally, e.g., through elastic and inelastic neutron scattering.
We characterize the textures through the local staggered magnetization, or Néel parameter, $\Delta_{\mathbf{r}} = (-1)^{r_x+r_y}\left(n_{\mathbf{r}\uparrow} -n_{\mathbf{r}\downarrow}\right)$,
where $r_x$ and $r_y$ are integers labeling the cartesian components of the position vector $\mathbf{r}$.
Starting from the uniform AFM at half filling, the electronic states organize into antiphase AFM domains separated by weakly magnetic interfaces.
The density of these interfaces increases with doping; here, $\Delta_{\mathbf{r}}$ changes sign continuously when crossing these interfaces, rather than abruptly, as discussed in the Supplementary Information. 
This behavior is reminiscent of the solitonic mass profile as suggested in Ref.~\cite{Poilblanc1989}, and connects naturally with the charged magnetic domain walls previously identified in mean-field descriptions of cuprate models~\cite{Zaanen1989Charged, Poilblanc1989, Machida1989Magnetism, Kato1990Soliton, Schulz1990Incommensurate, Verges1991}.
Inspection of the hole density reveals that the doped charge is preferentially localized at these interfaces, especially at lower doping.
This picture can be connected with phase separation scenarios in which hole-rich regions coexist with half-filled AFM regions~\cite{Grilli1991,Emery1993Frustrated}.
With increasing doping, the interfaces proliferate and a larger fraction of the holes flows into the AFM domains, progressively reducing the difference between interface and bulk.
A quantitative analysis of the hole content of the two regions is reported in the Supplementary Information.
This evolution is reflected in reciprocal space; the single AFM peak at $(\pi,\pi)$ obtained for the undoped AFM insulator broadens into a dense distribution of wavevectors as a consequence of hole doping, signalling AFM correlations of finite range.
With further doping, the distribution develops increasingly pronounced incommensurate stripe-like features peaked at $(\pi\pm\delta,\pi)$ and $(\pi,\pi\pm\delta)$, where $\delta$ defines the incommensurability with respect to the AFM order.
As detailed in the Supplementary Information, the AFM correlation length decreases with doping, while stripe-like correlations become progressively more relevant.
Unlike long-range ordered spiral or stripe solutions, characterized by one or a few ordering vectors, the present STs contain a dense set of folding vectors arising from a single solution hosting in the same supercell a variety of local environments.

\section*{Anatomy of the ST pseudogap}
\label{sec:pseudogap}

The phenomenology described in Fig.~\ref{fig:neel-charge} for a single representative solution is common to all the STs we obtained. 
As discussed later, the STs form a manifold of quasi-degenerate stationary states, differing among each other in the arrangement of their AFM domains and interfaces, but displaying closely similar distributions of magnetic ordering vectors.
\begin{figure}
    \centering
    \includegraphics[width=0.95\linewidth]{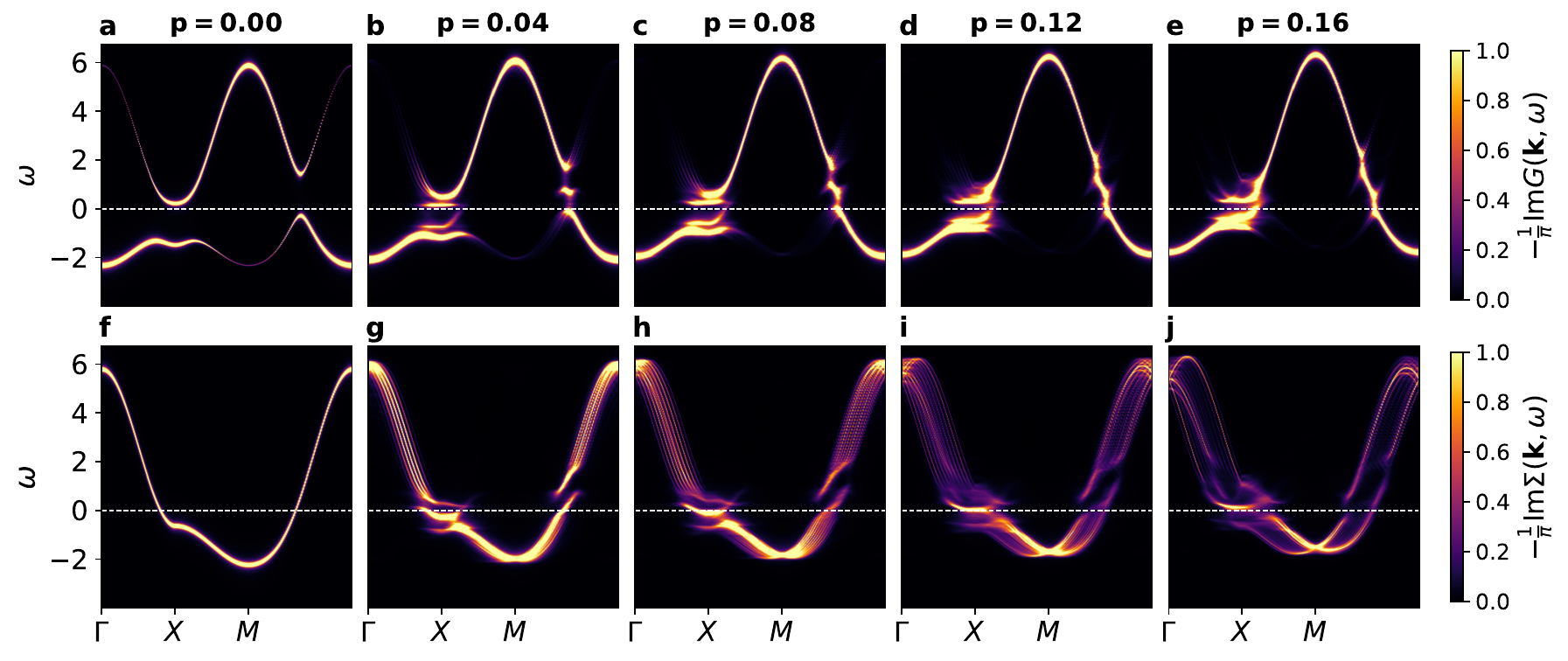}
    \caption{
     \textbf{Emergence of the nodal-antinodal dichotomy.}
     Evolution with hole-doping $p$ of the momentum-resolved spectral functions of the translationally invariant Green's functions (upper row) and self-energies (lower row) obtained from an ensemble average, with equal weights, of ten different spin textures, after they have been mapped to the primitive cell.
     The spectral functions are reported along a high-symmetry path in the primitive Brillouin zone,  including the antinodal region around the $X$ point ($0,\pi$) and the nodal sector in the $M (\pi,\pi)\rightarrow\Gamma(0,0)$ path.
     The Fermi level is at $\omega=0$ (white dashed line).
     (\textbf{a-e})~Green's function spectra: the fully gapped spectrum of the half-filled antiferromagnetic state develops a pronounced nodal-antinodal dichotomy upon hole doping.
     In-gap states emerge along the nodal direction, while the antinodal sector remains gapped.
     At larger doping, these states reconstruct the full nodal dispersion, whereas a smaller but persistent gap survives at the antinode.
     (\textbf{f-j})~Self-energy spectra:
     the coherent pole dispersion of the antiferromagnetic state is rapidly suppressed in the nodal sector upon doping, where a gap opens in the pole structure, while a strong, nearly dispersionless low-energy pole persists at the antinode.
    }
    \label{fig:gf-se-pseudogap}
\end{figure}
These real-space STs are then mapped to the translationally invariant self-energies and Green's functions in the primitive cell, by means of the embedding construction described above;
for the two-dimensional Hubbard model, the Hartree--Fock potential in Eq.~\eqref{eq:embedding-self-energy} for a given spin channel $\sigma$ reads $V_{\mathbf{q}\sigma}=Un_{\mathbf{q}\bar{\sigma}}$, where $n_{\mathbf{q}\bar{\sigma}}$ is the charge density of opposite spin.
The evolution of these spectral properties with hole doping is shown in Fig.~\ref{fig:gf-se-pseudogap}, where we report the ensemble average, with equal weights, of the spectral functions of a set of ten different STs. 
Here, each single ST encodes the same spectral features described below; the ensemble average, discussed in Methods and in the Supplementary Information, is used only as a tool to yield smoother spectra, suppressing configuration-dependent details and enhancing the features common to all states.
The Green's function evolves from the fully gapped spectrum of the half-filled AFM insulator into a strongly momentum-selective metallic state.
Upon doping, coherent in-gap excitations first appear along the nodal direction, where they progressively reconstruct the underlying noninteracting dispersion.
The antinodal sector, by contrast, remains gapped and develops additional incoherent-looking spectral weight.
Importantly, the nodal gap does not close through a continuous merging of the upper and lower AFM bands, but through the transfer of spectral weight to the in-gap states, consistently with cellular DMFT results~\cite{Kyung2006Pseudogap}.
As shown in the Supplementary Information, these states originate from the weakly magnetic interfaces separating the antiphase AFM domains, similarly to how solitonic profiles induce in-gap states~\cite{Poilblanc1989}.
The self-energy provides the complementary view of this evolution.
At half filling, the AFM solution is characterized by a single coherent-pole dispersion, due to the single ordering vector at $(\pi,\pi)$.
Upon doping, this structure develops a strong momentum dependence: a gap opens in the pole structure in the nodal sector, while a low-energy nearly dispersionless pole persists at the antinode.
With further doping, the self-energy
is progressively suppressed throughout the Brillouin zone, although antinodal scattering remains visible after the nodal structure has largely disappeared.
The incoherent features observed in the self-energy are tied to the dense distribution of ordering vectors in Fig.~\ref{fig:neel-charge}, which effectively couple a large number of momenta of the PBZ. 
Large supercells are therefore crucial to observe this phenomenon, allowing for denser distributions of the magnetic ordering vectors. 
Taken together, the Green's function and self-energy exhibit the characteristic nodal-antinodal dichotomy of the pseudogap.
 In this sense, the pseudogap emerges as the momentum-space spectral portrait of the real-space interplay between the AFM domains and their interfaces hosting the doped charges.
 This picture echoes the relevance of short-range spin fluctuations identified in a variety of other accurate nonperturbative methods~\cite{Schafer2021Tracking}.

\begin{figure}
    \centering
    \includegraphics[width=0.95\linewidth]{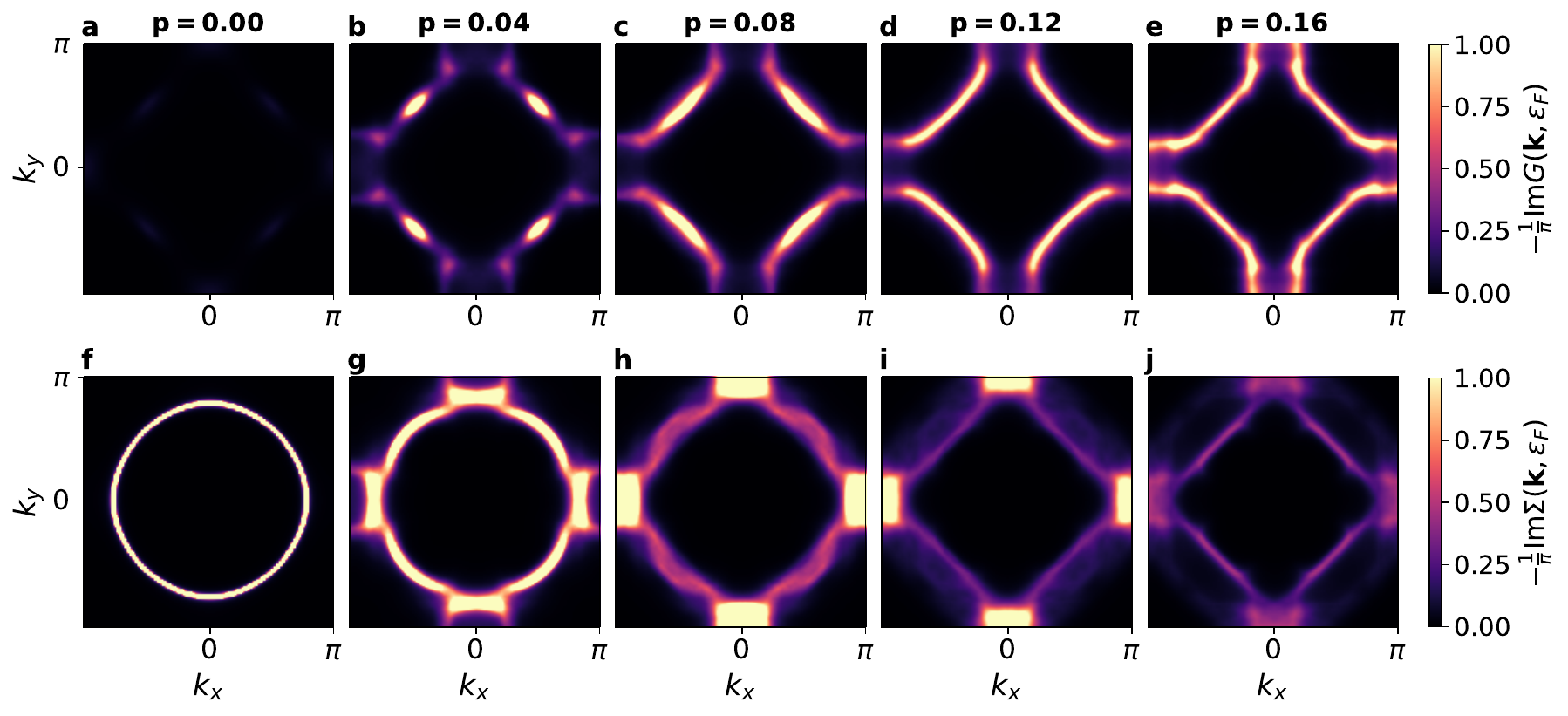}
    \caption{
    \textbf{Fermi and Luttinger surfaces.}
    The Fermi and Luttinger surfaces are obtained by plotting in the primitive Brillouin zone the imaginary part of the Green's functions and self-energies obtained from the ensemble average, with equal weights, of ten different spin textures for different values of hole doping $p$. 
    (\textbf{a-e})~Fermi surfaces obtained from the imaginary part of the Green's functions at the Fermi level.
    The half-filled antiferromagnetic state is fully gapped and therefore has no Fermi surface.
    Upon hole doping, spectral weight first emerges along the Brillouin-zone diagonals, forming Fermi arcs that progressively extend towards the antinodal regions at larger doping.
    (\textbf{f-j})~Luttinger surface obtained from the imaginary part of the self-energy at the Fermi level.
    At half filling, the antiferromagnetic self-energy produces a closed contour. Upon doping, this structure is rapidly suppressed along the Brillouin-zone diagonals, with strong scattering  confined in the antinodal sector.
    At larger doping, the Luttinger surface is attenuated throughout the Brillouin zone, reflecting the overall weakening of the self-energy.
    }
    \label{fig:panel-surf-pg}
\end{figure}
The momentum-selective character of the pseudogap is particularly noticeable at the Fermi level.
Figure~\ref{fig:panel-surf-pg} reports the spectral weight associated with the poles of the Green's function and its zeros, i.e., the poles of the self-energy, defining respectively the Fermi
and Luttinger surfaces~\cite{Dzyaloshinskii2003Some}.
As before, we report an ensemble average performed over ten different STs, where each ST independently displays the spectral features of a pseudogap.
At half filling, in the fully gapped AFM state no pole of the Green's function crosses the Fermi level.
Upon hole doping, spectral weight first reappears along the Brillouin-zone diagonals, forming incipient Fermi arcs.
With further doping, these arcs gradually extend towards the antinodal regions and eventually reconstruct a surface close to that of the noninteracting band.
The role of magnetic inhomogeneity is critical in shaping the Fermi arcs and it has also been discussed for stripe arrays with irregular spacing~\cite{Granath2002Distribution} and for antiferromagnetic domains with a distribution of gap sizes~\cite{Markiewicz2024Theory}; here, the arcs arise within each single self-consistent texture, without imposing stripes or gap distributions.
The Luttinger surface displays the complementary evolution.
At half filling, the pole dispersion of the self-energy obtained from the AFM solution produces a closed and sharp contour centered at the origin of the Brillouin zone.
Upon doping, this structure is rapidly suppressed in the nodal sector, while developing marked features at the antinodes, where the low-energy scattering removes spectral weight from the Fermi surface.
At larger doping, the Luttinger surface is attenuated throughout the Brillouin zone, reflecting the overall weakening of the self-energy.
The simultaneous growth of the Fermi arcs and disappearance of the Luttinger surface suggests a progressive transfer of spectral weight from zeros to poles of the Green's function.
Their momentum selectivity can be directly related to the magnetic textures: the Fourier components of the magnetization remain centered around the AFM ordering vector $(\pi,\pi)$, with a spread set by the finite AFM correlation length.
These scattering vectors couple the antinodal states of the noninteracting Hamiltonian much more effectively than the nodal ones, allowing coherent quasiparticles to appear along the zone diagonals while the antinodal gap persists to larger doping~\cite{Gunnarsson2015Fluctuation,Krien2020Fluctuation}.
The complementary evolution of the Fermi and Luttinger surfaces is consistent with recent analytical and diagrammatic descriptions~\cite{Worm2024Fermi}.

\begin{figure}
    \centering
    \includegraphics[width=0.95\linewidth]{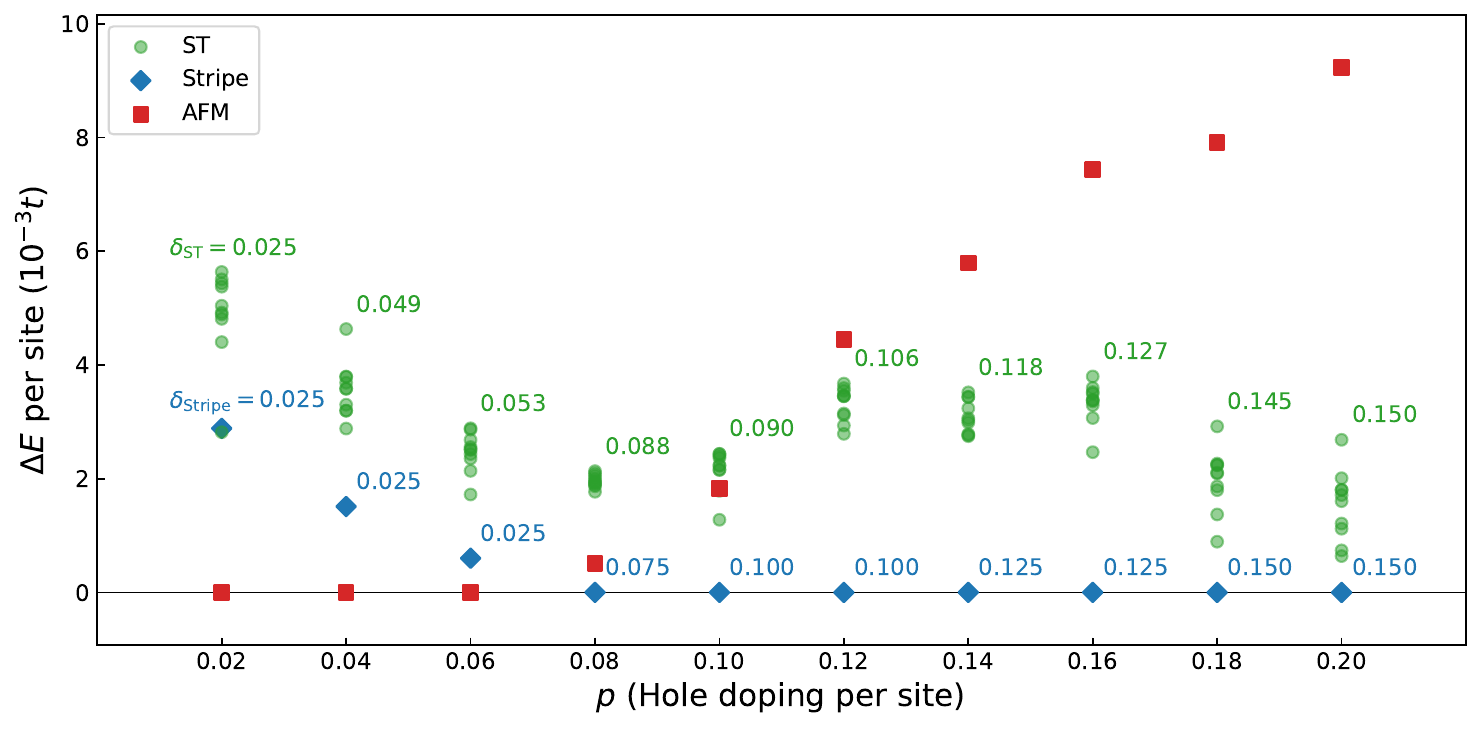}
    \caption{
    {\bf A low-energy landscape of competing orders}.
    Relative energy per site of AFM solutions (red squares), the lowest-energy stripe at each doping (blue diamonds), and individual spin textures (ST, green circles). Model parameters are $U=3t$ and $t'=-0.3t$. At each value of hole doping $p$, the lowest energy among the solutions considered defines $\Delta E=0$ and all calculations are performed in a $40\times 40$ supercell.
    The annotations report the characteristic incommensurability $\delta$, in units of $2\pi$, for the spin textures and stripes, defined as the shift of the peaks of the Fourier transform of the magnetization from the AFM ordering vector at $(\pi\pm\delta,\pi)$ and $(\pi,\pi\pm\delta)$.
    The AFM solution has the lowest energy through $p=0.06$, while stripes do from $p=0.08$ onward.
    The spin textures form a low-lying distribution whose characteristic incommensurability approaches that of the lowest energy stripe at higher doping.
    }
    \label{fig:energy-diagram}
\end{figure}

Last, we focus on energetics, showing how these STs define a manifold of nearly-degenerate states lying at energies per site close to the lowest energy solution found in our search, including long-range AFM solutions and stripes of different incommensurabilities, as reported in Fig.~\ref{fig:energy-diagram}.
The AFM state is the most stable solution for $p\leq 0.06$, while ordered stripes of different incommensurabilities prevail for $p\geq 0.08$.
The onset of stripe order agrees with the phase diagram of Ref.~\cite{Scholle2023Comprehensive}, whereas at lower doping that work finds a noncollinear spiral ground state, which is excluded from our collinear calculations.
While the energetic proximity among different orderings points to phase competition, the near energy-degeneracy of the STs, together with their configurational multiplicity, suggests the entropic dominance of these textures in finite-temperature ensemble averages.
Moreover, as hole doping increases, the characteristic incommensurability $\delta$ of the STs, i.e., the shift of the magnetic peaks from $(\pi,\pi)$, tracks that of the lowest-energy stripe.
%

%%%%%%%%%%%%%%%%%%%%%%%%%%%%%%%%%%%%%%%%%%
%  DISCUSSION
%%%%%%%%%%%%%%%%%%%%%%%%%%%%%%%%%%%%%%%%%%

\section*{Summary and outlook}
These results show that frequency- and momentum-dependent self-energies, qualitatively similar to those obtained by nonperturbative methods, can be extracted from mean-field solutions that break translational symmetry in physically meaningful ways.
In the doped Hubbard model this striking correspondence is realized by the spin textures:
although the underlying mean-field Hamiltonian remains static and the many-electron state is a single Slater determinant, the primitive-cell representation of a spin texture describes remarkably well the spectral phenomenology of a pseudogap, displaying the celebrated nodal-antinodal dichotomy and the formation of Fermi arcs.
Importantly, these properties are not tied to one specially selected spin texture: many stationary mean-field solutions with different microscopic realizations of the real-space pattern share closely similar distributions of ordering wavevectors and primitive-cell spectra,  indicating that the relevant spectral function is a property of the broader manifold of quasi-degenerate states.

A spontaneous interpretation follows from the multiconfigurational character of the many-body wavefunction of strongly correlated states.
The exact wavefunction involves several determinants with comparable weights; individual determinants may break symmetries of the Hamiltonian, while the exact state restores them through their coherent superposition.
Mean-field theories resolve this complexity by selecting a single broken-symmetry determinant as a stationary solution.
From this perspective, the quasi-degenerate spin textures found here may be viewed as mean-field representatives of the Slater determinants underlying the correlated state.
Since distinct textures differ macroscopically, their mutual overlaps and Hamiltonian matrix elements vanish in the thermodynamic limit, so that interference between them is negligible.
When the primitive-cell spectra are already similar across several states, an ensemble average over symmetry-related and quasi-degenerate mean-field solutions then provides a simple and effective tool to approximate the expectation value of translationally invariant observables; increasing the supercell size and averaging over representative configurations gives two
complementary routes towards the thermodynamic limit.
What is missing in this picture are rather the determinants obtained from each texture by few-particle excitations, which contribute off-diagonal Hamiltonian matrix elements and account for quantum fluctuations around each mean-field solution.

The same reasoning extends beyond model Hamiltonians, and sheds light on recent results showing that mean-field-like approaches such as Kohn-Sham DFT~\cite{Kohn1965Self-consistent} succeed in reproducing some features of strongly correlated materials~\cite{Trimarchi2018Polymorphous,
%Zhang2020Competing,
Perdew2021Interpretations, Zunger2026Symmetry, Carta2026Strong}; the present approach would provide a direct route to construct dynamical momentum-dependent self-energies in addition to unfolded band structures~\cite{Popescu2010Effective, Ku2010Unfolding}.
The idea that electronic correlations can be captured through broken symmetries is also at the root of the success of Koopmans functionals, whose orbital-dependent potentials are built by adding electrons to, or removing them from, localized orbitals~\cite{Dabo2010Koopmans, Nguyen2018Koopmans}: these potentials can be viewed as a quasiparticle approximation to a frequency-dependent self-energy~\cite{Ferretti2014Bridging}, and they include partial vertex corrections beyond perturbative schemes such as GW~\cite{Colonna2019Koopmans}.

The conclusion is therefore not that strong correlations are absent, but rather that a key part of the spectral phenomenology of strongly correlated systems, which is conventionally addressed within frequency-dependent many-body approaches, admits instead an alternative representation and even explanation in terms of static real-space patterns in supercells, mapped into frequency- and momentum-dependent self-energies in the primitive cell.

%

%%%%%%%%%%%%%%%%%%%%%%%%%%%%%%%%%%%
% METHODS
%%%%%%%%%%%%%%%%%%%%%%%%%%%%%%%%%%%

\section*{Methods}
\subsection*{Momentum-space representation and unfolding}
\begin{figure}
    \centering
    \includegraphics[width=0.95\linewidth]{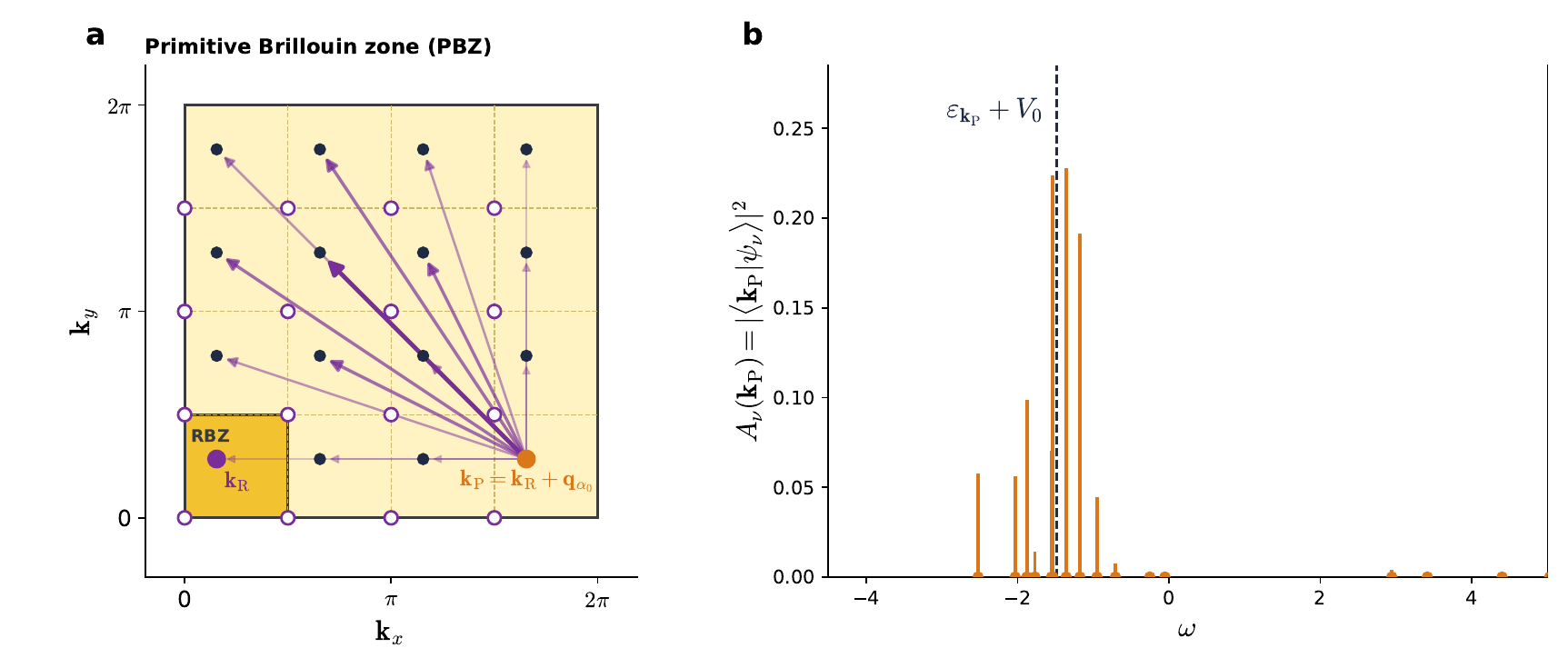}
    \caption{\textbf{Folding of the primitive Brillouin zone and fractionalization of the spectral weight.}
    \textbf{(a)} The noninteracting Hamiltonian is diagonal in the momenta $\mathbf{k}_\mathrm{P}$ of the primitive Brillouin zone (PBZ).
    A mean-field potential that is periodic in a supercell, here $4\times 4$ ($N_\mathrm{sc}=16$), folds the PBZ into the reduced Brillouin zone (RBZ) of the supercell, which tiles the PBZ $N_\mathrm{sc}$ times (dashed lines); open circles mark the reciprocal-lattice vectors of the supercell, i.e. the folding vectors $\mathbf{q}_\alpha$.
    Every PBZ momentum can be written uniquely as $\mathbf{k}_\mathrm{R}+\mathbf{q}_\alpha$ with $\mathbf{k}_\mathrm{R}$ in the RBZ (purple), and the potential couples all $N_\mathrm{sc}$ momenta of this set to one another.
    The arrows show the couplings of one particular member, $\mathbf{k}_\mathrm{P}=\mathbf{k}_\mathrm{R}+\mathbf{q}_{\alpha_0}$ (orange), to its $N_\mathrm{sc}-1$ folded copies $\mathbf{k}_\mathrm{P}+\mathbf{q}$, with width and opacity proportional to $|V_{\mathbf{q}}|$.
    For the Hartree--Fock Hubbard model used throughout ($U=3t$, $t'=-0.3t$), $V_{\mathbf{q}}$ is set by the Fourier components of the spin-resolved density; in this illustration the magnetization is a Gaussian distribution of components centred at $(\pi,\pi)$, so that the coupling is strongest for $\mathbf{Q}=(\pi,\pi)$ and decreases away from it.
    \textbf{(b)} Spectral function at $\mathbf{k}_\mathrm{P}$ obtained from Eq.~\eqref{eq:method_mero_green}: a sum of $N_\mathrm{sc}$ delta functions located at the eigenvalues $E_\nu$ of the block, with weights $A_\nu(\mathbf{k}_\mathrm{P})=|\langle\mathbf{k}_\mathrm{P}|\psi_\nu\rangle|^2$ that sum to one. The dashed line marks the bare level $\varepsilon_{\mathbf{k}_\mathrm{P}}+V_0$ of the uncoupled momentum. The eigenvalues $E_\nu$ are the same for all $N_\mathrm{sc}$ momenta of the set in panel (\textbf{a}); choosing a different $\mathbf{k}_\mathrm{P}$ changes only how the unit spectral weight is distributed among them. The spreading of the weight of a single bare level over many poles is what, in large supercells, produces the broadening and the satellites of the translationally invariant Green's function.}
    \label{fig:folding-map}
\end{figure}

We consider a noninteracting one-band Hamiltonian $\hat{h}_0 = \sum_{\mathbf{k}_\mathrm{P}}\varepsilon_{\mathbf{k}_\mathrm{P}}\hat{c}^{\dagger}_{\mathbf{k}_\mathrm{P}}\hat{c}_{\mathbf{k}_\mathrm{P}}$ where the operator $\hat{c}^{\dagger}_{\mathbf{k}_\mathrm{P}}$ creates an electron of crystal quasimomentum $\mathbf{k}_\mathrm{P}$ in the primitive Brillouin zone (PBZ).
For clarity of notation we omit the spin index $\sigma$, but in the following all the operators should be assumed as spin resolved.
We then treat the electron-electron interaction at the mean-field level, via a static local potential $V_{\mathbf{r}}$ coupled to the local charge density: 
\begin{equation}
\label{eq:methods_hamiltonian_second_quantization}
    \hat{H} = \hat{h}_0 + \hat{h}_\mathrm{int} = \sum_{\mathbf{k}_\mathrm{P}} \varepsilon_{\mathbf{k}_\mathrm{P}} \hat{c}^{\dagger}_{\mathbf{k}_\mathrm{P}}\hat{c}_{\mathbf{k}_\mathrm{P}} + \sum_{\mathbf{r}} V_{\mathbf{r}}\hat{c}^{\dagger}_{\mathbf{r}}\hat{c}_{\mathbf{r}},
\end{equation}
where the field operators transform as 
$\hat{c}_{\mathbf{r}} = 1/\sqrt{N_{\mathbf{k}_\mathrm{P}}} \sum_{\mathbf{k}_\mathrm{P}}e^{i\mathbf{k}_\mathrm{P}\cdot\mathbf{r}}\hat{c}_{\mathbf{k}_\mathrm{P}}$ and
$V_{\mathbf{r}}$ is allowed to break the translational symmetry of the lattice, being however periodic in a supercell containing $N_\mathrm{sc}$ repetitions of the primitive cell.
Its discrete Fourier expansion is defined on $N_\mathrm{sc}$ allowed folding vectors $\mathbf{q}_{\alpha}$, including $\mathbf{q}_{0}=\mathbf{0}$, corresponding to the $N_\mathrm{sc}$ reciprocal lattice vectors of the supercell lying in the PBZ:
\begin{equation}
    V_{\mathbf{r}} = \sum_{\mathbf{q}_\alpha} V_{\mathbf{q}_\alpha} e^{-i\mathbf{q}_\alpha \cdot \mathbf{r}},
\end{equation}
with the condition $V_{\mathbf{-q}_\alpha} = V_{\mathbf{q}_\alpha}^*$ to have a real-valued potential in real-space. 
The interaction term then becomes
\begin{equation}
\label{eq:methods_int_ham}
    \hat{h}_\mathrm{int} = \sum_{\mathbf{k}_\mathrm{P}}\sum_{\mathbf{q}_\alpha} V_{\mathbf{q}_\alpha} \hat{c}^{\dagger}_{\mathbf{k}_\mathrm{P}} \hat{c}_{\mathbf{k}_{\mathrm{P}}+\mathbf{q_{\alpha}}},
\end{equation}
coupling momenta $\mathbf{k}_\mathrm{P}$ of the PBZ differing by a folding vector $\mathbf{q}_\alpha$; as a consequence the Hamiltonian of Eq.~\eqref{eq:methods_hamiltonian_second_quantization} is off-diagonal in the Bloch basis of the primitive cell.
With a fixed convention for the zone boundaries, every PBZ momentum admits a unique representation $\mathbf{k}_\mathrm{P} = \mathbf{k}_{\mathrm{R}}+\mathbf{q}_\alpha$, where $\mathbf{k}_\mathrm{R}$ is a momentum in the reduced Brillouin zone (RBZ) of the supercell, as depicted in Fig.~\ref{fig:folding-map}~(\textbf{a}); sums of momenta are understood modulo a primitive reciprocal-lattice vector.
Substituting $\mathbf{k}_\mathrm{P}=\mathbf{k}_\mathrm{R}+\mathbf{q}_\alpha$, and using that, for fixed $\mathbf{q}_\alpha$, the vectors $\mathbf{q}_\alpha+\mathbf{q}_\beta$ run over all folding vectors, the Hamiltonian of Eq.~\eqref{eq:methods_hamiltonian_second_quantization} becomes, after relabeling indices:
\begin{equation}
    \hat{H}
    =
    \sum_{\mathbf{k}_\mathrm{R}}\sum_{\mathbf{q}_\alpha,\mathbf{q}_\beta}
    \left(
    \varepsilon_{\mathbf{k}_\mathrm{R}+\mathbf{q}_\alpha}\,\delta_{\alpha\beta}
    + V_{\mathbf{q}_\beta-\mathbf{q}_\alpha}\right)
    \hat{c}_{\mathbf{k}_\mathrm{R}+\mathbf{q}_\alpha}^{\dagger}\,
    \hat{c}_{\mathbf{k}_\mathrm{R}+\mathbf{q}_\beta},
\end{equation}
which is block-diagonal in $\mathbf{k}_\mathrm{R}$. 
Therefore, for a given RBZ momentum ${\mathbf{k}_\mathrm{R}}$, the matrix elements of the supercell Hamiltonian can be compactly written in the $\{\mathbf{k}_\mathrm{R}+\mathbf{q}_\alpha\}$ basis as
\begin{equation}
\label{eq:hamiltonian_extended}
    H_{\mathbf{q}_\alpha\mathbf{q}_\beta}({\mathbf{k}_\mathrm{R}}) = \delta_{\alpha\beta}\,\varepsilon_{{\mathbf{k}_\mathrm{R}}+\mathbf{q}_\alpha} + V_{\mathbf{q}_\beta-\mathbf{q}_\alpha}.
\end{equation}
Selecting a specific PBZ momentum \(\mathbf k_\mathrm{P}={\mathbf k}_\mathrm{R}+\mathbf
q_{\alpha_0}\), we relabel the set of the remaining states by \(\mathbf k_\mathrm{P}+\mathbf
q\), where \(\mathbf q\) ranges over \(\mathbf q_\beta-\mathbf q_{\alpha_0}\ne0\),
modulo primitive reciprocal-lattice vectors.
The Hamiltonian can then be partitioned into an upper-left $1\times 1$ block for $\ket{\mathbf{k}_\mathrm{P}}$, a lower-right block for the folded states $\ket{\mathbf{k}_\mathrm{P}+\mathbf{q}}$, and the coupling row and column between them.
This partition leads to the following form:
\begin{equation}
\label{eq:methods_block_matrix}
    \begin{pmatrix}
        \varepsilon_{\mathbf{k}_\mathrm{P}}+V_{\mathbf{0}} & V_{\mathbf{q}_1} & V_{\mathbf{q}_2} & \dots & V_{\mathbf{q}_{N_\mathrm{sc}-1}}\\
        V_{-\mathbf{q}_1} & \varepsilon_{\mathbf{k}_{\mathrm{P}}+\mathbf{q}_1}+V_{\mathbf{0}} & V_{\mathbf{q}_2-\mathbf{q}_1} & \dots & V_{\mathbf{q}_{N_\mathrm{sc}-1}-\mathbf{q}_1}\\
        V_{-\mathbf{q}_2} & V_{\mathbf{q}_1-\mathbf{q}_2} & \varepsilon_{\mathbf{k}_\mathrm{P}+\mathbf{q}_2}+V_{\mathbf{0}} & \ddots &V_{\mathbf{q}_{N_\mathrm{sc}-1}-\mathbf{q}_2}\\
        \vdots & \vdots & \ddots & \ddots & \vdots\\
        V_{-\mathbf{q}_{N_\mathrm{sc}-1}} & V_{\mathbf{q}_1-\mathbf{q}_{N_\mathrm{sc}-1}} & V_{\mathbf{q}_2-\mathbf{q}_{N_\mathrm{sc}-1}} & \dots & \varepsilon_{\mathbf{k}_\mathrm{P}+\mathbf{q}_{N_\mathrm{sc}-1}}+V_{\mathbf{0}}
         
    \end{pmatrix} = 
    \begin{pmatrix}
        \varepsilon_{\mathbf{k}_\mathrm{P}} +V_{\mathbf{0}} & {V}_{\mathbf{q}}\\
        {V}^{\dagger}_{\mathbf{q}} & {H}_{\mathbf{k_\mathrm{P}+q}}
    \end{pmatrix},
\end{equation}
which we write compactly as a two-by-two block matrix on the right-hand side of Eq.~\eqref{eq:methods_block_matrix}. 
Here, $V_{\mathbf{0}}$ is the translationally invariant component of the mean-field potential and the upper-right coupling row, written compactly as $V_{\mathbf{q}}$, has entries \(V_{{\mathbf{q}_\beta} -\mathbf q_{\alpha_0}}\), while the
lower-left column, written compactly as $V^\dagger_{\mathbf{q}}$ has entries \(V_{-(\mathbf{q}_\beta-\mathbf q_{\alpha_0})}=V_{\mathbf{q}_\beta-\mathbf
q_{\alpha_0}}^{*}\).
The sector of the folded $\ket{\mathbf{k}_\mathrm{P}+\mathbf{q}}$ states is written as  $H_{\mathbf{k_\mathrm{P}+q}}$; this shifts all the noninteracting eigenenergies by $V_\mathbf{0}$ and includes couplings among them. 
For a one-band model with a local interaction potential, the couplings are the same for all the momenta $\mathbf{k}_\mathrm{P}$ of the primitive cell. 
The $\mathbf{k}_\mathrm{P}$-diagonal component of the supercell Green's function, having the same translational symmetry of the lattice, obeys a Dyson equation $G(\mathbf{k}_\mathrm{P},\omega) = [\omega - \varepsilon_{\mathbf{k}_\mathrm{P}}-V_{\mathbf{0}} - \Sigma(\mathbf{k}_\mathrm{P},\omega)]^{-1}$ in terms of a momentum- and frequency-dependent embedding self-energy obtained by eliminating the folded sector by a Schur complement:
\begin{equation}
\label{eq:emb_self_energy}
    \Sigma(\mathbf{k}_\mathrm{P},\omega) = \bra{\mathbf{k}_\mathrm{P}}{V}_{\mathbf{q}} \left(\omega - {H}_{\mathbf{k_\mathrm{P}+q}}\right)^{-1}{V}^\dagger_{\mathbf{q}}\ket{\mathbf{k}_\mathrm{P}}.
\end{equation}
Notably, since the Hamiltonian of Eq.~\eqref{eq:hamiltonian_extended} is Hermitian, the self-energy naturally inherits a dynamical structure having real poles and Hermitian nonnegative residues
\begin{equation}
\label{eq:mero_sel_energy}
    \Sigma(\mathbf{k}_\mathrm{P},\omega) = \sum_{s}\frac{|v_{s}(\mathbf{k}_\mathrm{P})|^2}{\omega - \Omega_{s}(\mathbf{k}_\mathrm{P})},\quad\quad v_{s}(\mathbf{k}_\mathrm{P})
    =
    \sum_{\mathbf{q}_\alpha\neq 0}V_{\mathbf{q}_\alpha}\braket{\mathbf{k}_\mathrm{P}+\mathbf{q}_\alpha}{s\mathbf{k}_\mathrm{P}}
\end{equation}
where $\Omega_{s}(\mathbf{k}_\mathrm{P})$ and $\ket{s\mathbf{k}_\mathrm{P}}$ are the eigenvalues and eigenvectors of ${H}_{\mathbf{k_\mathrm{P}+q}}$.
In our calculations the self-energy is computed by diagonalizing the folded sector $H_{\mathbf{k}_\mathrm{P}+\mathbf{q}}$ and combining its eigenvalues and eigenvectors as in Eq.~\eqref{eq:mero_sel_energy}.
The Green's function obtained from the Dyson equation involving the self-energy of Eq.~\eqref{eq:emb_self_energy} corresponds to the unfolded Green's function obtained by projecting the supercell resolvent in the Bloch basis of the primitive cell
\begin{equation}
    G(\mathbf{k}_\mathrm{P},\omega) = \bra{\mathbf{k}_\mathrm{P}} \left(\omega  - \hat{H}\right)^{-1}\ket{\mathbf{k}_\mathrm{P}},
\end{equation}
which can also be expressed as a meromorphic function with real poles and nonnegative residues:
\begin{equation}
\label{eq:method_mero_green}
    G(\mathbf{k}_\mathrm{P},\omega) = \sum_{\nu} \frac{|A_\nu(\mathbf{k}_\mathrm{P})|^2}{\omega-E_{\nu}(\mathbf{k}_\mathrm{P})},
    \quad\quad
    A_\nu(\mathbf{k}_\mathrm{P})=\braket{\mathbf{k}_\mathrm{P}}{\psi_\nu(\mathbf{k}_\mathrm{P})},
\end{equation}
where $E_\nu(\mathbf{k}_\mathrm{P})$ and $\ket{\psi_\nu(\mathbf{k}_\mathrm{P})}$ are the eigenvalues and eigenvectors of the Hamiltonian of Eq.~\eqref{eq:methods_block_matrix}. 
The Green's function of Eq.~\eqref{eq:method_mero_green} has the usual structure obtained in unfolding methods~\cite{Ku2010Unfolding, Popescu2010Effective}.
Since the eigenvectors $\ket{\psi_\nu(\mathbf{k}_\mathrm{P})}$ of the Hamiltonian in Eq.~\eqref{eq:methods_block_matrix} are orthonormal and form a complete basis, the unfolded weights sum to unity $\sum_\nu|A_\nu(\mathbf{k}_\mathrm{P})|^2=1$.
The corresponding unfolded spectral function is given by
\begin{equation}
    A(\mathbf{k}_\mathrm{P},\omega) = -\frac{1}{\pi}\lim_{\eta\rightarrow0^+} \,\mathrm{Im} \, G(\mathbf{k}_\mathrm{P},\omega +i\eta)=\sum_{\nu}|A_\nu(\mathbf{k}_\mathrm{P})|^2\delta(\omega-E_\nu(\mathbf{k}_\mathrm{P})),
\end{equation}
where a finite broadening $\omega+i\eta$ is considered to plot the spectral function, with $\eta$ a small positive value.
The integral of the spectral function over the momenta of the PBZ reconstructs the supercell density of states, while the expectation values of translationally invariant operators, i.e., operators that are diagonal in the Bloch basis of the primitive cell $\hat{O}=\sum_{\mathbf{k}_\mathrm{P}}O_{\mathbf{k}_\mathrm{P}}\hat{c}^{\dagger}_{\mathbf{k}_\mathrm{P}}\hat{c}_{\mathbf{k}_\mathrm{P}}$ can be evaluated directly from $A(\mathbf{k}_\mathrm{P},\omega)$ as
\begin{equation}
    \langle{\hat{O}}\rangle =\sum_{\mathbf{k}_\mathrm{P}}\int d\omega \,O_{\mathbf{k}_\mathrm{P}} A(\mathbf{k}_\mathrm{P},\omega) f(\omega)
\end{equation}
where $f(\omega)$ is the Fermi-Dirac distribution.
The Galitskii--Migdal expression
\begin{equation}
\label{eq:methods-galitskii-migdal}
    E = \frac{1}{2}\sum_{\mathbf{k}_\mathrm{P}} \int d\omega \left(\omega + \varepsilon_{\mathbf{k}_\mathrm{P}}\right) A(\mathbf{k}_\mathrm{P},\omega) f(\omega)
\end{equation}
therefore takes the same value whether evaluated with the supercell or with the unfolded Green’s function, since it only involves terms which are traces of operators diagonal in the Bloch basis of the primitive cell.

\subsection*{Self-consistent Hartree-Fock calculations}
The unrestricted Hartree-Fock Hamiltonian of the two-dimensional Hubbard model on the square lattice reads, for a given spin channel $\sigma$:
\begin{equation}
\label{eq:hf_ham_hub_2d}
    \hat{H}_\sigma = \sum_{\mathbf{k}}\varepsilon_{\mathbf{k}}\hat{c}^{\dagger}_{\mathbf{k}\sigma}\hat{c}_{\mathbf{k}\sigma} + U\sum_{\mathbf{k}}\sum_{\mathbf{q}}n_{\mathbf{q}\bar{\sigma}}\hat{c}^{\dagger}_{\mathbf{k}\sigma}\hat{c}_{\mathbf{k+q}\sigma},
\end{equation}
with $\mathbf{k},\mathbf{q}$ as in Eq.~\eqref{eq:methods_int_ham} and where $\bar{\sigma}$ is the opposite spin channel and, for a supercell of $N_{\mathrm{sc}}$ sites, $n_{\mathbf{q}\sigma}$ is the Fourier transform of the spin-resolved charge density as $n_{\mathbf{q}\sigma}=1/N_{\mathrm{sc}} \sum_{\mathbf{r}}\langle \hat{n}_{\mathbf{r}\sigma}\rangle e^{i\mathbf{q}\cdot \mathbf{r}}$.
The interaction term has the same structure as Eq.~\eqref{eq:methods_int_ham}, where the couplings are identified as $V_{\mathbf{q}\sigma}=Un_{\mathbf{q}\bar{\sigma}}$.
As a representative parameter set to study the solutions of the doped model, we focus on a moderate value of the interaction strength $U=3t$, where $t$ is the nearest-neighbor hopping. 
This value of $U$ is appropriate for mean-field approaches, where the absence of dynamical screening would otherwise lead to an excessive overestimation of gap scales~\cite{Scholle2023Comprehensive}.
At the same time, it remains large enough to stabilize nontrivial broken-symmetry textures upon hole doping.
We also include a next-nearest-neighbor hopping along the lattice diagonals, setting $t'=-0.3t$, a value in the range commonly used in cuprate-motivated Hubbard models~\cite{Scholle2023Comprehensive}, which partially frustrates the long-range AFM ordering, preventing perfect nesting of the ordering vector at the Fermi surface.
The noninteracting dispersion reads 
\\$\varepsilon_{\mathbf{k}}=-2t(\cos k_x + \cos k_y) -4t' \cos k_x \cos k_y$.

We solve self-consistently the unrestricted Hartree-Fock equations for the Hubbard model in periodic supercells.
The supercell Hamiltonian is diagonal in the momenta ${\mathbf{k}_\mathrm{R}}$ of the reduced Brillouin zone (RBZ).
The momentum sampling is defined from a uniform grid of $320\times 320$ momenta $\mathbf{k}$ in the primitive Brillouin zone (PBZ).
For the $40\times 40$ supercells used in the converged calculations, this mesh is represented by a $8\times 8$ mesh in the RBZ where each momentum ${\mathbf{k}_\mathrm{R}}$ is associated to the $40 \times 40$ folded momenta ${\mathbf{k}_\mathbf{R}}+\mathbf{q}$ of the PBZ.
Starting from an initial spin-resolved density $n_{\mathbf{\mathbf{r}}\sigma}$, the Hartree-Fock Hamiltonian of Eq.~\eqref{eq:hf_ham_hub_2d} is diagonalized for each spin and RBZ momenta.
The output density at the {\it{i}}-th iteration is reconstructed as
\begin{equation}
    n^{\mathrm{out},i}_{\mathbf{r}\sigma} = \frac{1}{N_{{\mathbf{k}_\mathrm{R}}}} \sum_{\nu{\mathbf{k}_\mathrm{R}}} |u_{\nu{\mathbf{k}_\mathrm{R}}\sigma}(\mathbf{r})|^2 f(E_{\nu\sigma}({\mathbf{k}_\mathrm{R}})),
\end{equation}
where $E_{\nu\sigma}({\mathbf{k}_\mathrm{R}})$ and $u_{\nu{\mathbf{k}_\mathrm{R}}\sigma}(\mathbf{r})$ are respectively the eigenvalues and supercell-periodic part of the eigenstates of the supercell Hamiltonian, and $f(E_{\nu\sigma}({\mathbf{k}_\mathrm{R}}))$ is the Fermi-Dirac occupation function.
The results reported in the paper are obtained at zero temperature.
The chemical potential is computed at each iteration to enforce the conservation of the total particle number $\sum_{\mathbf{r}\sigma}n_{\mathbf{r}\sigma} = N_\mathrm{e}$.
To stabilize the self-consistent cycle we adopt a linear mixing procedure for the charge density $n^{i+1}_{\mathbf{r}\sigma} = \alpha n^{\mathrm{out},i}_{\mathbf{r}\sigma} + (1-\alpha)n^i_{\mathbf{r}\sigma}$.
The value of $\alpha=0.25$ has been used for calculations; these are considered converged when all Fourier components of the spin-resolved density change less than a given threshold 
\begin{equation}
    \max_{\mathbf{q},\sigma} |n_{\mathbf{q}\sigma}^{i} - n_{\mathbf{q}\sigma}^{i-1}| < \epsilon.
\end{equation}
In the results reported here we have considered $\epsilon=10^{-4}$.
In the case of the doped model convergence can require between 1000 and 2000 iterations.
The solver for the mean-field equations will converge to different stationary states depending on the initial configuration of the charge density.
Ordered states such as the antiferomagnet or long-range ordered stripe solutions are obtained by initializing the charge density with the corresponding pattern.
The spin texture solutions are instead obtained initializing the density in both spin channels with a uniform random distribution centered around the target mean-value of the density $n_0$ and with a width of $n_0/2$.
The effect of disorder is only included at the level of the initialization of the charge density, and the Hamiltonian contains no quenched disorder coming from external potentials.
Different spin textures are obtained from different seeds of the random number generator.
The total energy per site of each stationary solution is given by
\begin{equation}
\label{eq:method-hf-total-energy}
    E = \frac{1}{N_{{\mathbf{k}_\mathrm{R}}}N_\mathrm{sc}}\sum_{\nu\sigma{\mathbf{k}_\mathrm{R}}} E_{\nu\sigma}(\mathbf{{k}_\mathrm{R}})f(E_{\nu\sigma}({\mathbf{k}_\mathrm{R}})) - \frac{U}{N_{\mathrm{sc}}}\sum_{\mathbf{r}}n_{\mathbf{r}\uparrow}n_{\mathbf{r}\downarrow}.
\end{equation}
For the Hartree–Fock potential of Eq.~\eqref{eq:hf_ham_hub_2d}, the Galitskii--Migdal expression in Eq.~\eqref{eq:methods-galitskii-migdal} coincides with the total energy of Eq.~\eqref{eq:method-hf-total-energy}.
The spectral functions reported in Fig.~\ref{fig:gf-se-pseudogap} and Fig.~\ref{fig:panel-surf-pg} are given by an ensemble average, with equal weights, of the spectral functions of the different spin textures.
In the ensemble average we also symmetrize the spectra applying the operations of the symmetry group of the square lattice $D_4$. 
This comes from the fact that spin textures obtained from one another by applying a symmetry operation of the underlying Bravais lattice share the same energy, and therefore contribute equally to the ensemble average.
Moreover the spectral function reported is obtained summing over the spin index the spectral functions of the single spin textures.
A broadening $\eta=0.015\,t$ is used to plot the spectral functions in Fig.~\ref{fig:gf-se-pseudogap} and Fig.~\ref{fig:panel-surf-pg}.

% ================
% Acknowledgements
% ================
\subsection*{Acknowledgements}
We would like to thank G. Caldarelli, M. Caserta, S. Cigagna, A. Polyukhin, N. Colonna and especially A. Carta for many insightful discussions. 

% ====================
% Author contributions
% ====================
\subsection*{Author contributions}
N.M. conceptualized the idea motivating this work and supervised the project. 
M.Q. developed the theoretical methodology and performed the calculations.
M.Q., M.C. and N.M. analyzed the results and wrote the first draft of the manuscript. 
All authors participated in the review, discussion, and editing of the manuscript.

% =======
% Funding
% =======
\subsection*{Funding}
M.Q., A.P., T.C., and N.M. acknowledge financial support from the Swiss National Science Foundation (SNSF -- project number 213082). M.C. acknowledges financial support from the National Recovery and Resilience Plan PNRR MUR Project No. CN00000013-ICSC and MUR Project No. PE0000023-NQSTI.

% ===================
% Competing interests
% ===================
\subsection*{Competing interests}
The authors declare no competing interests.

%%%%%%%%%%%%%%%%%%%%%%%%%%%%%%%%%%%%%%%%%%%%%%%%%
% BIBLIOGRAPHY
%%%%%%%%%%%%%%%%%%%%%%%%%%%%%%%%%%%%%%%%%%%%%%%%%


\clearpage

% ====================
% SUPPLEMENTARY INFORMATION STARTS HERE
% ====================

\clearpage
\setcounter{page}{1}
\setcounter{section}{0}
\setcounter{figure}{0}
\setcounter{table}{0}
\setcounter{equation}{0}
\renewcommand{\thesection}{S\arabic{section}}
\renewcommand{\thefigure}{S\arabic{figure}}
\renewcommand{\thetable}{S\arabic{table}}
\renewcommand{\theequation}{S\arabic{equation}}
% hyperref anchors must be unique as well (they use \theH..., not \the...)
\renewcommand{\theHsection}{S\arabic{section}}
\renewcommand{\theHsubsection}{S\arabic{section}.\arabic{subsection}}
\renewcommand{\theHfigure}{S\arabic{figure}}
\renewcommand{\theHtable}{S\arabic{table}}
\renewcommand{\theHequation}{S\arabic{equation}}
\renewcommand{\thepage}{S\arabic{page}}
\pdfbookmark[0]{Supplementary Information}{si-start}   % bookmark for the SI start

% ---- Supplementary title block (plain text, editable) ----
\begin{center}
  {\Large\bfseries Supplementary Information for\\[3pt]
   ``Spectral signatures of spin textures and the emergence of a pseudogap''\par}
  \vspace{1.2em}
  {\large Matteo Quinzi$^{1}$, Andrea Pintus$^{1}$, Tommaso Chiarotti$^{2}$,
   Massimo Capone$^{3,4}$, Nicola Marzari$^{1,5,6,*}$\par}
  \vspace{0.9em}
  {\small
   $^{1}$Theory and Simulation of Materials (THEOS), and National Centre for Computational Design and
   Discovery of Novel Materials (MARVEL), \'Ecole Polytechnique F\'ed\'erale de Lausanne, Lausanne, 1015, Switzerland.\\
   $^{2}$Department of Applied Physics and Materials Science, California Institute of Technology, Pasadena, 91125, California, USA.\\
   $^{3}$SISSA, Scuola Internazionale Superiore di Studi Avanzati, Trieste, 34136, Italy.\\
   $^{4}$CNR-IOM, Istituto Officina dei Materiali, Trieste, 34136, Italy.\\
   $^{5}$PSI Center for Scientific Computing, Theory and Data, Paul Scherrer Institute, Villigen, 5232, Switzerland.\\
   $^{6}$Theory of Condensed Matter, Cavendish Laboratory, University of Cambridge, Cambridge, CB3 0US, United Kingdom.\\[4pt]
   $^{*}$Corresponding author. E-mail: nm901@cam.ac.uk\par}
\end{center}
\vspace{1em}

\tableofcontents
\clearpage

%%%%%%%%%%%%%%%%%%%%%%%%%%%%%%%%%%%%%%%
%  HALF-FILLED HUBBARD MODEL
%%%%%%%%%%%%%%%%%%%%%%%%%%%%%%%%%%%%%%%

\section{Mapping real-space mean-field solutions into frequency-dependent self-energies}
\label{sec:hubbard}
\begin{figure}
    \centering
    \includegraphics[width=0.95\linewidth]{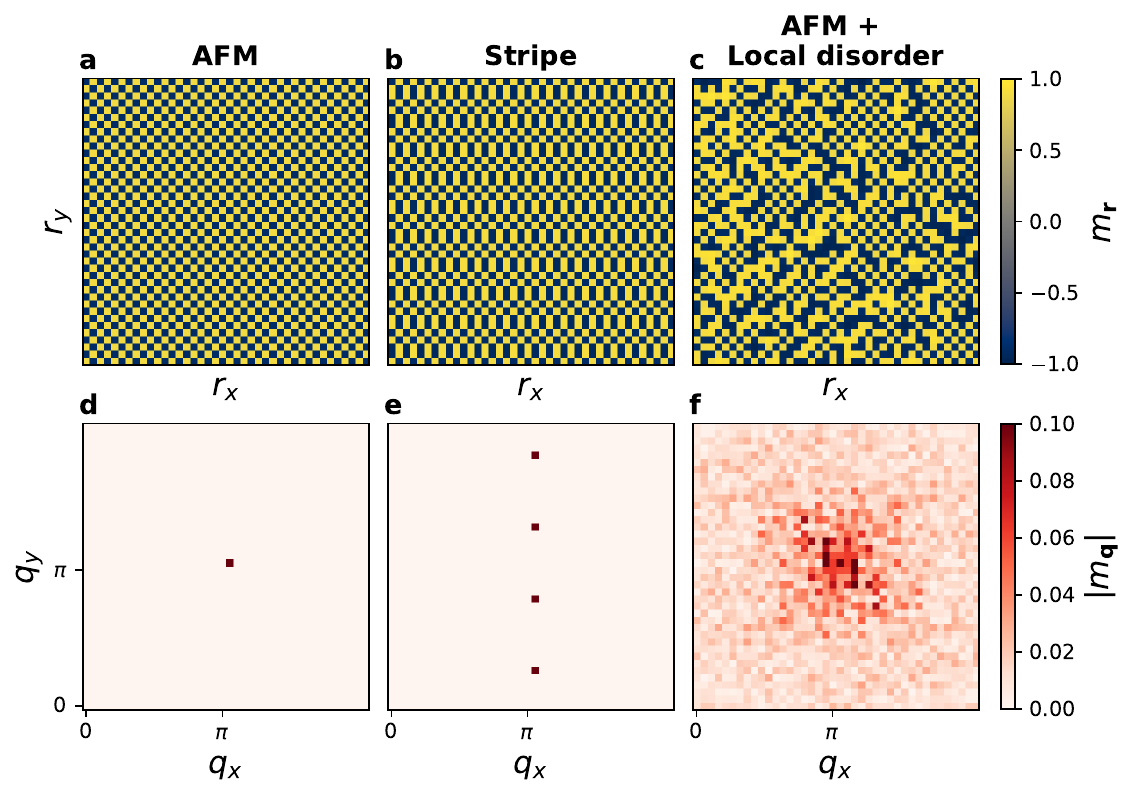}
    \caption{
    Symmetry-broken mean-field solutions obtained for the half-filled and particle-hole symmetric Hubbard model on the square lattice with $U=8t$ in a $40\times 40$ supercell.
    Upper row: real space pattern of the magnetization $m_{\mathbf{r}}=n_{\mathbf{r}\uparrow} - n_{\mathbf{r}\downarrow}$ for different symmetry-broken solutions.
    Lower row: distribution of the absolute value of the Fourier components of the magnetization in reciprocal space.
    \textbf{(a,d)}~Antiferromagnetic solution. The magnetic order parameter is characterized by a single delta peak at $(\pi,\pi)$.
    \textbf{(b,e)}~Stripe solution with a spatial repetition pattern of $2\times8$. The magnetic order parameter is now characterized by four isolated delta peaks centered at $(\pi,\frac{\pi}{4})$, $(\pi,\frac{3\pi}{4})$, $(\pi,\frac{5\pi}{4})$ and $(\pi,\frac{7\pi}{4})$.
    \textbf{(c,f)}~Antiferromagnetic background with local magnetic disorder, converged from an initial random initialization of the spin-resolved charge density. The magnetic order parameter is here characterized by a broad and dense distribution centered around $(\pi,\pi)$, compatible with short range AFM correlations.
    }
    \label{fig:panel-mag-fourier-q}
\end{figure}
 We derive explicitly the self-energies associated with different mean-field solutions of the half-filled two-dimensional Hubbard model. Even if, for this filling, the antiferromagnetic (AFM) solution is stable, we consider different kind of real space solutions in order to gain insight on the consequences of the mapping between real-space static potentials and dynamical self-energies.

%The single-band Hubbard model is widely regarded as a prototypical platform to study strong electronic correlations, and its two-dimensional version is believed to be central to high-temperature superconductors~\cite{Qin2022Computational, Arovas2022Hubbard}.
%
%In the context of this work, it also provides us with one of the simplest settings where the consequences of this map between static symmetry-broken potentials and dynamical self-energies can be made fully explicit. 
%
In this section we consider only nearest-neighbour hopping, so that  the noninteracting energy dispersion is given by $\varepsilon_\mathbf{k}=-2t\,(\cos k_x +\cos k_y)$ and the local Hubbard interaction can be written as a sum of processes where an electron with spin $\sigma$ scatters off a density modulation of opposite spin $\bar{\sigma}$ carrying momentum $\mathbf{q}$, with a strength set to $U$:
\begin{equation}
    \hat{h}_{\mathrm{int}}^{\mathrm{Hub}} = U\sum_{\mathbf{r}}\hat{n}_{\mathbf{r}\bar{\sigma}} \hat{n}_{\mathbf{r}\sigma} = U\sum_{\mathbf{q}}\sum_\mathbf{k} \hat{n}_{\mathbf{q}\bar{\sigma}} \hat{c}_{\mathbf{k}\sigma}^{\dagger} \hat{c}_{\mathbf{k+q}\sigma} 
\end{equation}
Within a mean-field approximation the density operator $\hat{n}_{\mathbf{q}\sigma}$ is then traded by its expectation value $n_{\mathbf{q}\sigma}$, neglecting fluctuations.
Thereby, a nonzero component at wavevector $\mathbf{q}$
implies a supercell commensurate with the ordering vector, breaking the translational
symmetry of the lattice.
The mean-field couplings in the case of the Hubbard interaction are given by 
\begin{equation}
\label{eq:coupling-hubb}
 V_{\mathbf{q}\sigma}
 = U\,n_{\mathbf{q}\bar{\sigma}}
 = \frac{U}{2} (\rho_{\mathbf{q}}+s_{\bar{\sigma}}m_{\mathbf{q}})   
\end{equation}
where we $\rho_{\mathbf{q}} = n_{\mathbf{q}\uparrow}+n_{\mathbf{q}\downarrow}$ and $m_{\mathbf{q}}=n_{\mathbf{q}\uparrow}-n_{\mathbf{q}\downarrow}$, and $s_{\uparrow}=+1$ and $s_{\downarrow}=-1$.
Eq.~\eqref{eq:coupling-hubb} exposes the direct relation between the mean-field potential and the effective bosonic modes of the system.
Since the solutions of the half-filled model are dominated by the magnetic channel, their nature and complexity can be directly characterized by the distribution in reciprocal space of the Fourier components of the magnetization $m_{\mathbf{q}}$.
We consider the value of $U=8t$, for which different magnetic configurations can be stabilized by changing the initial spin density.
Our purpose here is not to determine their relative thermodynamic stability~\cite{Scholle2023ComprehensiveSupp}, but to show how the reciprocal-space structure of the magnetization reshapes the primitive-cell spectra.

We begin considering the antiferromagnetic (AFM) mean-field solution, which provides the simplest example of translational symmetry breaking, involving only one ordering vector $\mathbf{q}=(\pi,\pi)$ as reported in Fig.~\ref{fig:panel-mag-fourier-q}~(\textbf{a,d}), and illustrates how the embedding viewpoint in this case reduces to the familiar spin-density wave (SDW) description~\cite{Altshuler1998LuttingerSupp, Scheurer2018TopologicalSupp, Wu2017PseudogapSupp}.
%MC 

\begin{figure}
    \centering
    \includegraphics[width=0.95\linewidth]{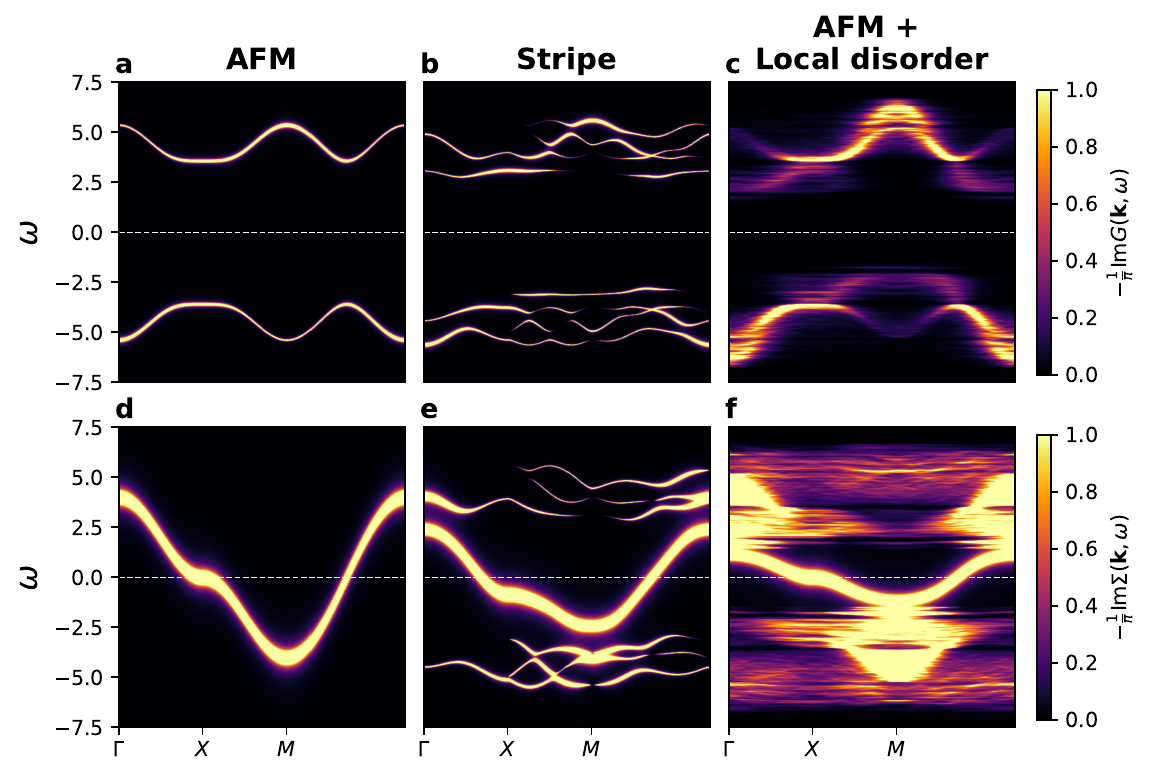}
    \caption{
    Momentum-resolved spectral functions of the translationally invariant Green's function (upper row) and self-energy (lower row), obtained by mapping the different self-consistent Hartree--Fock solutions of Fig.~\ref{fig:panel-mag-fourier-q} from the supercell to the primitive unit cell.
    The Fermi level is at $\omega=0$ (white-dashed energy).
    (\textbf{a,d}) The antiferromagnetic state is characterized by a single coherent self-energy pole, which opens a spectral gap splitting the Green's function spectrum
    into two sharp bands.
    (\textbf{b,e}) In the stripe state, several discrete, but still coherent, self-energy poles generate additional gaps, further fragmenting the upper and lower spectral branches.
    (\textbf{c,f}) The large-supercell state characterized by a finite antiferromagnetic correlation length maps instead onto dense, almost dispersionless self-energy structures in addition to the fundamental coherent pole.
    These yield broadened bands and satellites in the Green's function, recovering features typically associated with strong electronic correlations despite the static Hartree--Fock origin of the underlying state.
    }
    \label{fig:spectral-bands-q}
\end{figure}

The momentum-dependence of the poles of the Green's function and the self-energy obtained from the embedding construction discussed in the main text can be visualized from the momentum-resolved spectral functions of the two operators, which in this one-band model are just proportional to their imaginary part as shown in Fig.~\ref{fig:spectral-bands-q}~(\textbf{a,d}).
The former gives the ARPES spectrum of the system (neglecting the effect of ARPES matrix elements);
the AFM solution is already capable of opening a spectral gap, splitting the original one-particle dispersion into two separate bands, each one carrying half of the spectral weight.

We thus recover the well known result for the AFM solution, proving that a single Fourier component in the order parameter is sufficient to open a spectral gap, but it lacks any incoherence effects.
The next stage in complexity is offered by stripe order, a ubiquitous instability of the two-dimensional Hubbard model characterized by a more structured spatial pattern~\cite{White1998DensitySupp, Jiang2020StripesSupp, Zhang2020CompetingSupp, Wietek2021StripesSupp, Simkovic2024OriginSupp}.
We consider a stripe pattern with an $8\times2$ periodicity in Fig.~\ref{fig:panel-mag-fourier-q}~(\textbf{b,e}).
Instead of a single AFM wavevector, the magnetization now displays four isolated Fourier peaks, each one corresponding to a different scattering channel.
When this state is mapped back to the primitive cell, these additional channels induce other momentum-dependent splitting of the bands. 
As shown in Fig.~\ref{fig:spectral-bands-q}~(\textbf{b,e}), the coherent bands of the AFM solution split into several branches, with additional gaps opening where the folded dispersion hybridize with the noninteracting one.
As a consequence, the self-energy acquires additional poles.
While the spectrum is richer than in the AFM case, it remains essentially coherent: a finite set of isolated Fourier components delivers additional structure, without leading to broad incoherent features observed in the spectra of correlated systems.
Amplifying the complexity of the spatial pattern, a qualitatively different regime is reached when the broken-symmetry solution is no longer described by a small number of isolated Fourier components of the magnetization.
To access this limit, we initialize a $40\times40$ supercell from a random uniform distribution of the spin-resolved charge density and let the mean-field equations converge self-consistently.
The resulting state, shown in Fig.~\ref{fig:spectral-bands-q}~(\textbf{c,f}), retains an AFM background, but now hosting local magnetic disorder, with different regions of the supercell realizing inequivalent magnetic environments.
This change is most clearly emphasized in reciprocal space, where the Fourier components of the magnetization are no longer isolated peaks, as in the AFM or stripe solutions.
Instead, they form a dense distribution centered around $(\pi,\pi)$, identifying, for the model parameters used here ($U=8t$ and no second neighbor hopping), AFM correlations with a finite correlation length, which is estimated to be of a few lattice sites based on a fit of the antiferromagnetic correlation function as reported in Sec.~\ref{sec:afm-corr-length-half-filled}.
This distribution of ordering vectors is expected to become continuous in the thermodynamic limit.
%

%We also notice that a solution characterized by the presence of localized magnetic moments with vanishing AFM correlation length is a realization of the Mott-Hubbard insulator in infinite dimensions~\cite{Metzner1989Correlated, Georges1992Hubbard, Georges1996Dynamical, Schafer2015Fate}, for which a completely disordered symmetry-broken mean-field solution provides an approximate description.
%MC La commento per il momento, poi magari ci torniamo sopra

When this real-space pattern is mapped back to the primitive cell, the spectral properties change dramatically, now appearing like those obtained in beyond-mean field approaches~\cite{Dahnken2004VariationalSupp, Mai2026TwistingSupp}.
As shown in Fig.~\ref{fig:spectral-bands-q}~(\textbf{c,f}), the Green's function no longer displays only coherent features: the upper and lower bands acquire an effective broadening, accompanied by satellite structures.
The reconstructed self-energy mirrors this change, developing broad and weakly dispersive poles in addition to the fundamental one around the Fermi level.

As reported in detail in Sec.~\ref{subsec:quasi-degenerate-half-fill}, there are many quasi-degenerate states with different real-space textures; they all display similar distributions of the Fourier components of the magnetization. 
Strikingly, all these states share the same spectral features when mapped to the unit cell, proving the robustness of the spectral functions reported in Fig.~\ref{fig:spectral-bands-q}~(\textbf{c,f}) against the arbitrariness of choosing a particular symmetry-broken pattern. 
As we will discuss later, we interpret these multiplicity of quasi-degenerate states as stemming from the many-body wavefunction which in strongly correlated regimes is characterized by many symmetry-broken Slater determinants appearing here as competing stationary states of the mean-field equations.

Overall, the progression from a distribution of few and isolated to many and dense ordering vectors marks the crossover from simple long-range order to a regime characterized by the coexistence of multiple local environments, with the onset of short-range correlations.
In the former case, symmetry breaking mainly opens gaps while preserving a coherent quasiparticle dispersion.
In the latter, the dense set of scattering channels produces a correspondingly dense self-energy structure, yielding broadened bands and incoherent spectral features, which are typically recognized as signatures of strong correlation and attributed to intrinsically dynamical self-energies.
%

%%%%%%%%%%%%%%%%%%%%%%%%%%%%%%%%%%%%%%%%%%%%%
% DEGENERACY OF THE SYMMETRY-BROKEN SOLUTIONS
%%%%%%%%%%%%%%%%%%%%%%%%%%%%%%%%%%%%%%%%%%%%%

\section{Quasi-degeneracy of the mean-field solutions}
\subsection{Solutions of the half-filled model}
\label{subsec:quasi-degenerate-half-fill}
\begin{figure}[h]
    \centering
    \includegraphics[width=0.6\linewidth]{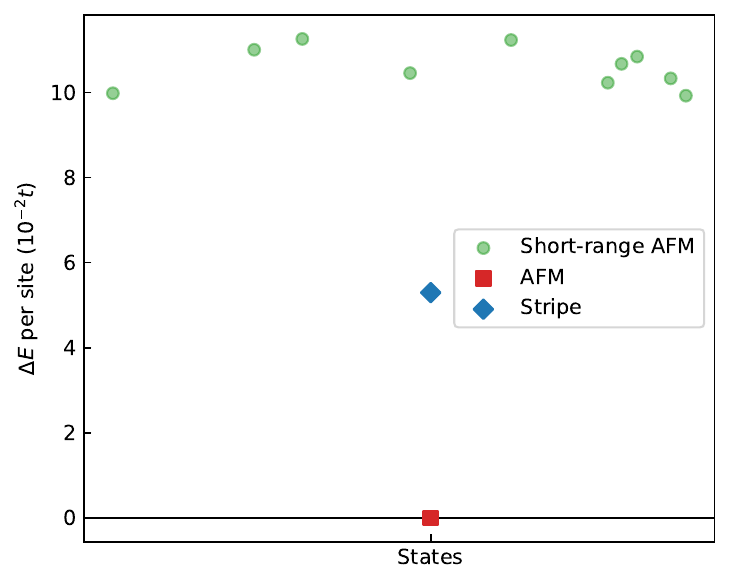}
    \caption{Energy per site differences, with respect to the AFM ground states, of the solutions reported in Tab.~\ref{tab:energy-hub-half-fill} and of the stripe pattern reported in Fig.~\ref{fig:panel-mag-fourier-q} for the half-filled and particle-hole symmetric Hubbard model at $U=8t$. 
    The set of solutions obtained from a random initialization of the charge-density constitutes a manifold of quasi-degenerate states which is well separated from the stripe and AFM solutions at lower energies. A horizontal shift is added to the set of quasi-degenerate states to better resolve their energy differences.}
    \label{fig:energy-spread-half-filled}
\end{figure}

\begin{figure}[h]
    \centering
    \includegraphics[width=0.7\linewidth]{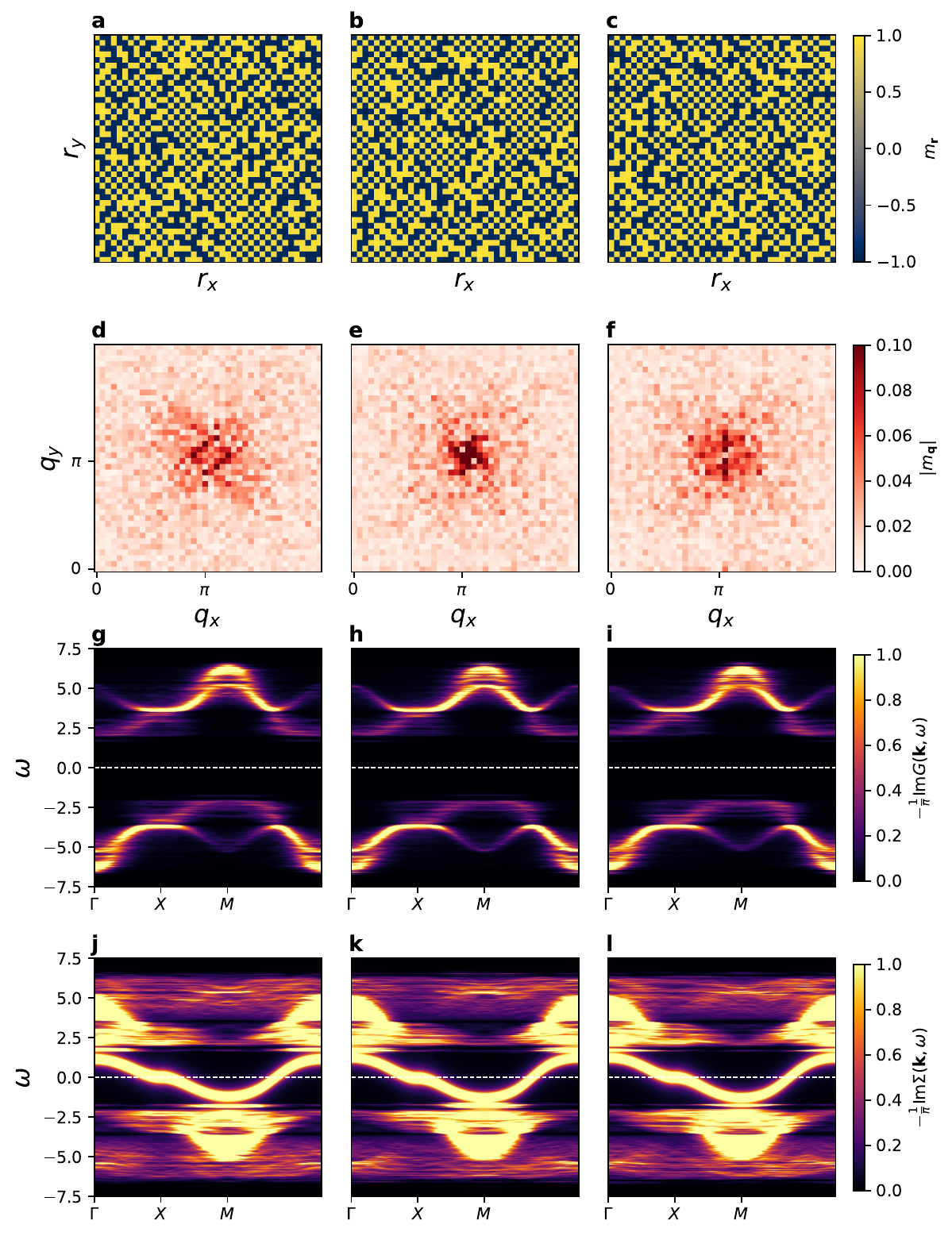}
    \caption{
    Spectral properties of the quasi-degenerate stationary states of the half-filled model.
    We report three of the symmetry-broken configurations in the set of the quasi-degenerate states of Tab.~\ref{tab:energy-hub-half-fill}.
    (\textbf{a-c}): real-space magnetization. 
    These states display different magnetization patterns that can not be related to each other via simple symmetry operations of the symmetry group of the Bravais lattice.
    (\textbf{d-f}): Fourier transform of the magnetization.
    Despite the different real-space pattern, the stationary solutions share similar distributions of the Fourier modes of the magnetization. 
    All distributions are compatible with short-range antiferromagnetic correlations.
    (\textbf{g-i}): Spectra of the Green's function along a high-symmetry path in the primitive Brillouin Zone.
    All spectral functions are characterized by a fundamental spectral gap and show closely similar incoherent features.
    (\textbf{j-l}): Spectra of the self-energy along the same high-symmetry path. All spectra are characterized by a sharp pole dispersion around the Fermi level, followed by incoherent structures at higher energies. All spectra are nearly indistinguishable.
    }
    \label{fig:panel-degeneracy-half-filled}
\end{figure}

In this section we report different mean-field solutions obtained from different initializations of the spin-resolved charge density in the case of the half-filled and particle-hole symmetric Hubbard model with $U=8t$.
The spin-resolved charge density is initialized with a uniform random distribution using different seeds for the random number generator.
The total energy per site of the different configurations is reported in Tab.~\ref{tab:energy-hub-half-fill} in units of the hopping term $t$.
The configurations examined share similar energies with a total spread between the highest and the lowest values found of $\Delta E\simeq  10^{-2}\,t$, which is small compared to the energy difference between these states and the AFM and stripe solutions.
The energy differences with respect to the AFM ground state of these solutions are reported in Fig.~\ref{fig:energy-spread-half-filled}.
This result shows that these symmetry-broken configurations constitute a set of nearly degenerate mean-field solutions, well distinguished with respect to the stripes and AFM states found at lower energies.
\begin{table*}[h]
    \centering
    \begin{tabular}{c c}
    \hline 
    Configuration & $E$ per site\\
    \hline
    AFM    &  -0.46520096 \\
    Stripe &  -0.41224339 \\
       rand 1   &  -0.35285300 \\
       rand 2   &  -0.36593888 \\
       rand 3   &  -0.36289109 \\
       rand 4   &  -0.35514332 \\
       rand 5   &  -0.35259679 \\
       rand 6   &  -0.36187406 \\
       rand 7   &  -0.36537078 \\
       rand 8   &  -0.35675145 \\
       rand 9   &  -0.35845627 \\
       rand 10  &  -0.36063181 
    \end{tabular}
    \caption{Total energy per site, in units of the hopping parameter $t$, of the symmetry-broken solutions found in the half-filled and particle-hole symmetric Hubbard model for $U=8t$.
    Calculations are performed in a $40\times 40$ supercell.
    Configurations labeled by `rand' refer to stationary mean-field solutions obtained starting from a random initialization of the spin-resolved charge density from a uniform random distribution.
    Different configurations are obtained from different seeds of the random number generator.
    These latter constitute a manifold of quasi-degenerate states with similar energies compared to the AFM and stripe solutions reported in Fig.~\ref{fig:panel-mag-fourier-q}.}
    \label{tab:energy-hub-half-fill}
\end{table*}
The spatial pattern of the magnetization and the distribution of the Fourier components is reported in the first two rows of Fig.~\ref{fig:panel-degeneracy-half-filled} for the first three random configurations of Tab.~\ref{tab:energy-hub-half-fill}.
All these solutions are characterized by different real-space patterns of the magnetization, where local antiferromagnetic domains are separated by regions of locally disordered magnetic moments.
We stress that in this case the magnetic disorder does not depend on an external potential in the Hamiltonian, but it is self-generated from the stationary Hartree-Fock solutions.
Despite the differences in real-space, the distributions of the Fourier components are however quite similar, and are all characterized by a broadened peak centered at $(\pi,\pi)$.
The last two rows of Fig.~\ref{fig:panel-degeneracy-half-filled} show the spectral functions of the Green's functions and self-energies obtained after mapping the real-space solutions of the supercell to the unit cell.
This map reveals that the spectral features of these solutions within the unit cell are basically indistinguishable from one another. 
The Green's functions are all characterized by the same spectral gap, with the band dispersion characterized by a finite broadening and satellite structures.
Complementary, the self-energy develops a main pole within the spectral gap and two broad and almost dispersionless structures in the upper and lower bands, responsible for the finite broadening of the band dispersion.
The set of quasi-degenerate stationary states of the mean-field equations maps then to the same one-particle spectral features within the unit cell.
In this sense, the spectral features obtained from the $\mathbf{k}$-diagonal Green's functions and self-energies are robust against a particular choice of the symmetry-broken configuration. 
\clearpage
\subsection{Solutions of the doped model}
\label{subsec:quasi-degenerate-doped}
We report different spin texture solutions of the hole-doped Hubbard model obtained starting from different random initialization of the spin-resolved charge density for three values of hole doping $p=0.08$, $p=0.10$ and $p=0.16$.
For $U=3t$ and $t'=-0.3t$, the spin textures upon hole doping are characterized by antiphase antiferromagnetic domains, with metallic walls at the interface.
Similarly to the half-filled case, the particular realization of this pattern is dictated by the initial random configuration. 
We track the evolution upon doping of ten different spin textures obtained from different random initialization of the spin-resolved charge density, and we compare their stability with respect to the AFM solution.
We also report the lowest energy stripe pattern whose periodicity is allowed within the $40\times 40$ supercell.
Their total energies per site are reported in Tab.~\ref{tab:energy-hub-doped} in units of the hopping parameter $t$.
\begin{table*}
    \centering
    \begin{tabular}{c ccc}
        \hline
        \multirow{2}{*}{Configuration}
        & \multicolumn{3}{c}{Energy per site} \\
        & $p=0.08$ & $p=0.10$ & $p=0.16$\\
        \hline
        AFM    & -0.98713 & -0.99386 &  -1.00644 \\
        STRIPE & -0.98764 & -0.99569 &  -1.01387 \\
        rand 1      & -0.98568 & -0.99441 &  -1.01081 \\
        rand 2      & -0.98571 & -0.99390 &  -1.01141 \\
        rand 3      & -0.98564 & -0.99332 &  -1.01038 \\
        rand 4      & -0.98568 & -0.99353 &  -1.01035 \\
        rand 5      & -0.98551 & -0.99353 &  -1.01058 \\
        rand 6      & -0.98559 & -0.99345 &  -1.01007 \\
        rand 7      & -0.98577 & -0.99325 &  -1.01049 \\
        rand 8      & -0.98554 & -0.99346 &  -1.01028 \\
        rand 9      & -0.98575 & -0.99325 &  -1.01051 \\
        rand 10     & -0.98587 & -0.99328 &  -1.01049 \\
        \hline
    \end{tabular}
    \caption{Energy per site of the different symmetry-broken stationary solutions at hole doping $p=0.08$, $p=0.10$ and $p=0.16$. The model parameters are $U=3t$ and $t'=-0.3t$.
    Calculations are performed in a $40\times 40$ supercell.
    Energies are reported in units of the hopping parameter $t$.
    We also report the energy of a homogeneous AFM state and that of the lowest energy stripe converged in the supercell. We discuss in Sec.~\ref{sec:energy-diagram} how these solutions are converged. We stress that the incommensurability of the lowest-energy stripe changes with hole doping, as also discussed in the main text.}
    \label{tab:energy-hub-doped}
\end{table*}
%
%In Fig.~\ref{fig:energy-spread-doped} we give a visualization of these relative energy differences, with respect to the ground state, within a scatter plot.
%
%\begin{figure}
%    \centering
%    \includegraphics[width=0.6\linewidth]{in_preparation/images_supp/configuration_energy_spread.pdf}
%    \caption{Relative total energy per site, with respect to the ground state, for different values of hole dopings.
%    The solutions with antiphase AFM domains are the same reported in Tab.~\ref{tab:energy-hub-doped}.
%    These solutions form a quasi-degenerate manifold whose energy spread is smaller then the typical energy difference between AFM and stripe states. A horizonatal spread is added to the antiphase domains to better resolve them visually.}
%    \label{fig:energy-spread-doped}
%\end{figure}
%
%Notably, the solutions characterized by the antiphase AFM domains have a compact spread in energy, of the order of $10^{-3}t$ for the different values of doping. 
%Moreover, their energy spread is smaller than the typical energy difference between AFM and stripe states.
%
%For $p=0.10$ we also found that one of the examined solution has an energy lower then the AFM state.
%Fig.~\ref{fig:energy-spread-doped} also shows that upon hole doping the ground state changes from AFM to stripe.
%

\begin{figure}[t]
    \centering
    \includegraphics[width=0.7\linewidth]{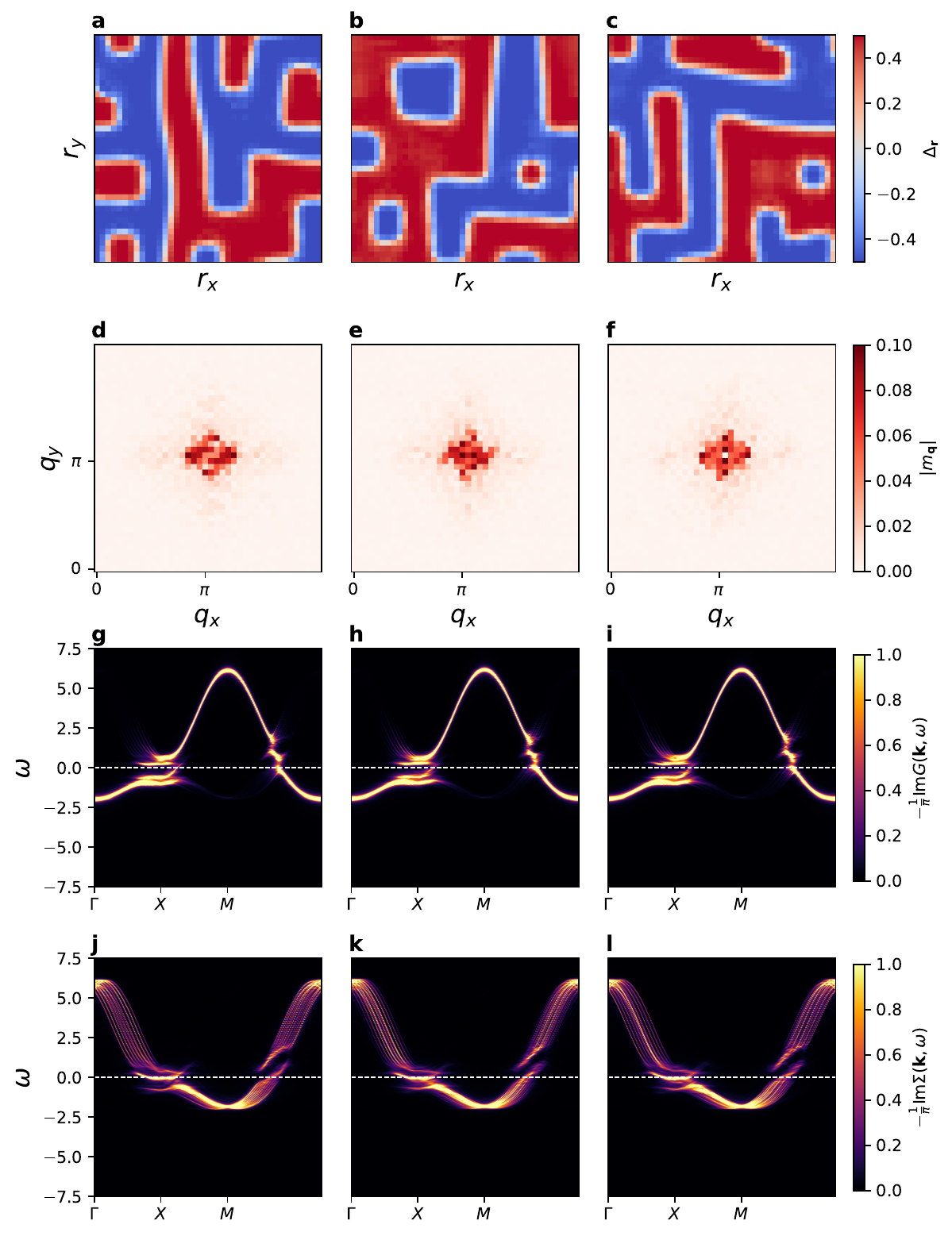}
    \caption{
    Spectral properties of the quasi-degenerate stationary states of the hole-doped model.
    We report three representative spin textures out of the set of the quasi-degenerate states of Tab.~\ref{tab:energy-hub-doped}.
    The states are selected for hole doping $p=0.08$.
    (\textbf{a-c}): real-space Néel parameter. 
    These states are characterized by different arrangements of the AFM domains and of the interfaces, realizing different spin textures.
    (\textbf{d-f}): Fourier transform of the magnetization.
    Despite the different real-space pattern, the stationary solutions share similar distributions of the Fourier modes of the magnetization. 
    All distributions are peaked around the AFM ordering vector $(\pi,\pi)$ with a finite broadening.
    (\textbf{g-i}): spectra of the Green's function along a high-symmetry path in the primitive Brillouin Zone.
    All spectra display the nodal-antinodal dichotomy, with in-gap states appearing in the nodal sector, while the antinodal one remains gapped.
    (\textbf{j-l}): spectra of the self-energy along the same high-symmetry path. In all cases the same dichotomy is reflected in the self-energies, which display a nearly dispersionless pole in the antinodal sector while the nodal dispersion has a gap in the pole structure.
    }
    \label{fig:panel-degeneracy-doped}
\end{figure}

We show in Fig.~\ref{fig:panel-degeneracy-doped} (first and second rows) the real-space pattern of the Néel parameter together with the Fourier transform of the magnetization, for three configurations obtained at hole-doping $p=0.08$.
The same considerations apply for the other configurations and for other values of doping.
These symmetry-broken configurations are characterized by different patterns of the Néel parameter, realizing different arrangements of the AFM domains and of their interfaces.
However, the Fourier transforms of the magnetization share similar features, where in all cases the distribution of Fourier modes is centered at the AFM ordering vector $(\pi,\pi)$ with similar spread, compatible with a finite AFM correlation length.
In Fig.~\ref{fig:panel-degeneracy-doped}~(third and fourth row) we report the spectral function of the Green's function and self-energy obtained mapping these solutions to the unit cell.
Despite the differences in the real-space patterns, all the solutions map to closely similar one-particle spectral features.
Crucially, the pseudogap phenomenology and, in particular, the nodal-antinodal dichotomy of the spectra, is present in each single symmetry-broken configuration, with only minor differences related to the particular configuration examined.
As a consequence, ensemble averages over these quasi-degenerate states do not create the pseudogap phenomenology, but only contributes to smooth out the spectra.
\subsection{Ensemble averages}
We discuss the protocol used to perform the ensemble averages of the quasi-degenerate stationary mean-field solutions in the case of the doped Hubbard model and which are shown in Fig.~3 and Fig.~4 of the main text.
As discussed in Section~\ref{subsec:quasi-degenerate-doped} these quasi-degenerate solutions map to similar one-particle spectral features in the unit cell, each one displaying the characteristic signatures of the pseudogap.
The ensemble average is used to obtain a smooth spectral function, independent of the particular details of each configuration.
\begin{figure}
    \centering
    \includegraphics[width=0.7\linewidth]{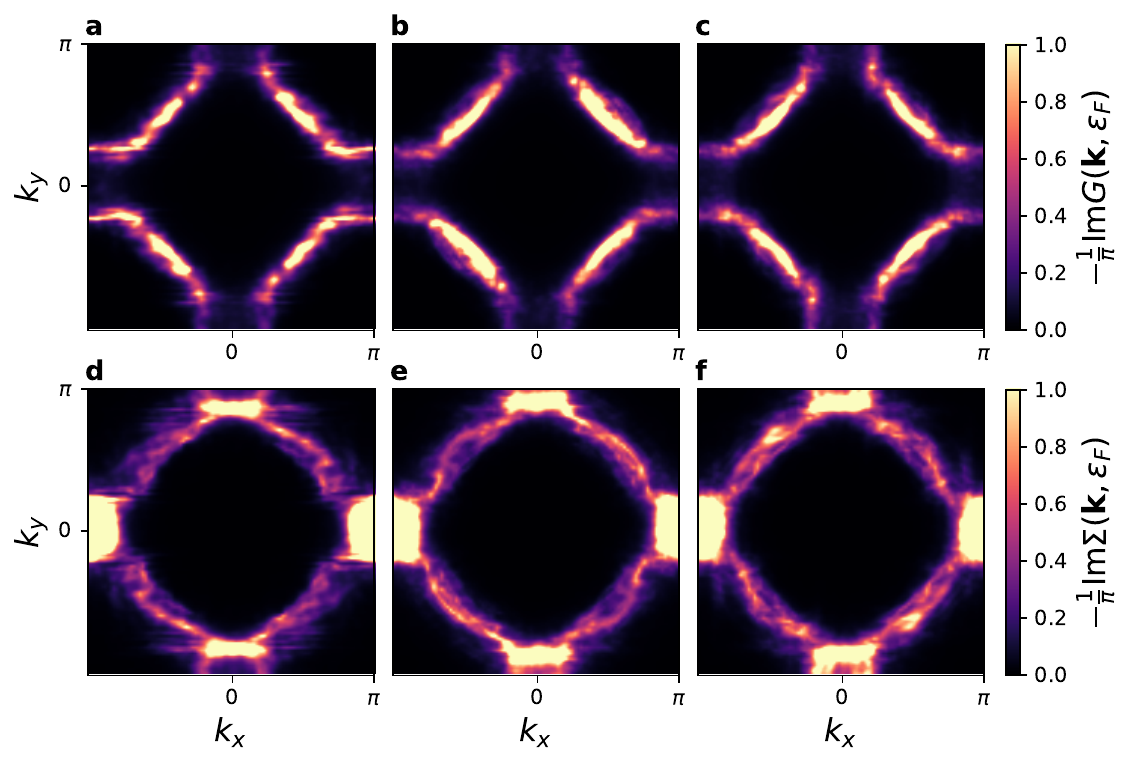}
    \caption{
    Fermi and Luttinger surface in the primitive Brillouin Zone obtained from the three spin textures reported in Fig.~\ref{fig:panel-degeneracy-doped} (hole doping $p=0.08$ and model parameters $U=3t$ and $t'=-0.3t$).
    (\textbf{a-c}): Fermi surfaces obtained from the imaginary part of the Green's function at the Fermi level. All solutions show coherent spectral weight in the nodal region, while suppression of spectral weight in the antinodal one.
    (\textbf{d-f}): Luttinger surfaces obtained from the imaginary part of the self-energy at the Fermi level.
    Complementary to the Fermi surfaces, all these Luttinger surfaces are characterized by coherent spectral weight in the antinodal region and suppression of the spectral weight in the nodal one.
    }
    \label{fig:panel-surf-repeated}
\end{figure}
In Fig.~\ref{fig:panel-surf-repeated} we show the Fermi and Luttinger surfaces of the three solutions reported in Fig.~\ref{fig:panel-degeneracy-doped} (with model parameters $U=3t$ and $t'=-0.3t$).
All these solutions share the same nodal-antinodal dichotomy at the Fermi surface, but the momentum-dependence of the spectral functions display minor differences due to the particular symmetry-broken configuration.
Furthermore we notice that individual solutions also break the $C_{4}$ rotational symmetry, due to differences in the vertical and horizontal extension of the real-space textures.
However, real-space solutions that are related to each other by a symmetry operation of the D$_4$ group are still stationary solutions of the mean-field Hamiltonian with the same total energy.
To remove configuration-dependent effects of the set of the quasi-degenerate state we perform an ensemble average, where the ensemble spectral function is obtained as a weighted sum of the single spectral functions:
\begin{equation}
    A_\mathrm{ens}(\mathbf{k},\omega) = \frac{\sum_{i} w_i\,A_i(\mathbf{k},\omega)}{\sum_i w_i}.
\end{equation}
We consider equal weights, corresponding to a large-temperature limit of the thermal average with Boltzmann factors.
Furthermore, since textures obtained from each other by a symmetry operation of the point group of the square lattice ($D_4$) have the same energy, we account for their contribution by symmetrizing the spectra in momentum space.

The ensemble average is performed for the spectral function of both the Green's function and the self-energy.
We stress that we apply this operation to smooth out the representative features of both the Fermi and Luttinger surfaces, and that the Green's function and self-energy defined from the ensemble average are not related by a Dyson equation:
\begin{equation}
    G_\mathrm{ens}(\mathbf{k},\omega) \neq [\omega - \varepsilon_{\mathbf{k}}-\Sigma_{\mathrm{ens}}(\mathbf{k},\omega)]^{-1}.
\end{equation}

\subsection{In-gap states induced by antiphase AFM domains}
\begin{figure}
    \centering
    \includegraphics[width=0.7\linewidth]{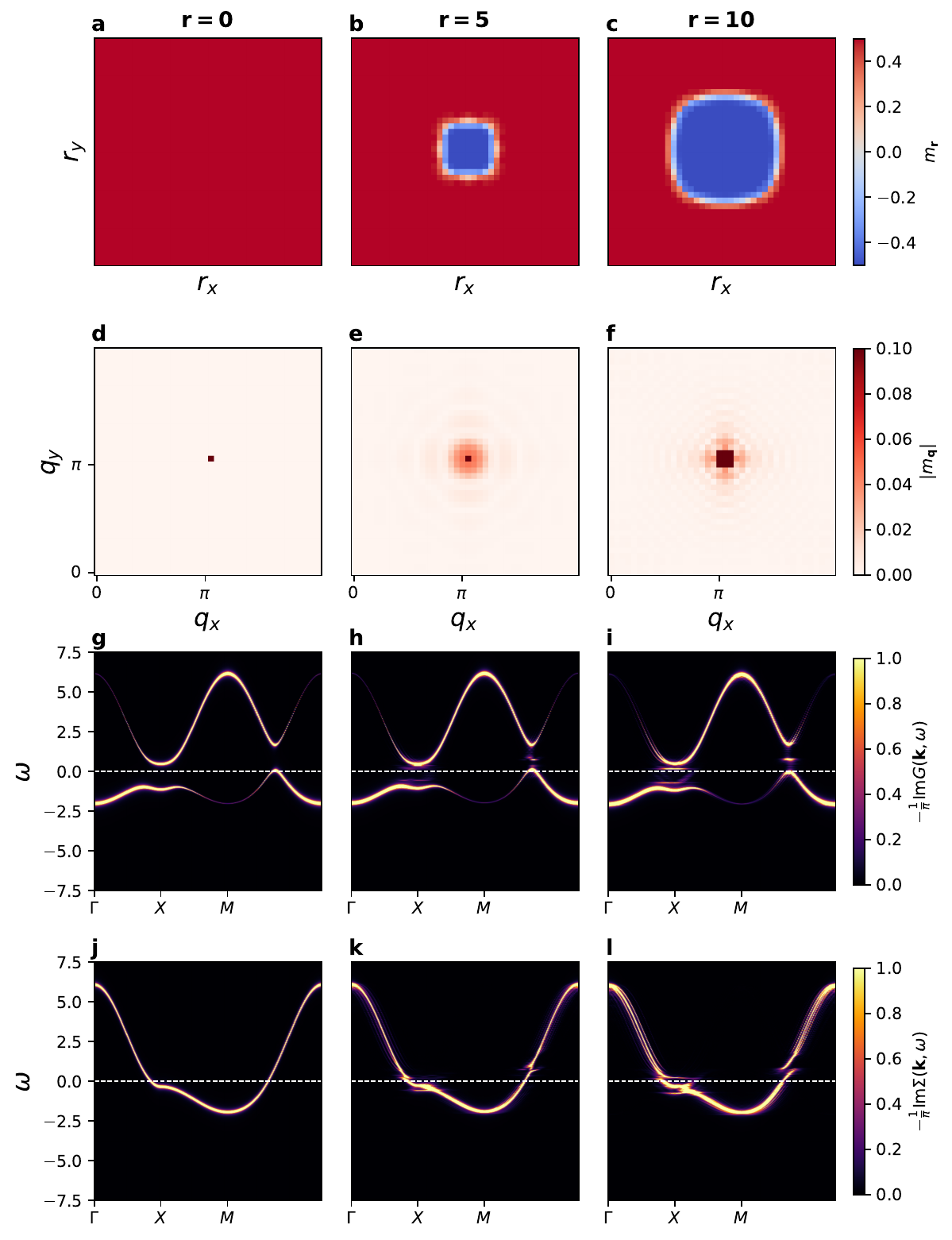}
    \caption{
    Spectral signatures of circular antiphase AFM domains.
    Self-consistent Hartree--Fock solutions for $U=3t$, $t'=-0.3t$, and hole-doping $p=0.02$, initialized from circular AFM domains embedded in a background of opposite phase.
    The three columns correspond to domain radii $r=0$, $r=5$, and $r=10$ lattice sites. Here, calculations are performed in a $40\times 40$ supercell.
    (\textbf{a-c}): real space Néel parameter. 
    (\textbf{d-f}): Fourier amplitudes of the magnetization. All solutions are characterized by a peak around ($\pi,\pi$), with different broadening in their distributions. 
    (\textbf{g-i}): spectra of the Green's function along a high-symmetry path in the primitive Brillouin Zone.
    Due to the presence of hole doping the Fermi level crosses the top of the valence band in the nodal region ($M\rightarrow\Gamma$), while the dispersion is otherwise characterized by a fundamental gap in both the nodal and antinodal sectors.   
    The presence of a nonmagnetic interface where the N\'eel parameter changes sign introduces in-gap states. These appear in the middle of the gap in the nodal region and as split-off bands at the band edges of the antinodal region.
    These states become more dispersive with longer interfaces.
    (\textbf{j-l}): spectra of the self-energy along the same high-symmetry path. The antiferromagnetic solution is characterized by a coherent single-pole dispersion. The nonmagnetic interface introduces a flat pole in the antinodal region, while the self-energy spectrum is gapped in the nodal sector.
    }
    \label{fig:panel-interface-spectral}
\end{figure}

\begin{figure}
    \centering
    \includegraphics[width=0.95\linewidth]{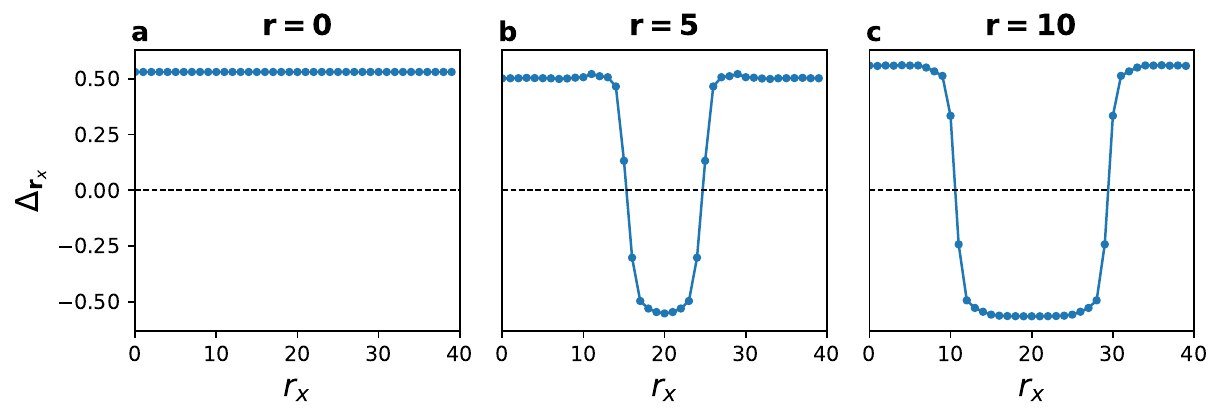}
    \caption{
    Soliton-like profile of the Néel parameter for circular AFM domains.
    (\textbf{a-c}) Horizontal cuts of the real-space N\'eel parameter through the center of the configurations reported in Fig.~\ref{fig:panel-interface-spectral}~(\textbf{a-c}).
    The uniform AFM state display a constant N\'eel parameter, whereas the circular antiphase domains display a smooth reversal of the parameter.
    The change of sign is not abrupt, but it takes place over approximately $4$-to-$5$ lattice sites, independently of the radius of the circular domain.
    }
    \label{fig:panel-soliton}
\end{figure}

We show that a reversal of the sign of the Néel order parameter between antiphase AFM domains is sufficient to generate nodal in-gap states once the symmetry-broken texture is mapped to the $\mathbf{k}$-diagonal Green's functions and self-energies of the primitive unit cell.
We first consider self-consistent Hartree--Fock solutions initialized from idealized circular AFM domains embedded in an AFM background characterized by a N\'eel parameter of opposite sign.
The calculations are performed for $U=3t$, $t'=-0.3t$, and hole doping $p=0.02$.
Fig.~\ref{fig:panel-interface-spectral} compares the uniform AFM solution ($r=0$) with circular antiphase domains of radii $r=5$ and $r=10$ lattice sites.
The real-space Néel parameter shows two antiphase AFM regions separated by a closed, weakly magnetic boundary.
Correspondingly, the magnetic Fourier components broaden around $(\pi,\pi)$, reflecting the finite spatial extent of the AFM domains.
The smaller domain produces the broader momentum-space distribution.
After mapping to the primitive unit cell, both antiphase textures display nodal in-gap spectral weight that is absent in the uniform AFM solution.
The effect is more pronounced for the larger domain, which contains a longer antiphase boundary.
The self-energy shows the complementary evolution, with the low-energy pole structure suppressed in the nodal sector as the in-gap states emerge in the Green's function.
To characterize the boundary profile, Fig.~\ref{fig:panel-soliton} reports a horizontal cut of the Néel parameter through the center of the same configurations.
The self-consistent boundaries therefore have a smooth, soliton-like profile rather than an abrupt sign change.
Their finite width, however, is not essential to the appearance of the in-gap states: the relevant feature is the reversal of the Néel order parameter between antiphase regions. 
We have checked that in-gap states are also present for non-self-consistent solutions characterized by a sharp change of the sign of the N\'eel parameter.

\begin{figure}
    \centering
    \includegraphics[width=0.95\linewidth]{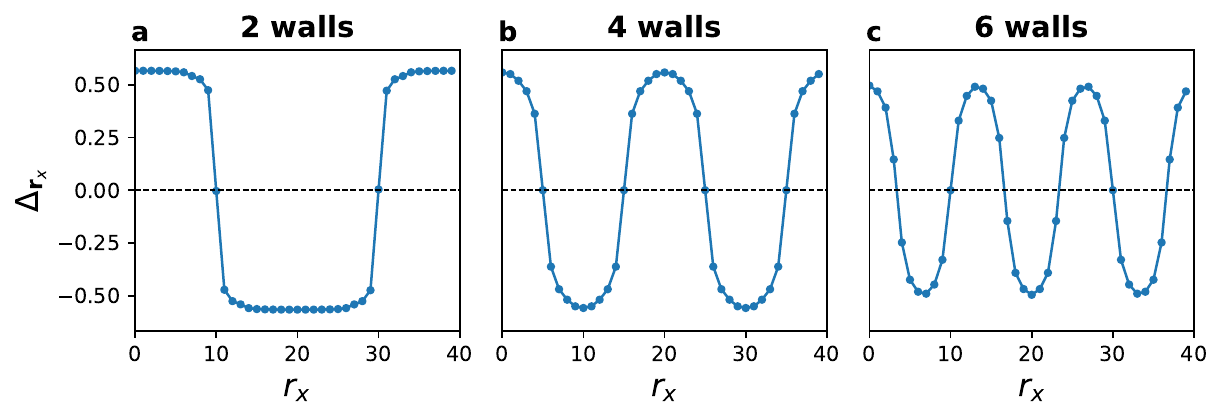}
    \caption{
    Evolution of the Néel profile with increasing boundary density for stripe solutions.
    (\textbf{a-c}) Horizontal cuts of the real-space Néel parameter for stripe textures containing two, four, and six antiphase boundaries across the supercell, reported in Fig.~\ref{fig:panel-interface-spectral-stripes}~(\textbf{a-c}).
    For distant domain walls the profile of the interface is closely similar to that of an isolated circular domain. 
    As the separation between adjacent boundaries decreases, the profile evolves from well-separated domain walls towards an approximately sinusoidal modulation. 
    The maxima of the N\'eel parameter, in absolute value, are not substantially modified by the number of interfaces.
    }
    \label{fig:panel-soliton-stripes}
\end{figure}

\begin{figure}
    \centering
    \includegraphics[width=0.7\linewidth]{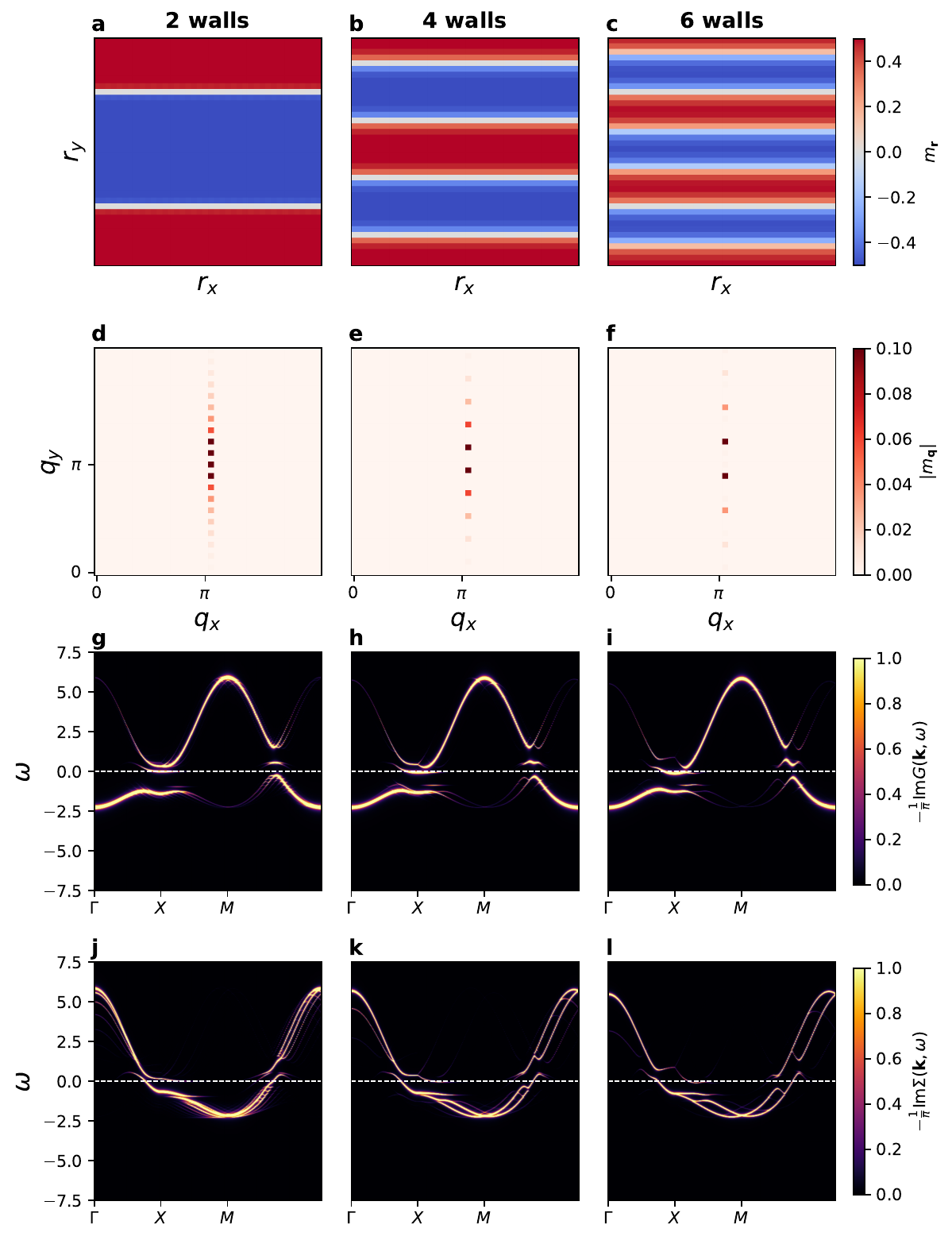}
    \caption{
    Evolution of the in-gap states with increasing antiphase-boundary density, obtained from self-consistent stripe textures containing two, four, and six antiphase boundaries.
    Calculations are performed for $U=3t$ and $t'=-0.3t$ in a $40\times 40$ supercell at hole-doping $p=0.02$.
    (\textbf{a-c}) real-space Néel parameter for the stripes of different periodicities.
    (\textbf{d-f}): absolute value of the Fourier amplitudes of the magnetization. The initial pattern of the magnetization is initialized with the smallest Fourier component allowed in the supercell, but self-consistency introduces higher harmonics.
    (\textbf{g-i}): spectra of the Green's functions obtained mapping the stripe pattern in the unit cell and plotted along a high-symmetry path in the primitive Brillouin Zone. 
    All spectra are characterized by in-gap states in the nodal sector; they become more dispersive with increasing number of domain walls, consistent with stronger hybridization between the states localized on the interfaces.
    (\textbf{j-l}): spectra of the self-energy reported along the same high-symmetry path. All spectra are characterized by a gap in the nodal region. 
    The antinodal sector is characterized by a low-energy pole just below the Fermi level. At variance with the circular walls reported in Fig.~\ref{fig:panel-interface-spectral}~(\textbf{j-l}) whose pole is more dispersive.
    }
    \label{fig:panel-interface-spectral-stripes}
\end{figure}

We next consider periodic stripe textures with increasing numbers of antiphase boundaries.
The corresponding Néel profiles are shown in Fig.~\ref{fig:panel-soliton-stripes}.
As the density of boundaries increases, the isolated domain-wall profiles increasingly overlap and the Néel parameter evolves to an approximately sinusoidal modulation. 
Fig.~\ref{fig:panel-interface-spectral-stripes} shows the corresponding reciprocal-space distributions and primitive-cell spectra.
Although the textures are initialized using a single cosine modulation for the profile of the magnetization, that amounts to an incommensurability with respect to the AFM solution defined by the fundamental harmonic of the cosine, self-consistency generates higher harmonics.
Since at small incommensurability the Néel parameter saturates within broad antiphase domains, and reverses over comparatively narrow boundaries, it produces a square-like profile described by several Fourier components.
With increasing boundary density, neighbouring domain walls overlap, the modulation becomes increasingly sinusoidal, and the spectral weight concentrates in the fundamental harmonic.
This is qualitatively different with respect to the closed circular domain of Fig.~\ref{fig:panel-interface-spectral} where the Fourier components are dense around $(\pi,\pi)$.
All stripe textures generate nodal in-gap states.
As the number of antiphase boundaries increases, the interface states overlap more strongly and acquire an increasingly pronounced dispersion, eventually forming a broader in-gap band. 
We also notice that the spectral features of these stripe states are qualitatively similar to those of the circular AFM boundary, and are thus less sensitive to the precise shape of the interface between the AFM domains.
Additionally, circular patterns introduce incoherent features that are absent for stripes with short real-space period.
Together, the circular-domain and stripe calculations show that the nodal in-gap states are robustly associated with reversals of the sign of the Néel order parameter.
The detailed boundary profile mainly affects the details of the pole structure of the self-energy, whereas the density and separation of the antiphase boundaries control the dispersion of the resulting in-gap states.
\clearpage

%%%%%%%%%%%%%%%%%%%%%%%%%%%%%%%%%%%%%%%%%%%%%%%
% AFM CORRELATION LENGTH
%%%%%%%%%%%%%%%%%%%%%%%%%%%%%%%%%%%%%%%%%%%%%%%

\section{Analysis of the antiferromagnetic correlation length}
In this section we analyze the antiferromagnetic spatial correlations in the Hartree-Fock symmetry-broken solutions of the Hubbard model.
We characterize the magnetic textures through the normalized autocorrelation function of the local Néel parameter.
Specifically, we define
\begin{equation}
    \label{eq:supp-C-afm}
    C_{\mathrm{AFM}}(\mathbf{R}) = \frac{\sum_{\mathbf r} \Delta_{\mathbf r} \Delta_{\mathbf r+\mathbf R}}
    {\sum_{\mathbf r}\Delta_{\mathbf r}^{2}},
\end{equation}
where periodic boundary conditions are used in the displacement $\mathbf R$.
We do not subtract the spatial average of $\Delta_{\mathbf r}$.
With this convention, a perfectly ordered AFM texture has $C_{\mathrm{AFM}}(\mathbf R)=1$ for every displacement $\mathbf R$.
To obtain a compact measure of the spatial decay, we further perform a radial average over displacement vectors with the same radial distance from the origin.
On the lattice this is performed considering bins of finite width $\Delta r$ corresponding to one lattice site:
\begin{equation}
    C^{\mathrm{rad}}_{\mathrm{AFM}}(r) = \frac{1}{N_r} \sum_{\mathbf{R}:|\mathbf{R}|\in\{r,r+\Delta r\}} C_{\mathrm{AFM}}(\mathbf{R}),
\end{equation}
and where $N_r$ is the number of sites in the bin with distance from the origin in the interval $\{r,r+\Delta r\}$.

We adopt two fitting forms for the radial propagator motivated by the Ornstein-Zernike (OZ) description of magnetic correlations.
In momentum space, the static susceptibility close to the AFM ordering wavevector is conventionally written as
\begin{equation}
    \label{eq:supp-OZ-afm}
    \chi_{\mathrm{AFM}}(\mathbf q) = \frac{A}{(\mathbf{q}
    -\mathbf{Q})^2+\xi_{\mathrm{AFM}}^{-2}},
\end{equation}
where $\mathbf Q=(\pi,\pi)$ is the AFM ordering vector and $\xi_{\mathrm{AFM}}$ is the corresponding antiferromagnetic correlation length.
Similar expressions are commonly used to extract magnetic correlation lengths from the static spin susceptibility of the Hubbard model~\cite{Schafer2021TrackingSupp}.
In particular, within the OZ form one has $\chi(\mathbf Q)\propto \xi^2$.

The two-dimensional Fourier transform of Eq.~\eqref{eq:supp-OZ-afm} is proportional to the modified Bessel function of the second kind, $C_{\mathrm{OZ}}(r)\propto K_0(r/\xi_{\mathrm{AFM}})$.
Since our autocorrelation function may not decay to zero at infinite distance, we allow for a nonzero long-distance component and fit the commensurate AFM correlations with the following asymptotic form
\begin{equation}
    \label{eq:supp-C-afm-fit}
    C_{\mathrm{AFM}}^{\mathrm{fit}}(r) =
    C_{\infty} + A \frac{e^{-r/\xi_{\mathrm{AFM}}}}{r^{1/2}}.
\end{equation}
Stripe-like correlations, which are relevant in the case of the doped model, are instead described by allowing the magnetic response to be centered at incommensurate wavevectors displaced from $\mathbf Q$ by $\boldsymbol{\delta}$.
The corresponding OZ expression is
\begin{equation}
    \label{eq:supp-OZ-stripe}
    \chi_{\mathrm{Stripe}}(\mathbf{q})
    =
    \frac{A_{\mathrm{Stripe}}}
    {(\mathbf q-\mathbf{Q}-\boldsymbol\delta)^2+\xi_{\mathrm{Stripe}}^{-2}}
    +
    \frac{A_{\mathrm{Stripe}}}
    {(\mathbf q - \mathbf{Q} +\boldsymbol\delta)^2+\xi_{\mathrm{Stripe}}^{-2}} .
\end{equation}
The real-space counterpart contains both an envelope, controlled by $\xi_{\mathrm{Stripe}}$, and an oscillatory factor, controlled by the incommensurability $\boldsymbol\delta$ with $C_{\mathrm{Stripe}}(\mathbf R) \propto K_0(R/\xi_{\mathrm{Stripe}}) \cos(\boldsymbol\delta\cdot\mathbf R)$.
For the stripe-like fit we adopt then the following phenomenological asymptotic form
\begin{equation}
\label{eq:supp-C-stripe-fit}
    C_{\mathrm{Stripe}}^{\mathrm{fit}}(r) =
    C_{\infty} + A \frac{e^{-r/\xi_{\mathrm{Stripe}}}}{r^{1/2}}
    \cos(\delta\cdot r+\phi).
\end{equation}
This expression captures the oscillatory sign changes associated with incommensurate antiphase domains.
Here, the phase $\phi$ is included as an effective parameter accounting for the effects of the radial average and the irregular shape of the real-space domains pattern.
In this case, the parameter $\xi_{\mathrm{Stripe}}$ measures the decay length of the stripe envelope, while $\delta$ determines the real-space period of the oscillations and the displacement of the magnetic peaks from $\mathbf{Q}$.
In practice, the fits are performed via a least-square algorithm.
\subsection{AFM correlation length in the solutions of the half-filled model}
\label{sec:afm-corr-length-half-filled}
\begin{figure}
    \centering
    \includegraphics[width=0.95\linewidth]{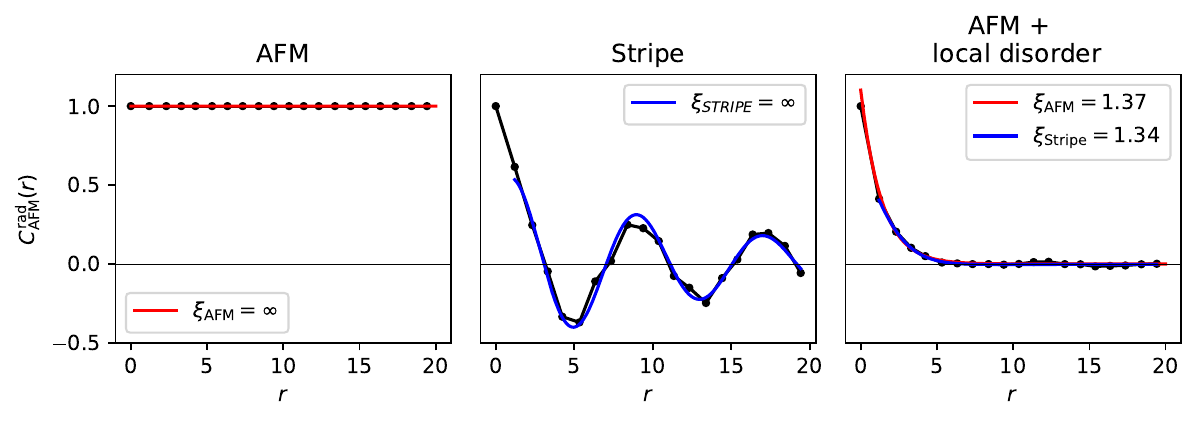}
    \caption{
     Radially averaged autocorrelation function (black dots) together with the results of the fit, for the three solutions reported in Fig.~\ref{fig:panel-mag-fourier-q}~(\textbf{a-c}) for the half-filled and particle-hole symmetric Hubbard model at $U=8t$ in a $40\times 40$ supercell.
     The fit of the AFM state (red line) is performed with the exponentially decaying envelope of Eq.~\eqref{eq:supp-C-afm-fit}.
     The fit of the stripe state is performed with the expression involving a cosine in Eq.~\eqref{eq:supp-C-stripe-fit} (blue line).
     The fit of the solution characterized by locally disordered AFM environments is done with both functions.
     Correlation lengths larger than $20$ sites are labelled as $\infty$ since they are larger than half the size of the supercell considered.
     }
    \label{fig:C-afm-half-filled}
\end{figure}

We report the values of the AFM correlation length for the solutions of the particle-hole symmetric and half-filled Hubbard model with $U=8t$.
We consider first the three solutions reported in Sec.~\ref{sec:hubbard} (Fig.~\ref{fig:panel-mag-fourier-q}), namely the AFM solution with infinite correlation length, a stripe pattern with periodicity $8\times2$ and an inhomogeneous solution characterized by a short-range AFM correlations.
The AFM autocorrelation function of these three states is reported in Fig.~\ref{fig:C-afm-half-filled}, together with their respective fits.
The fit of the AFM states has been done using the expression of Eq.~\eqref{eq:supp-C-afm-fit}, while the fit of the stripe state has been done using Eq.~\eqref{eq:supp-C-stripe-fit}.
As expected, the AFM state is characterized by a constant autocorrelation function, leading to an infinite AFM correlation length.
The stripe state is instead characterized by periodic oscillations of the autocorrelation function, due to the fact that the Néel parameter changes sign periodically.
The amplitude of the oscillations is slowly damped due to the fact that we are fitting the radial average of the autocorrelation function instead of a linear component parallel to the stripe direction.
The inhomogeneous solution is instead characterized by a decay of the AFM autocorrelation function. 
The fit with Eq.~\eqref{eq:supp-C-afm-fit} reveals a finite autocorrelation length of $\xi\simeq1.3$.
We also considered a fit with the stripe profile of Eq.~\eqref{eq:supp-C-stripe-fit} which gives a correlation length compatible with that of the AFM fit.
This has to be expected, since in both cases the correlation length dictates the decay in space of the profile of $C^\mathrm{rad}_\mathrm{AFM}(r)$.
Similar values of the AFM autocorrelation length are reported for the other quasi-degenerate inhomogeneous solutions discussed in Section~\ref{subsec:quasi-degenerate-half-fill}.
The values of the AFM correlation length extracted from these patterns and evaluated with the fit of Eq.~\eqref{eq:supp-C-afm-fit} are reported in Tab.~\ref{tab:xi-afm-hub-half-fill}.
\begin{table*}[h]
    \centering
    \begin{tabular}{c c}
    \hline 
    Configuration & $\xi_\mathrm{AFM}$\\
    \hline
       rand 1   &  1.24 \\
       rand 2   &  2.77 \\
       rand 3   &  1.37 \\
       rand 4   &  1.21 \\
       rand 5   &  1.10 \\
       rand 6   &  1.76 \\
       rand 7   &  1.31 \\
       rand 8   &  1.22 \\
       rand 9   &  1.74 \\
       rand 10  &  1.42 \\
    \end{tabular}
    \caption{
    Antiferromagnetic correlation length computed for the quasi-degenerate solutions of the half-filled and particle-hole symmetric Hubbard model at $U=8t$ reported in Tab.~\ref{tab:energy-hub-half-fill}.
    The correlation length is extracted from the fit of the radial autocorrelation function using the function in Eq.~\eqref{eq:supp-C-afm-fit}.
    }
    \label{tab:xi-afm-hub-half-fill}
\end{table*}

\subsection{AFM correlation length in the solutions of the hole-doped model}
\label{subsec:afm_correlation_lenght_doped}
Mean-field calculations in the doped Hubbard model are performed for the parameters $U=3t$ and $t'=-0.3t$.
For each value of hole doping, the correlation lengths and incommensurabilities are obtained from the ensemble averages of ten different spin textures obtained from random initialization of the spin-resolved density in a $40\times 40$ supercell, as discussed in Sec.~\ref{subsec:quasi-degenerate-doped}.
We consider values of hole doping ranging from $p=0.02$ to $p=0.20$.

\begin{figure}
    \centering
    \includegraphics[width=0.95\linewidth]{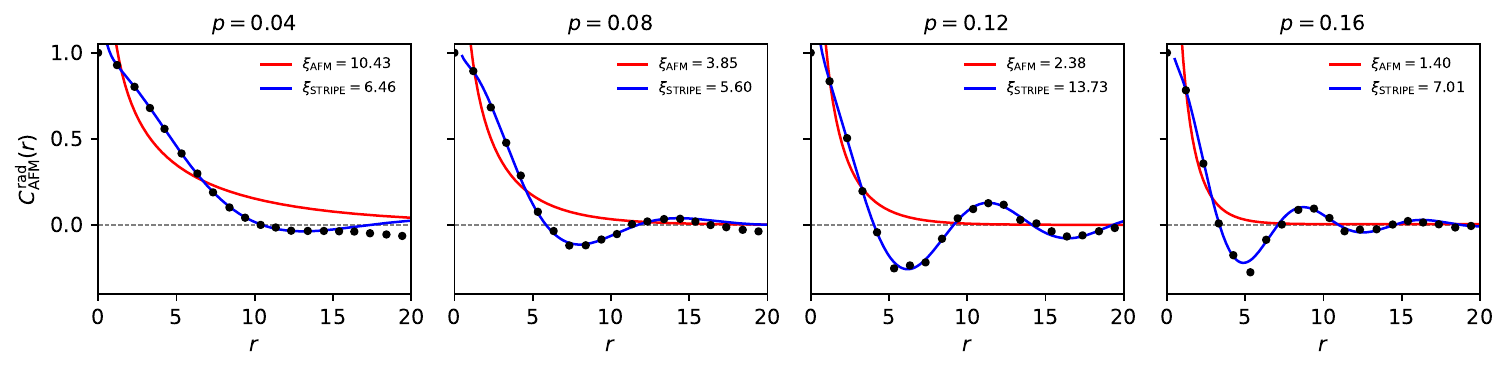}
    \caption{Radially averaged autocorrelation function $C^\mathrm{rad}_\mathrm{AFM}$ of the Néel parameter (black dots) for one spin texture at different values of hole doping $p$.
    Model parameters are $U=3t$ and $t'=-0.3t$.
    The curves show the commensurate AFM fit of Eq.~\eqref{eq:supp-C-afm-fit} (red line) and the stripe-like fit (blue line) of Eq.~\eqref{eq:supp-C-stripe-fit}.
    With increasing hole doping $C^\mathrm{rad}_\mathrm{AFM}$ develops wiggles associated to the onset of stripe-like correlations.}
    \label{fig:panel-C-afm}
\end{figure}

We show in Fig.~\ref{fig:panel-C-afm} the normalized radial autocorrelation functions for some of the doping values considered, together with the fits of Eqs.~\eqref{eq:supp-C-afm-fit} and \eqref{eq:supp-C-stripe-fit}, for one single symmetry-broken configuration.
The commensurate AFM fit of Eq.~\eqref{eq:supp-C-afm-fit} is performed as follows: the long-distance component $C_\infty$ is fixed to the average of $C^\mathrm{rad}_\mathrm{AFM}(r)$ over $15 \leq r \leq 20$, or set to zero when this average is negative; this removes the degeneracy between $C_\infty$ and $\xi_\mathrm{AFM}$ that affects a free three-parameter fit when the correlator does not decay to zero within the supercell.
The amplitude $A$ and the correlation length $\xi_\mathrm{AFM}$ are then obtained by a weighted least-squares fit over $1 \leq r \leq 15$, restricted to the first positive branch of the correlator: all values after the first sign change are set to $10^{-3}$.
The stripe-like fit of Eq.~\eqref{eq:supp-C-stripe-fit} is performed over $1 \leq r \leq 15$, with $
A$, $\xi_\mathrm{Stripe}$, $\delta$, $\phi$ and $C_{\infty}$ as free parameters.
At zero doping (not shown) the self-consistent solution is a uniform AFM state, and the normalized correlator is equal to one at all distances.
Upon doping, the correlator decays and develops oscillations related to presence of antiphase AFM domains.
%When the radial correlator becomes negative, the commensurate AFM fit of Eq.~\eqref{eq:supp-C-afm-fit} is restricted to the first positive branch: to perform the fit all the values of the autocorrelation function after the first negative one are set to a small positive value; here we set $10^{-3}$.
%The stripe-like fit of Eq.~\eqref{eq:supp-C-stripe-fit} is instead performed over distances sufficiently large, in this case $1\leq r \leq 15$ sites, where the asymptotic approximation used in Eq.~\eqref{eq:supp-C-stripe-fit} is justified.
%
We note that Eq.~\eqref{eq:supp-C-afm-fit} is the large-distance limit of the Ornstein–Zernike form Eq.~\eqref{eq:supp-OZ-afm}; when $\xi_\mathrm{AFM}$ becomes comparable to half the supercell, as for $p\leq0.04$, the fit is applied outside its range of validity and $\xi_\mathrm{AFM}$ should be read as an indicative scale rather than a measured length.
The value of $\xi_\mathrm{AFM}$ extracted from Eq.~\eqref{eq:supp-C-afm-fit} should therefore be interpreted as an effective commensurate AFM length: it measures the distance over which the local Néel parameter remains positively correlated, identifying the typical size of the AFM domains in the supercell. By contrast, $\xi_\mathrm{Stripe}$ measures the decay of the stripe envelope after the onset of the oscillations.

\begin{figure}
    \centering
    \includegraphics[width=0.9\linewidth]{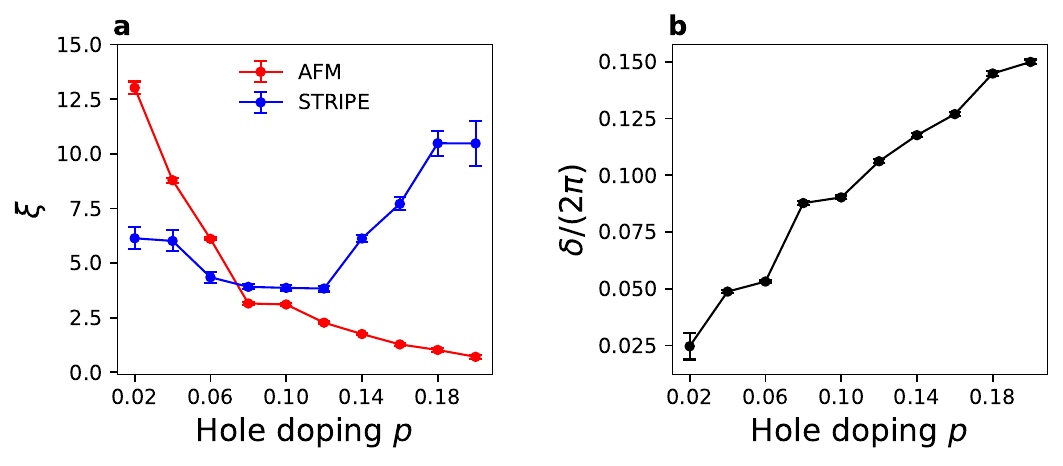}
    \caption{Doping dependence of the fit parameters, for both the AFM fit of Eq.~\eqref{eq:supp-C-afm-fit} and the stripe-like fit of Eq.~\eqref{eq:supp-C-stripe-fit}.
    The values are obtained from a weighted average of the parameters obtained from the fit of ten different spin textures for each value of hole doping $p$. 
    The parameters of each spin texture are obtained as shown in Fig.~\ref{fig:panel-C-afm}.
    The weights used in the average are the inverse variance of the parameters obtained from a least-square fit.
    (\textbf{a}) Effective commensurate AFM correlation length $\xi_{\mathrm{AFM}}$ (red dots) and stripe-envelope correlation length $\xi_{\mathrm{Stripe}}$ (blue dots).
    The error bar is obtained propagating the fit uncertainties with the same weights.
    (\textbf{b}) Incommensurate shift $\delta$, in units of $2\pi$, of the stripe peaks away from $\mathbf Q=(\pi,\pi)$.
    The error bar is obtained propagating the fit uncertainties with the same weights of the ensemble average.
    }
    \label{fig:panel-C-afm-corr-lengths}
\end{figure}

We report in Fig.~\ref{fig:panel-C-afm-corr-lengths} the evolution of the fitted parameters as a function of hole doping for values ranging from $p=0.02$ to $p=0.20$.
For each doping, the reported value is obtained as a weighted average over the different symmetry-broken configurations, $\bar x=\sum_i w_i x_i/\sum_i w_i$, where $x_i$ are the fitted quantities from the single fits.
The uncertainty on the ensemble average is obtained by propagating the fitting uncertainties $\sigma_i$ of the individual configurations using the same weights, $\sigma_{\bar x}=[\sum_i w_i^2\sigma_i^2]^{1/2}/\sum_iw_i$.
We use inverse-variance weights $w_i=1/\sigma_i^2$, such that poorly representative fits contribute less to both the ensemble average and its uncertainty.

The correlation lengths are shown in Fig.~\ref{fig:panel-C-afm-corr-lengths}~(\textbf{a}).
The effective commensurate AFM correlation length decreases rapidly with hole doping.
By contrast, the stripe-envelope length increases at larger dopings, consistent with the emergence of more coherent stripe-like textures.
This indicates that the dominant magnetic correlations evolve from commensurate AFM correlations at half filling to incommensurate stripe-like correlations in the doped solutions.

\begin{figure}
    \centering
    \includegraphics[width=0.95\linewidth]{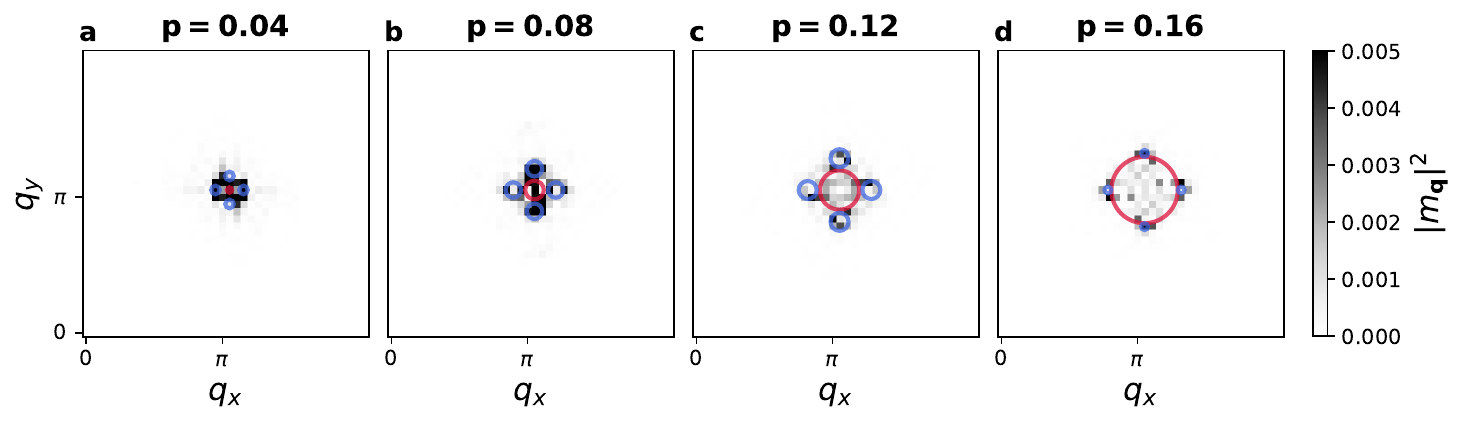}
    \caption{Magnitude of the square values of the Fourier components of the magnetization $|m_{\mathbf{q}}|^2$ for the spin texture reported in Fig.~\ref{fig:panel-C-afm} at different values of hole doping $p$.
    Model parameters are $U=3t$ and $t'=-0.3t$.
    Red circles indicate the inverse commensurate AFM correlation length $\xi_{\mathrm{AFM}}^{-1}$ around $\mathbf Q=(\pi,\pi)$ extracted from the fit of Eq.~\eqref{eq:supp-C-afm-fit}.
    Blue circles are obtained from the fit of Eq.~\eqref{eq:supp-C-stripe-fit}: their centers are displaced from $\mathbf Q$ by $\pm\boldsymbol\delta$ and have radii given by $\xi_{\mathrm{Stripe}}^{-1}$.
    In all cases the AFM correlation length identifies a region in reciprocal space where the dominant components of $|m_\mathbf{q}|^2$ lie, while the blue circles identify the dominant peaks shifted away from $(\pi,\pi)$ and their characteristic spread.}
    \label{fig:panel-C-afm-mag-overlay}
\end{figure}

In Fig.~\ref{fig:panel-C-afm-corr-lengths}~(\textbf{b}) we report the evolution of the incommensurate shift $\delta$ extracted from the stripe-like fit.
The displacement from the AFM ordering vector grows with doping in steps set by the discrete incommensurabilities allowed in the $40\times 40$ supercell.
To assess if the parameters extracted from the fit give a faithful representation of the magnetic correlation, we analyze the Fourier transform of the square values of the magnetization, since the autocorrelation function is proportional to the magnetic structure factor, $S(\mathbf q)\propto|m_{\mathbf q}|^2$.
In Fig.~\ref{fig:panel-C-afm-mag-overlay} we report the distribution of $|m_{\mathbf q}|^2$ together with circles indicating the momentum scales inferred from the real-space fits for one single symmetry-broken configuration at different values of hole doping $p$.
Red circles are centered at $\mathbf Q=(\pi,\pi)$ and have radius $\xi_{\mathrm{AFM}}^{-1}$.
They therefore represent the width associated with a commensurate AFM interpretation of the magnetic correlations.
Blue circles are centered at the incommensurate stripe wavevectors $\mathbf Q\pm\boldsymbol\delta$, along either the horizontal or vertical direction, and have radius $\xi_{\mathrm{Stripe}}^{-1}$.
The stripe-like parameters obtained from the radial correlator match exactly the positions of the maxima of 
$|m_{\mathbf q}|^2$, signaling that the phenomenological form used in Eq.~\eqref{eq:supp-C-stripe-fit} gives a reasonable description of the stripe-like patterns.
Complementary, the inverse AFM length identifies a momentum scale around $\mathbf Q$ on which the dominant Fourier components lie.
Thus the same suppression of weight at $\mathbf Q$ can be interpreted either as a short commensurate AFM correlation length or as a displacement of the dominant peaks away from $\mathbf Q$.

This equivalence can be seen directly from the OZ expressions.
For a commensurate AFM response,
\begin{equation}
    \label{eq:supp-chi-Q-afm}
    \chi_{\mathrm{AFM}}(\mathbf Q) = A_{\mathrm{AFM}} \xi_{\mathrm{AFM}}^2 .
\end{equation}
For a stripe response with two symmetric peaks at $\mathbf Q\pm\boldsymbol\delta$, Eq.~\eqref{eq:supp-OZ-stripe} gives
\begin{equation}
    \label{eq:supp-chi-Q-stripe}
    \chi_{\mathrm{Stripe}}(\mathbf Q) = \frac{2A_{\mathrm{Stripe}}}{\delta^2+\xi_{\mathrm{Stripe}}^{-2}}.
\end{equation}
Equating Eqs.~\eqref{eq:supp-chi-Q-afm} and \eqref{eq:supp-chi-Q-stripe} defines an effective commensurate AFM correlation length for the stripe state:
\begin{equation}
    \label{eq:supp-xi-effective-general}
    \xi_{\mathrm{AFM,eff}}^{-2}=\frac{A_{\mathrm{AFM}}}{2A_{\mathrm{Stripe}}}\left(\delta^2+\xi_{\mathrm{Stripe}}^{-2}\right).
\end{equation}
If the prefactors are absorbed in the normalization of the two fit functions, or if one is interested only in the characteristic momentum scale, this reduces to
\begin{equation}
    \label{eq:supp-xi-xi}
    \xi_{\mathrm{AFM,eff}}^{-2} \simeq \delta^2+\xi_{\mathrm{Stripe}}^{-2}.
\end{equation}
In the limit of a highly coherent stripe state, $\xi_{\mathrm{Stripe}}\rightarrow\infty$ and one obtains $\xi_{\mathrm{AFM,eff}}^{-1}\simeq \delta$.
Thus an ordered stripe state, whose true magnetic peaks are sharp but displaced from $\mathbf Q$, can appear as a short-range commensurate AFM state if one probes only the value of the magnetic response at $\mathbf Q=(\pi,\pi)$.
In this sense, the apparent reduction of commensurate AFM correlations may indicate either a genuine shortening of the AFM correlation length or the emergence of incommensurate stripe correlations.
%

%%%%%%%%%%%%%%%%%%%%%%%%%%%%%%%%%%%%%%%%%%
% INTERFACE HOLE CONTENT
%%%%%%%%%%%%%%%%%%%%%%%%%%%%%%%%%%%%%%%%%%

\section{Analysis of the charge content of the interfaces in the solutions of hole-doped model}
\label{sec:interface-analysis}
In this section we analyze the evolution of the antiferromagnetic domains and interfaces in the doped Hubbard-model solutions for $U=3t$ and $t'=-0.3t$, for hole dopings ranging from $p=0$ to $p=0.20$.
As discussed in the main text, the doped solutions are characterized by antiphase antiferromagnetic domains, which can be visualized through the local Néel parameter:
\begin{equation}
    \label{eq:supp-neel-parameter}
    \Delta_{\mathbf r} = (-1)^{r_x+r_y} m_{\mathbf r},
\end{equation}
where $m_{\mathbf r}=n_{\mathbf r\uparrow}-n_{\mathbf r\downarrow}$ is the local magnetization.
We also define the local hole density as
\begin{equation}
    h_{\mathbf r}=1-\rho_{\mathbf r},
\end{equation}
where $\rho_{\mathbf r}=n_{\mathbf r\uparrow}+n_{\mathbf r\downarrow}$ is the local charge density.
The spatial average of $h_{\mathbf r}$ gives the average hole doping $p$.

\begin{figure}
    \centering
    \includegraphics[width=0.95\linewidth]{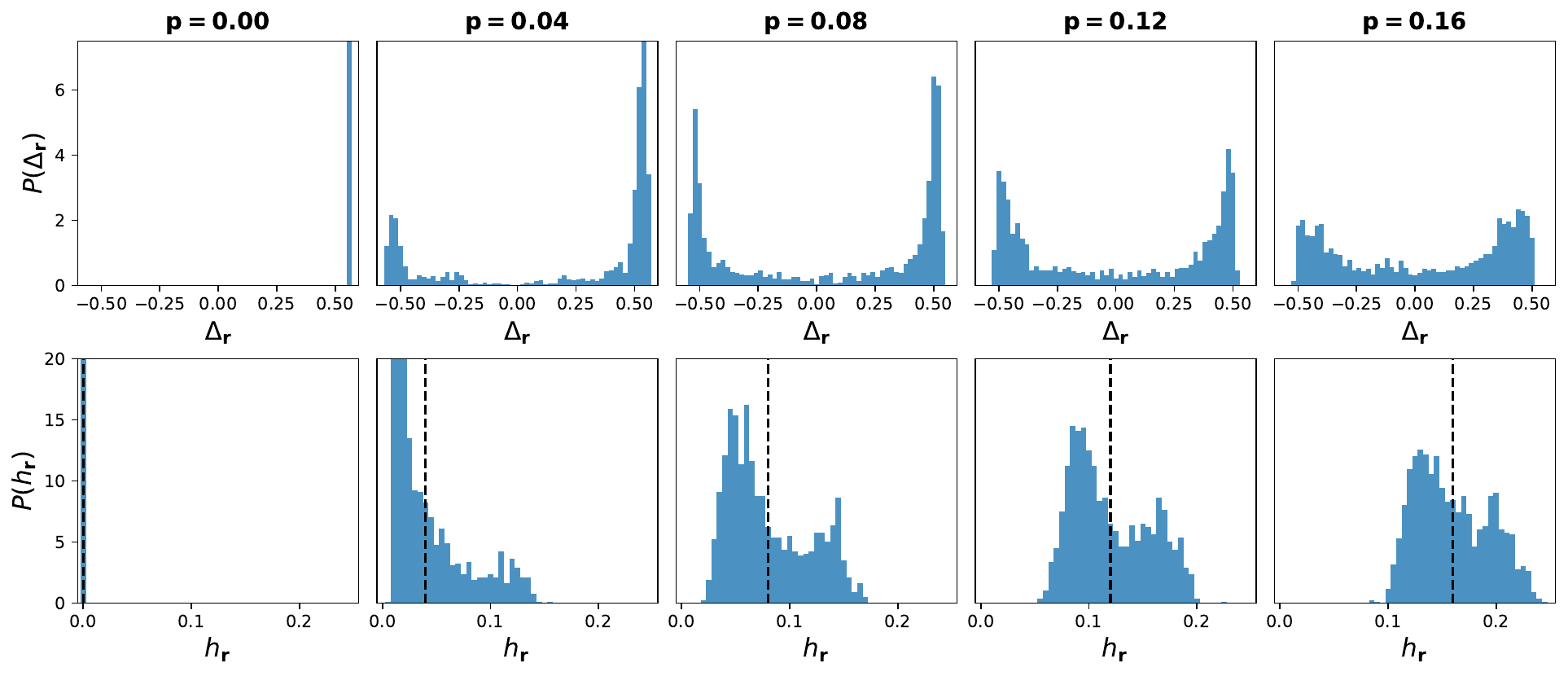}
    \caption{
    Normalized probability distributions of the local Néel parameter $\Delta_{\mathbf r}$ (upper row) and of the local hole density $h_{\mathbf r}$ (lower row) for a single spin texture at different values of hole doping $p$.
    Model parameters are $U=3t$ and $t'=-0.3t$.
    The histograms are obtained from the distributions of the values of the local real-space N\'eel parameter and hole density in a $40\times 40$ supercell.
    In the half-filled antiferromagnet the N\'eel parameter is characterized by a sharp delta peak.
    With increasing hole-doping a second peak develops and it is associated to values of the N\'eel parameter with opposite sign. At larger doping the distribution goes continuously from positive to negative values.
    This is consistent with the nucleation of antiphase AFM domains upon doping.
    Complementary, the hole density distribution develops a two-peak structure with increasing doping, displaying a larger peak at lower values of $h_{\mathbf{r}}$ and a smaller peak at larger values.
    This is consistent with holes occupying preferentially the nonmagnetic interfaces between AFM domains, realizing a phase separation with a hole-rich interface and hole-poor AFM domains.
    }
    \label{fig:neel-histogram}
\end{figure}

In Fig.~\ref{fig:neel-histogram} we report the normalized probability distributions $P(\Delta_{\mathbf{r}})$ of the local Néel parameter $\Delta_{\mathbf r}$ and of the local hole density $h_{\mathbf r}$ obtained from the distribution of values within one single symmetry-broken supercell at different values of the average hole doping $p$.

At $p=0$, the distribution of $\Delta_{\mathbf r}$ is sharply peaked at a positive value, consistent with a homogeneous AFM texture.
Upon hole doping, the distribution develops two main maxima.
The larger positive contribution corresponds to the dominant antiferromagnetic texture, while the smaller negative contribution signals the nucleation of antiphase AFM domains.
With increasing doping, the relative height of the two peaks is reduced, indicating that the two antiferromagnetic textures occupy similar fractions of the system.
Importantly, the distribution does not develop an isolated peak at $\Delta_{\mathbf r}=0$.
Instead, it evolves continuously between positive and negative values, indicating that the Néel parameters does not change sign in a sharp way, but it crosses a continuous distribution of values.
Moreover, the extrema of the distribution do not shrink appreciably towards the center, revealing that there are local regions where the Néel parameter is as strong as in the undoped AFM state at half-filling, signaling a persistence of AFM correlations even at large values of hole doping. 

The distribution of the local hole density $h_{\mathbf{r}}$ provides complementary information.
At small doping it is dominated by a peak close to $h_{\mathbf r}=0$, corresponding to regions that remain close to half filling.
With increasing doping, this peak is progressively shifted towards finite values and depleted, with a second contribution appearing at larger values of $h_{\mathbf r}$.
This shows that the doped solutions contain regions with enhanced hole density occupying a smaller fraction of the supercell, i.e. the interfaces, coexisting with more extended regions of lower hole density, i.e. the AFM domains.
Overall, the continuous distributions of $\Delta_{\mathbf r}$ and $h_{\mathbf r}$ prove that these symmetry-broken supercell solutions realize a genuine real-space ensemble of local environments, rather than a simple arrangement of a few sharply idealized textures.
Furthermore the increase in hole doping leads to a diversification of these local environments within the supercell, manifested in broader distributions of $\Delta_{\mathbf{r}}$ and $h_{\mathbf{r}}$.

\begin{figure}
    \centering
    \includegraphics[width=0.95\linewidth]{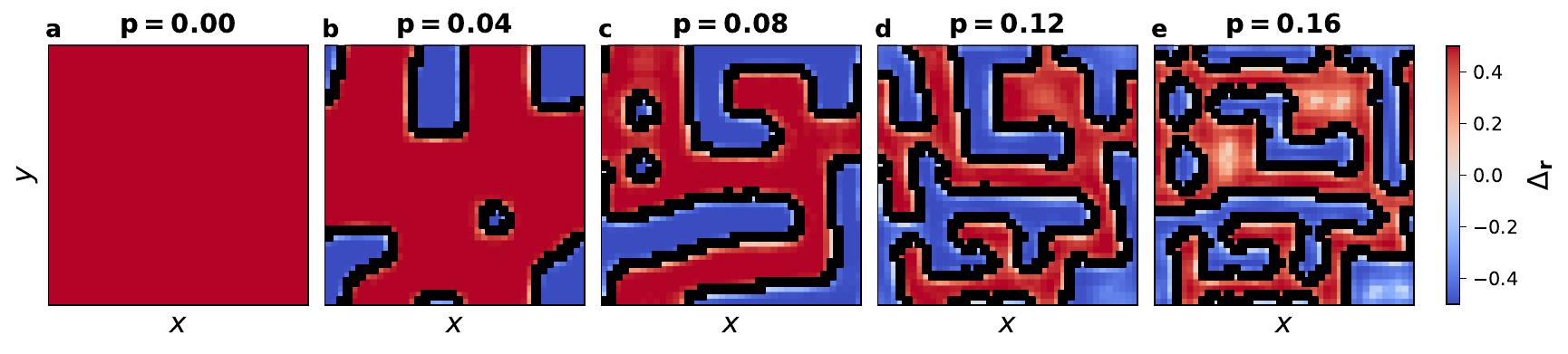}
    \caption{
    Real-space maps of the local Néel parameter $\Delta_{\mathbf r}$ for the same spin texture reported in Fig.~\ref{fig:neel-histogram} at different values of hole doping $p$.
    Model parameters are $U/t=3$ and $t'=-0.3t$; calculations are converged in a $40\times 40$ supercell.
    Black markers identify the sites classified as interfaces between antiphase antiferromagnetic domains following the procedure defined in Sec.~\ref{sec:interface-analysis}.
    In particular, the procedure identifies interface sites yielding smooth interfaces.
    }
    \label{fig:neel-interface}
\end{figure}

We next analyze the extension of the antiphase AFM domains and the proliferation of their interfaces with hole doping.
To this end, we assign each site of the supercell to one of three classes: $\mathrm{AFM}_{+}$, $\mathrm{AFM}_{-}$, or interface.
A site is marked as belonging to an interface if two conditions are met:
first, the local Néel parameter is suppressed below the mean value of the distribution, taken in absolute value $|\Delta_{\mathbf{r}}|<\mathrm{mean}(|\Delta_{\mathbf{r}}|)$;
second, the site lies next to a sign change of the Néel parameter along either the horizontal or the vertical directions.
In Fig.~\ref{fig:neel-interface} we show the real-space texture of the local Néel parameter for different hole dopings.
The sites identified as interfaces by the above criterion are highlighted with black markers.
This construction effectively yields smooth contours separating regions with opposite values of the Néel parameter.
We stress that this definition is meaningful only when the antiphase domains are separated by regions where the local magnetic moments are suppressed and it is therefore designed to identify nonmagnetic or weakly magnetic domain walls, rather than sharp sign changes.
Sites not assigned to the interface are classified as AFM$_{+}$ or AFM$_{-}$ according to the sign of the local Néel parameter.
We adopt the convention of assigning a positive value of the N\'eel parameter to the domains occupying a larger fraction of the supercell.
Interface sites are excluded from this second classification in order to avoid double counting.

\begin{figure}
    \centering
    \includegraphics[width=0.8\linewidth]{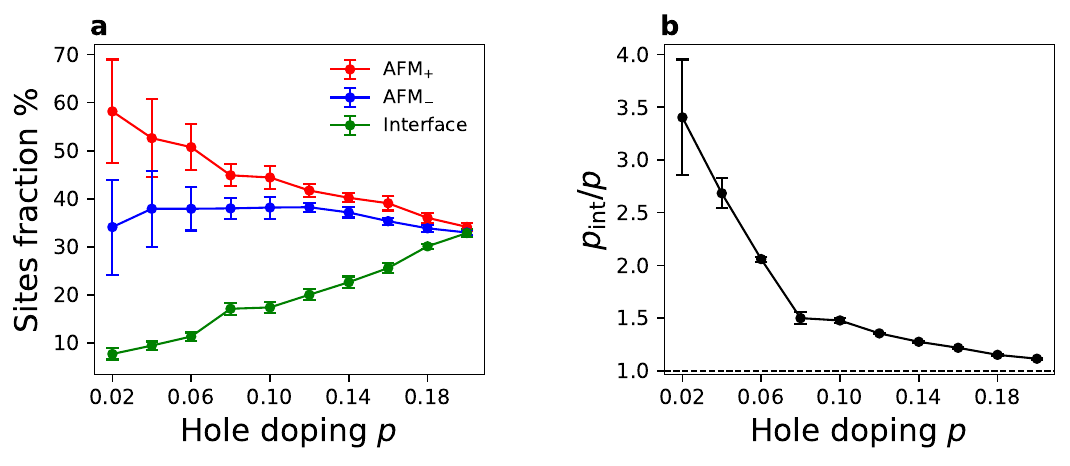}
    \caption{
    Analysis of the magnetic-domain and interface content of the spin textures found in the doped Hubbard model.
    For each value of hole doping $p$ we consider ten different spin textures. Quantities of interest are obtained from ensemble averages with equal weights, over the set of spin textures.
    Error bars are obtained from the standard deviation of the distribution and are thus representative of the spread of values.
    We observe that spin textures at lower values of hole doping are characterized by distributions of values with larger spread.
    \textbf{(a)} Site fraction of the $\mathrm{AFM}_{+}$, $\mathrm{AFM}_{-}$ and interface regions as a function of hole doping. At larger values of hole doping the three regions share similar fractions of the whole available supercell.
    \textbf{(b)} Ratio between the average hole density on the interface and the global hole doping $p$.
    Values larger than one indicate that holes are localized on the interface relative to a spatially uniform distribution. 
    The localization of holes is stronger at smaller values of doping, while at larger values the average density of holes at the interfaces approaches that of the holes in the AFM domains.
    }
    \label{fig:neel-analysis}
\end{figure}

For each solution, we then compute the site fraction of each class as $f_\alpha =N_\alpha/N_{\mathrm{sites}}$ with $\alpha \in \{\mathrm{AFM}_{+},\mathrm{AFM}_{-},\mathrm{int}\}$ and
where $N_\alpha$ is the number of sites belonging to class $\alpha$ and $N_{\mathrm{sites}}$ is the total number of sites in the supercell.
We then perform an ensemble average, with equal weights, over ten different spin textures, and for each value of hole doping we take as error bar the standard deviation of the distribution of values obtained from the ensemble.
The corresponding percentages are reported in Fig.~\ref{fig:neel-analysis}~(\textbf{a}).
With increasing doping, the fraction of the dominant AFM texture (here $\mathrm{AFM}_{+}$) decreases.
At the same time, the fraction of the antiphase texture (here $\mathrm{AFM}_{-}$) increases, indicating their progressive nucleation and growth.
The fraction of interface sites increases approximately linearly with hole doping.
Interestingly, for large values of hole doping the three classes occupy one third each of the supercell sites.

Finally, we quantify the relation between the interface regions and the distribution of doped charge.
As discussed in the main text, the doped holes preferentially accumulate on the interfaces between antiphase domains.
To make this statement quantitative, we compute the average hole density on the interface,
\begin{equation}
    p_{\mathrm{int}}
    =
    \frac{1}{N_{\mathrm{int}}}
    \sum_{\mathbf r\in\mathrm{int}} h_{\mathbf r},
\end{equation}
where $N_{\mathrm{int}}$ is the number of interface sites and the summation is restricted to the sites marked as interface.
We then compare this quantity with the average value of the hole density $p$ per site by considering their ratio $R_\mathrm{int}=p_{\mathrm{int}}/p$. This measures the localization of holes on the interfaces relative to a spatially uniform distribution. 
The result is shown in Fig.~\ref{fig:neel-analysis}~(\textbf{b}), where again we report the ensemble average of ten different spin textures for each doping, adopting as error bar the standard deviation of the distribution.
For all doped solutions considered here, $R_{\mathrm{int}}>1$, demonstrating that the holes are preferentially localized on the interface sites, especially at lower doping.
At larger doping, $R_{\mathrm{int}}$ decreases towards one, indicating that the hole density becomes progressively more homogeneous across the supercell.
These results confirm that the second peak observed in the distribution in Fig.~\ref{fig:neel-histogram} at larger values of $h_{\mathbf{r}}$ is indeed associated to the hole rich interfaces that proliferate with hole doping.
%

%%%%%%%%%%%%%%%%%%%%%%%%%%%%%%%%%%%%%%%%%
% HOW TO CONSTRUCT THE PHASE DIAGRAM
%%%%%%%%%%%%%%%%%%%%%%%%%%%%%%%%%%%%%%%%%

\section{Construction of the energy diagram}
\label{sec:energy-diagram}
We discuss in detail the construction of the energy diagram reported in Fig.~5 of the main text, where we plot the total energy per site differences among the states explored in our search.
We explore three classes of states as a function of hole doping $p$: homogeneous antiferromagnet (AFM), long-range ordered stripes and spin textures (STs).
To ensure consistency, all calculations are performed in a $40\times 40$ supercell with the same sampling of $k$ points (we consider a $8\times 8$ uniform mesh of $k$-points in the reduced Brillouin Zone, corresponding to a uniform mesh of $320\times 320$ points in the primitive Brillouin Zone).
We discuss how each class of solutions is converged and analyzed:
\begin{itemize}
    \item The AFM solutions are obtained initializing the magnetization with a checkerboard pattern $m(x,y)=(-1)^{x+y}m_0$, where $m_0$ is an initial guess, and letting the charge density to relax in the Hartree-Fock self-consistent loop.
    In this case the self-consistent solution is always a homogeneous AFM state.
    \item Stripes are obtained initializing the real-space magnetization with a cosinusoidal profile along one of the two directions of the supercell:
    \begin{equation}
        m(x,y) = (-1)^{x+y} m_0 \cos(2\pi\delta \cdot x),  
    \end{equation}
    where the incommensurability $\delta$ is chosen from those allowed in the periodic supercell, i.e. $\delta_n = n/L$, where $L$ is the lateral size of the supercell and $n$ is an integer smaller than $L$. 
    In our case $L=40$ and we choose values of $n$ between $1$ and $10$. An example of the spatial profile of these stripes is reported in Fig.~\ref{fig:panel-interface-spectral-stripes}.
    Then we let the charge relax within a self-consistent Hartree-Fock calculation for each stripe. 
    In this case self-consistency can introduce higher order harmonics, which are multiples of the chosen $\delta_n$, in the Fourier spectrum.
    We still identify the incommensurability of the final self-consistent pattern from the fundamental harmonic $\delta_n$ used in the initialization of the magnetization.
    For each value of hole doping we then select the lower energy stripes found in our search.
    For all the doped configurations we only consider initialization of the stripe pattern where the domain wall is site-centred.
    The stripe pattern reported in Sec.~\ref{sec:hubbard} has instead a bond-centred domain wall.
    \item Spin textures are obtained initializing the real-space spin-resolved charge density $n_{\mathbf{r}\sigma}$ with a uniform random distribution. For each spin texture we set a different seed in the random number generator.
    The charge density is then relaxed with a self-consistent Hartree-Fock loop, resulting in the patterns described in the main text, which are characterized by nonmagnetic domain walls separating AFM regions where the N\'eel parameter changes sign when crossing an interface. The arrangement of the domain walls and AFM domains changes depending on the initial guess for the spin resolved density.
    We consider ten different STs and, for each value of hole doping $p$ we report all their energies to display the spread of the energy distribution.
    For each ST we compute the incommensurability $\delta$ adopting the procedure discussed in Sec.~\ref{subsec:afm_correlation_lenght_doped}. At fixed doping the STs share similar values of the incommensurability, therefore we extract a representative value with a weighted average that accounts for the uncertainties introduced by the fitting procedure, as discussed in Sec.~\ref{subsec:afm_correlation_lenght_doped}; the extracted values are the same reported in Fig.~\ref{fig:panel-C-afm-corr-lengths}~(\textbf{b}).
\end{itemize}
These procedure yields, for each value of hole doping $p$, the energy of the homogeneous AFM state, the energy of the lowest-energy stripe and its incommensurability $\delta_\mathrm{Stripe}$, the energies of the manifold of STs with their representative incommensurability $\delta_\mathrm{ST}$.
For each value of doping, we then plot the difference between the energies collected for each class and the lowest energy found in our search. 
We also report with an annotation the incommensurability of the stripe and of the manifold of STs. 
%

%%%%%%%%%%%%%%%%%%%%%%%%%%%%%%%%%%%%%%%
% FIXED DOPING, DIFFERENT U
%%%%%%%%%%%%%%%%%%%%%%%%%%%%%%%%%%%%%%%
%
\section{Solutions of the hole-doped Hubbard model for different values of $U$}
\begin{figure}[h]
    \centering
    \includegraphics[width=0.95\linewidth]{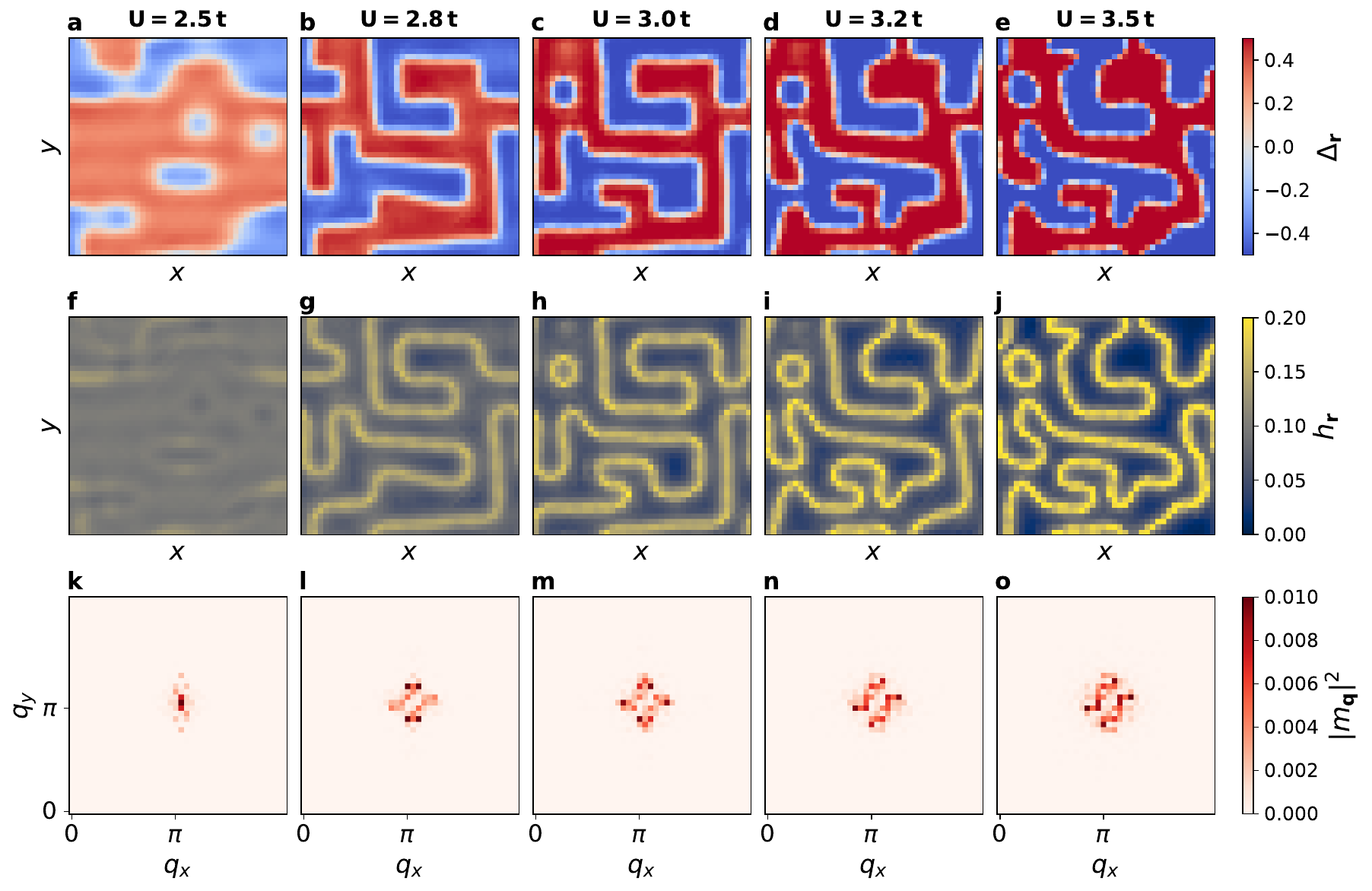}
    \caption{Solutions of the hole-doped model at different values of $U$.
    We report the results for a single spin texture.
    Calculations are performed at fixed hole doping $p=0.10$, with second neighbor hooping $t'=-0.3t$ in a $40\times 40$ supercell.
    (\textbf{a-e}) real-space Néel parameter $\Delta_{\mathbf{r}}$. All the solutions are characterized by antiphase domains separated by nonmagnetic interfaces. The interface length and the absolute value of the N\'eel parameter increase with $U$. 
    (\textbf{f-j}) hole density in real-space $h_{\mathbf{r}}$. The holes are spread in the supercell for small values of $U$, while they become strongly localized on the domain walls at larger $U$. 
    (\textbf{k-o}): absolute value of the Fourier components of the magnetization. The distribution of components in reciprocal space is centered around the AFM ordering vector ($\pi,\pi$) with a spread that grows with the value of $U$.}
    \label{fig:neel-charge-mag-change-U}
\end{figure}

\begin{figure}[h]
    \centering
    \includegraphics[width=0.95\linewidth]{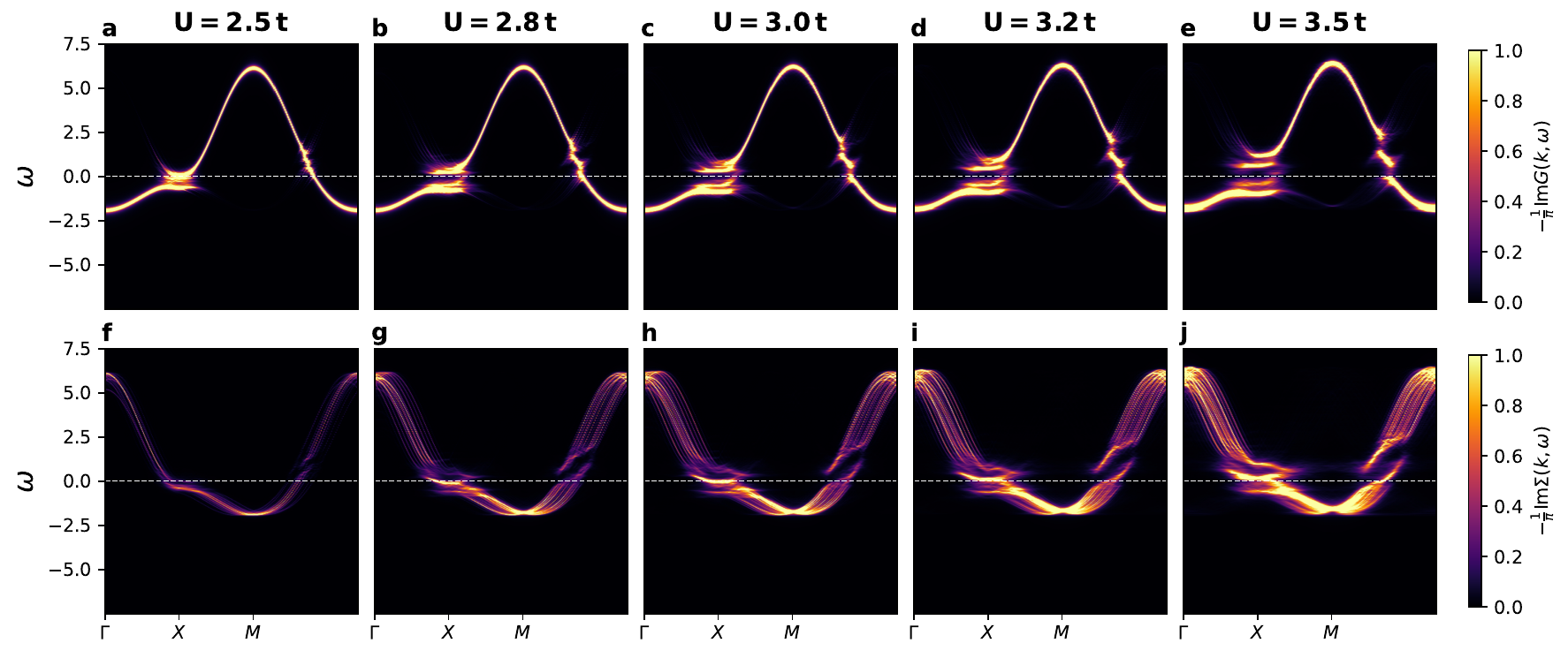}
    \caption{
    Nodal-antinodal dichotomy of the hole-doped solutions as a function of $U$.
    Spectral functions of the Green's function (upper row) and self-energy (lower row) obtained mapping the spin texture of Fig.~\ref{fig:neel-charge-mag-change-U} in the unit cell. Spectral functions are reported along a high-symmetry path in the primitive Brillouin Zone.
    (\textbf{a-e}) Spectra of the Green's function. Increasing the value of $U$ the system undegoes a metal-to-insulator transition. The antinodal sector is the first to be gapped out, while the nodal dispersion is fragmented with the presence of in-gap states at larger values of $U$.
    (\textbf{f-j}) Spectra of the self-energy. At smaller values of $U$ the self-energy is characterized by a dispersion with weak spectral weight. Increasing $U$ spectral weight appears first in the antinodal sector with a low-energy dispersionless pole. At larger values of $U$ the spectral weight is enhanced along all the path, while a gap persists in the nodal sector.
    }
    \label{fig:spectral-bands-change-U}
\end{figure}

\begin{figure}
    \centering
    \includegraphics[width=0.95\linewidth]{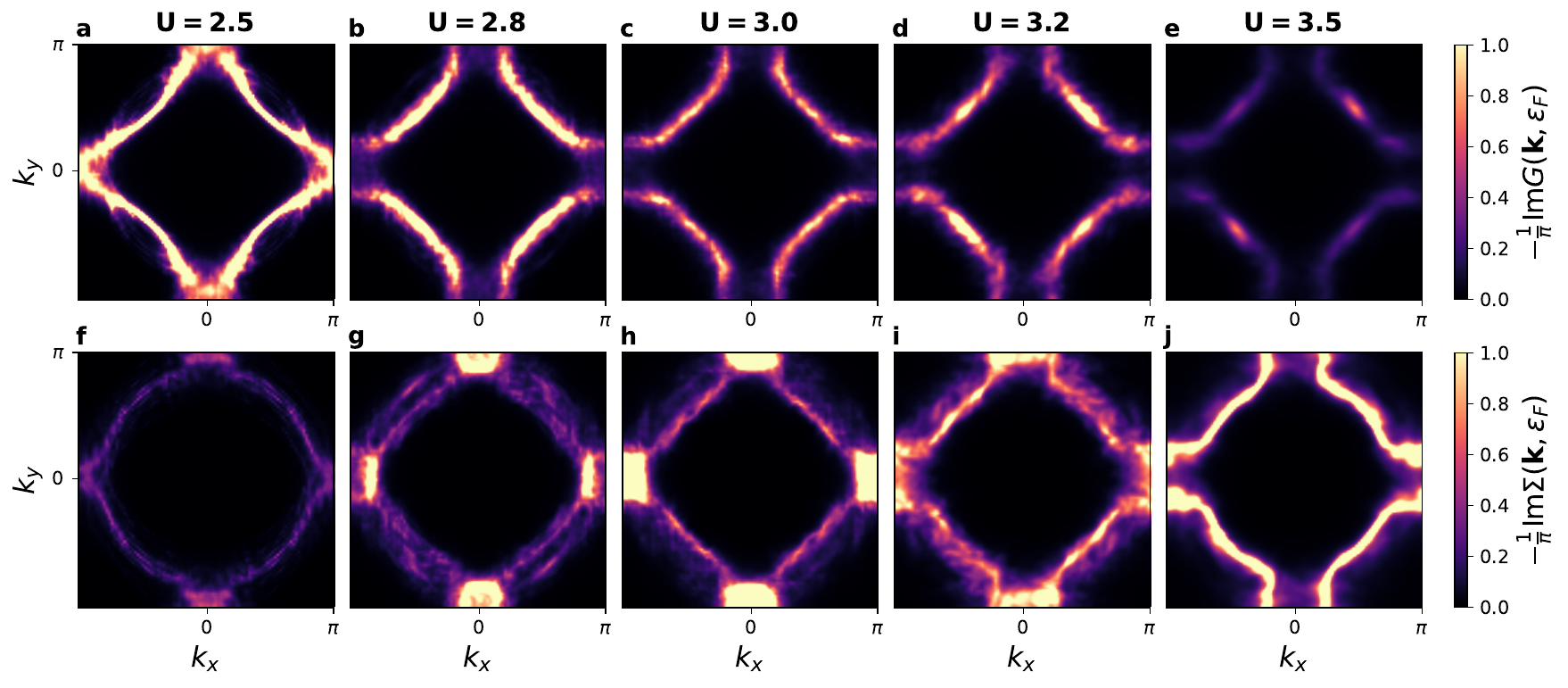}
    \caption{Fermi surface (upper row) and Luttinger surface (lower row) obtained from the spectral functions of Fig.~\ref{fig:spectral-bands-change-U} for a single spin texture at different values of $U$.
    (\textbf{a-e}) Fermi surface obtained from the imaginary part of the Green's function at the Fermi level. The evolution of the Fermi surface exposes the metal-to-insulator transition: at smaller values of $U$ the metallic dispersion is similar to that of the noninteracting system. Increasing the value of $U$ the metallic excitations are first gapped at the antinodal sector and finally also in the nodal region, causing the fading of the Fermi surface.
    (\textbf{f-j}) Luttinger surface obtained from the imaginary part of the self-energy at the Fermi level.
    The evolution with $U$ is complementary to that of the Fermi surface: at small values of $U$ the Luttinger surface has fading spectral weight everywhere in the primitive Brillouin Zone; increasing $U$ the spectral weight first emerges at the antinodes, and finally also in the nodal sector, outlining a closed contour.
    }
    \label{fig:surf-U}
\end{figure}

In this section we study the stability of the hole-doped solutions with respect to changes in the value of $U$, showing that the solutions characterized by phase separation of antiphase domains are present also for a range of values of $U$ around $U=3t$ when considering hole-doping.
In this case we consider a fixed hole doping $p=0.10$ and we study the solutions in the interval from $U=2.5t$ to $U=3.5t$ and we focus on a single symmetry-broken configuration obtained from a random initialization of the charge density as discussed in Sec.~\ref{subsec:quasi-degenerate-doped}.
In Fig.~\ref{fig:neel-charge-mag-change-U} we report the value of the Néel parameter, together with the hole density and the Fourier transform of the magnetization at different values of $U$.
Increasing the value of $U$ increases the percolation of the two antiphase AFM textures, increasing consequently the extension of the interface.
Moreover, at higher values of $U$ there is a stronger localization of the doped holes on the interfaces, while at smaller values of $U$ the holes are also delocalized in the AFM domains.
The Fourier transform of the magnetization does not show dramatic changes with $U$, besides developing a slightly larger distribution in momentum space at larger values of $U$.
In Fig.~\ref{fig:spectral-bands-change-U} we report the imaginary part of the Green's functions and self-energies extracted from the symmetry-broken mean-field solutions at different values of $U$.
Here we observe that going from small values of $U$ toward larger values, the system undergoes a metal-to-insulator transition.
In particular the spectrum of the Green's function in the antinodal sector is the first to be gapped, resulting in a pseudogapped solution at intermediate values of $U$.
The nodal states are then also eventually gapped at larger values of $U$, with the breakdown of the metallic dispersion of the noninteracting band along the nodal direction.
Complementary, this evolution is revealed in the spectrum of the self-energy, with an increasingly larger scattering in the antinodal sector with increasing $U$ and a gap in the nodal sector, revealing how the nodal-antinodal dichotomy persists in this range of values of the electronic interaction.
The metal-insulator transition is also evident by looking at the Fermi and Luttinger surfaces in Fig.~\ref{fig:surf-U}.
Here these surfaces are not averaged around quasi-degenerate symmetry-broeken configurations, but we only report those extracted from the chosen symmetry-broken solution.
The Fermi surface is still metallic for $U=2.5t$, becomes pseudogapped at intermediate values of $U$ and then eventually gets fully gapped at $U=3.5t$.
By contrast the Luttinger surface is coherent at larger values of $U$, loses coherence in the nodal regions at intermediate values of $U$ and eventually fades at lower values of $U$.
This set of solutions complements the analysis reported in the main text, where the solutions of the doped Hubbard model are studied at fixed value of $U$ and for varying doping.
Furthermore, these solutions clarify that the pseudogap features reported are associated to a set of different parameters of the model, not solely for the value of $U$ reported in the main text.
%
%%%%%%%%%%%%%%%%%%%%%%%%%%%%%%%%%%%%%%%
% FIXED DOPING, DIFFERENT t'
%%%%%%%%%%%%%%%%%%%%%%%%%%%%%%%%%%%%%%%
\section{Solutions of the hole-doped Hubbard model for different values of second neighbor hopping $t'$}
\label{sec:supp-different-tp}
We study spin texture solutions of the hole-doped Hubbard model for different values of second neighbor hopping $t'$ in a range between $t'=-0.05t$ and $t'=-0.25t$, at fixed value of the interaction strength $U=3t$, where $t$ is the first neighbor hopping, and at fixed value of hole doping $p=0.10$.
We track the properties of one spin texture by initializing the profile of the spin-resolved charge density with the same initial guess, taken from a uniform random distribution.
\begin{figure}
    \centering
    \includegraphics[width=0.95\linewidth]{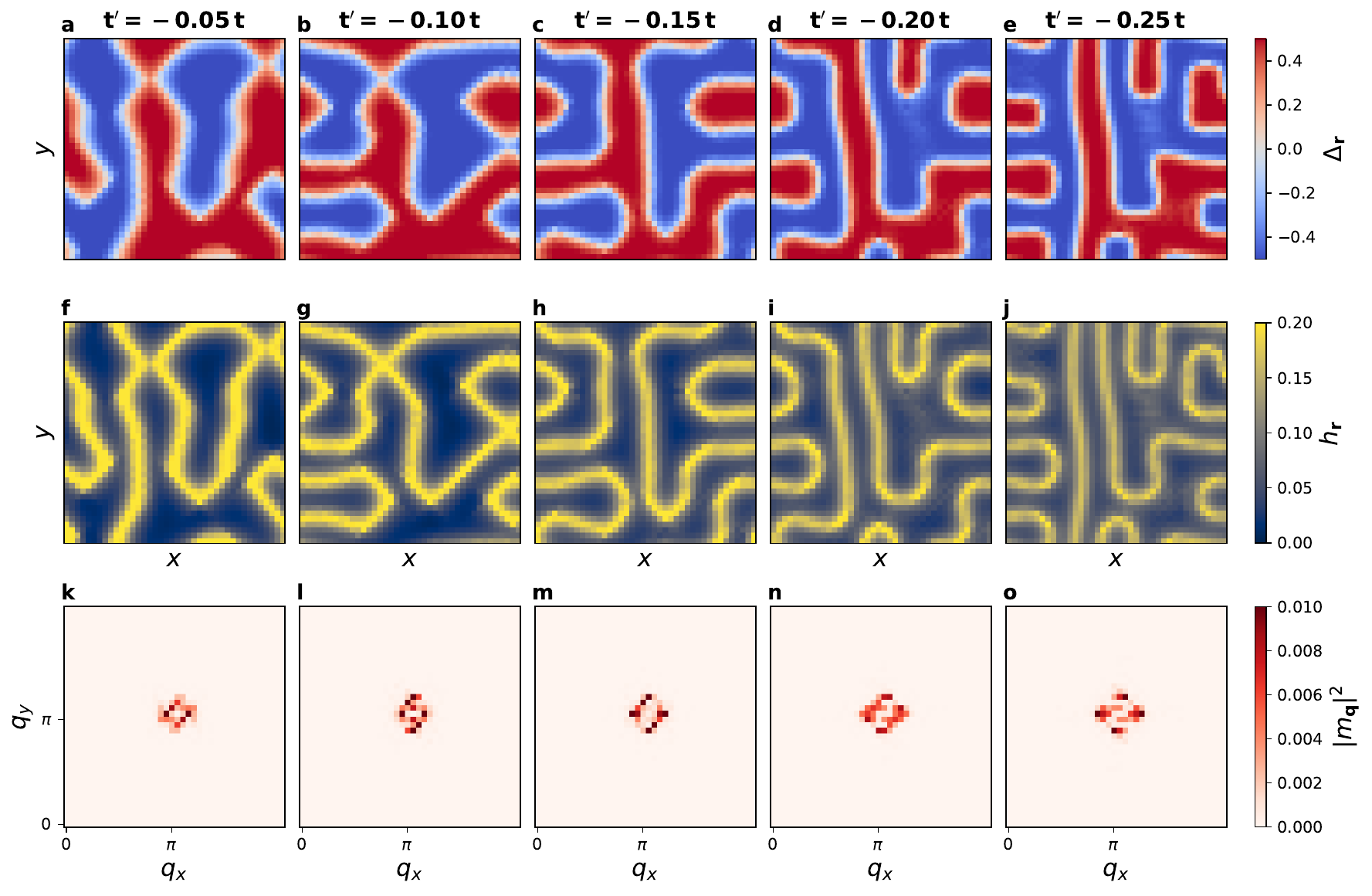}
    \caption{Evolution of a single spin texture solution of the hole-doped Hubbard model for different values of second-neighbor hopping $t'$. Model parameters are $U=3t$ and $p=0.10$. Calculations are performed in a $40\times 40$ supercell.
    \textbf{(a-e)}~Real-space N\'eel parameter $\Delta_{\mathbf{r}}$. The shape of the characteristic antiphase antiferromagnetic domains becomes progressively narrower and stripe-like at larger $|t'|$.
    \textbf{(f-j)}~Real-space hole density $h_{\mathbf{r}}$. The holes are more localized at the domain walls for smaller values of $|t'|$, while they progressively leak into the antiferromagnetic domains at larger $|t'|$.
    \textbf{(k-o)}~Reciprocal-space components of the magnetization $|m_\mathbf{q}|^2$. Increasing the value of $|t'|$, the Fourier components evolve from a distribution centered around the antiferromagnetic ordering vector $(\pi,\pi)$ towards a stripe-like distribution characterized by four incommensurate peaks.}
    \label{fig:panel-neel-charge-mag-different-tp}
\end{figure}
In Fig.~\ref{fig:panel-neel-charge-mag-different-tp} we report the N\'eel parameter, hole density and Fourier components of the squared magnetization as a function of $t'$.
The real-space N\'eel parameter reveals again that the spin texture solution is characterized by antiphase AFM domains separated by weakly magnetic interfaces.
Increasing $|t'|$ the AFM domains evolve from bubble-like structures to progressively narrower and stripe-like shapes. 
The distribution of holes is also influenced by $|t'|$: the holes are more tightly confined on the nonmagnetic domain walls for smaller values of $|t'|$, while they leak into the AFM domains at larger values.
The Fourier components of the magnetization also reveals a progressive evolution from short-range AFM correlations, characterized by a narrow distribution of Fourier components around $(\pi,\pi)$, to stripe-like correlations, characterized by four incommensurate peaks found at $(\pi\pm\delta,\pi)$ and $(\pi,\pi\pm\delta)$.
This result is expected, since the second neighbor hopping introduces AFM frustration, enhancing stripe correlations upon doping.

\begin{figure}
    \centering
    \includegraphics[width=0.95\linewidth]{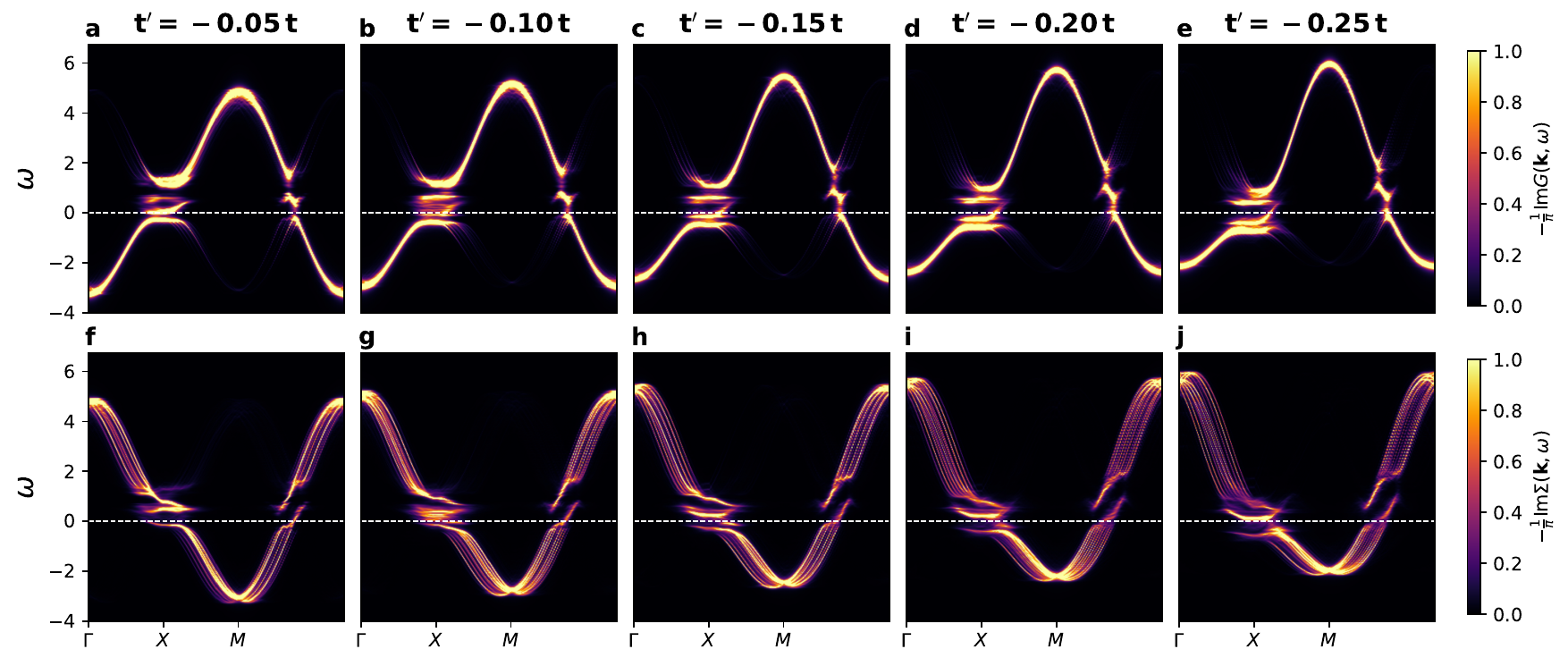}
    \caption{Spectral functions of the translationally invariant Green's functions and self-energies, obtained mapping the spin texture of Fig.~\ref{fig:panel-neel-charge-mag-different-tp} to the unit cell, for different values of the second neighbor hopping $t'$.
    The spectral functions are reported along a high-symmetry path in the primitive Brillouin Zone.
    The Fermi level is at $\omega=0$ (white-dashed line).
    \textbf{(a-e)} Green's function spectra. The antinodal region ($X$ point) is characterized by a fundamental spectral gap with the addition of two split-off bands at the band edges. Increasing $|t'|$, the Fermi level crosses the split-off bands and finally lies in the middle of the fundamental gap.
    The nodal sector ($M\rightarrow\Gamma$) is characterized by in-gap states that become more dispersive with $|t'|$.
    \textbf{(f-j)} Self-energy spectra. The antinodal sector displays an almost dispersionless pole that approaches the Fermi level from above with increasing $|t'|$. The nodal sector is instead characterized by a gap in the poles of the self-energy for all values of $|t'|$.
    }
    \label{fig:panel-spectral-bands-different-tp}
\end{figure}

We then map the single spin texture of Fig.~\ref{fig:panel-neel-charge-mag-different-tp} to the unit cell Green's functions and self-energies for each value of $t'$. This is shown in Fig.~\ref{fig:panel-spectral-bands-different-tp}.
We observe that the low-energy spectral weight of the Green's function is strongly influenced by the value of $t'$.
The antinodal sector is characterized by a fundamental gap, with the addition of two split-off dispersionless bands around the band edges.
The Fermi level crosses the lower split-off band at smaller values of $|t'|$, and progressively gets in the middle of the fundamental gap at larger values of $|t'|$.
The nodal sector is characterized by in-gap states due to the nonmagnetic interfaces. These states become more dispersive at larger values of $|t'|$.
The self-energy shows a similar sensitivity to the value of $t'$. 
In particular the antinodal sector develops a low-energy dispersionless pole; increasing $|t'|$ this pole approaches the Fermi level from above, gapping the Green's function in the antinodal sector.

\begin{figure}
    \centering
    \includegraphics[width=0.95\linewidth]{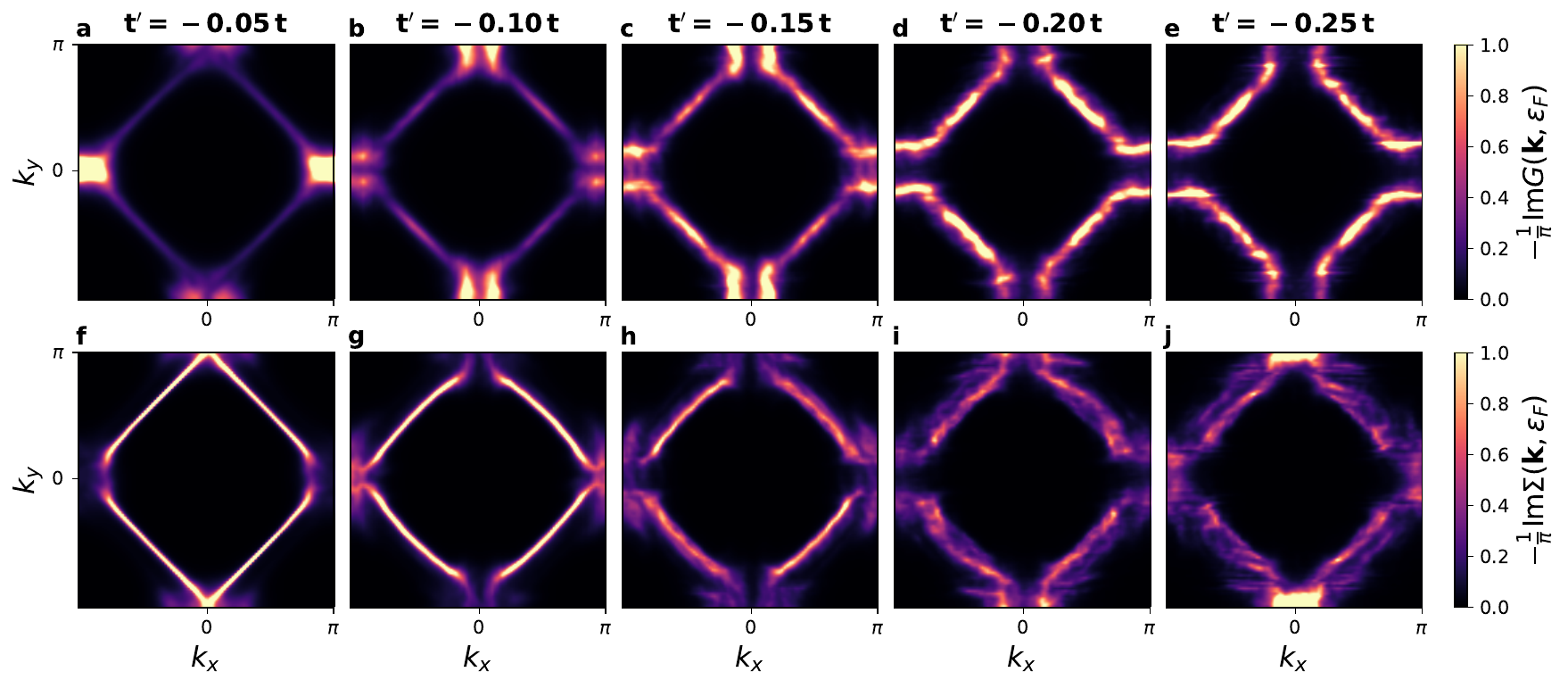}
    \caption{Fermi and Luttinger surfaces in the primitive Brillouin Zone obtained from the imaginary part of the Green's functions and self-energies of Fig.~\ref{fig:panel-spectral-bands-different-tp} for different values of the second neighbor hopping $t'$.
    \textbf{(a-e)} Fermi surface. We notice that when the single spin texture of Fig.~\ref{fig:panel-neel-charge-mag-different-tp} is mapped to the unit cell, it restores translational symmetry, but $C_4$ rotational symmetry remains broken; this is reflected in a different behaviour along the coordinate $k_x$ and $k_y$ axis in the Brillouin Zone. 
    For small values of $|t'|$ spectral weight is found in the antinodal regions, while the nodal sector becomes populated at larger $|t'|$.
    \textbf{(f-j)} Luttinger surface. Broken $C_4$ symmetry is similarly obseerved in the self-energy, more evidently at lower values of $|t'|$. The Luttinger surface shows a complementary behaviour with respect to the Fermi surface: increasing $|t'|$ the spectral weight disappears from the nodal sector and appears at the antinodal one.
    }
    \label{fig:panel-surf-different-tp}
\end{figure}

For each value of $|t'|$ we then inspect the spectral weight of the Green's functions and self-energies at the Fermi level, plotting the Fermi and Luttinger surfaces in the primitive Brillouin Zone in Fig.~\ref{fig:panel-surf-different-tp}.
The Fermi surfaces display a breaking of the $C_4$ rotational symmetry, which is particularly evident at smaller values of $|t'|$.
This is due to the fact that the embedding construction reported in the main text restores translational symmetry, but does not restore additional point group symmetries.
At smaller values of $|t'|$ the Fermi surface is characterized by metallic excitations confined in the antinodal sector: these are due to the split-off band crossing the Fermi level in Fig.~\ref{fig:panel-spectral-bands-different-tp}. 
Increasing $|t'|$ coherent metallic excitations disappear from the antinodal region and appear in the nodal one, outlining the emergence of the pseudogap.
The Luttinger surface also displays broken $C_4$ symmetry.
At lower values of $|t'|$ it is characterized by coherent spectral weight in the nodal region.
Increasing $|t'|$ this weight is transferred to the antinodal sector, complementary to the pseudogap in the Fermi surface.

\begin{figure}
    \centering
    \includegraphics[width=0.95\linewidth]{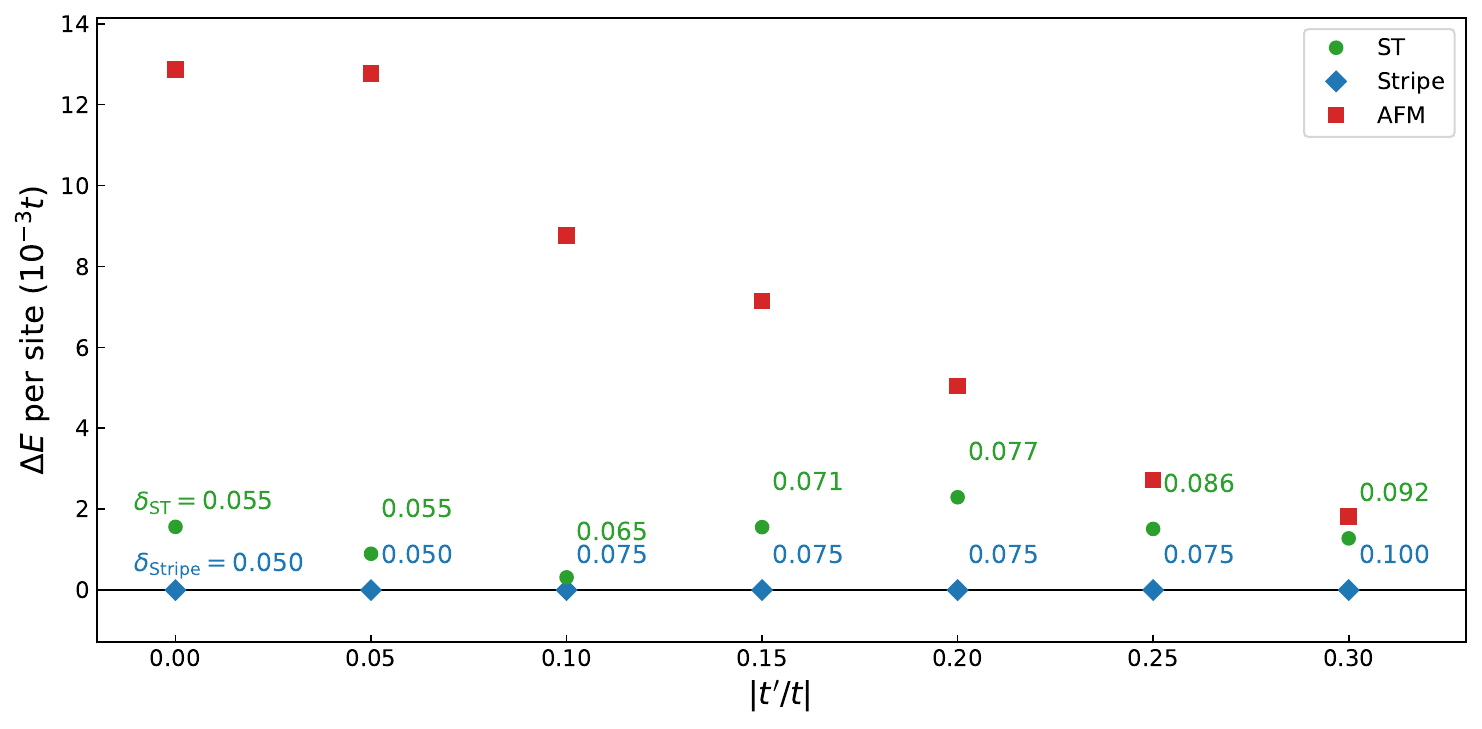}
    \caption{
    Energy differences (per site) between spin texture (ST), stripe and uniform antiferromagnet (AFM) as a function of $|t'|$.
    Model parameters are $U=3t$ and hole doping $p=0.10$.
    For each value of $t'$ the state with the lowest energy defines $\Delta E=0$. We also report for spin texture and stripes, in units of $2\pi$, the value of the typical incommensurability $\delta$ that defines the shift from the AFM ordering vector at $(\pi\pm\delta,\pi)$ and $(\pi,\pi\pm\delta)$.
    }
    \label{fig:energy-diagram-tp}
\end{figure}

Finally, we study how the energetics of this spin texture compares with respect to the uniform AFM state and to long-range ordered stripe states. 
For each value of $t'$ we consider stripes of different incommensurability $\delta$ allowed by the periodic boundary conditions of the $40\times 40$ supercell employed and we extract the lowest energy stripe.
In Fig.~\ref{fig:energy-diagram-tp} we report the energy differences between the spin texture solution and the AFM and stripe states, where for each value of $t'$ the lowest energy among the solutions included in our search defines $\Delta E = 0$.
We also report the typical incommensurabilities in units of $2\pi$ for both the spin texture and stripes. 
For stripes the incommensurabilities are attributed based on the initial cosine profile at the beginning of the self-consistent loop. 
For spin textures they are defined according to the fit based on Eq.~\eqref{eq:supp-C-stripe-fit}.
We observe that for this value of hole doping the stripes are always the lowest energy solution among those considered in our search. 
The energetics of the spin texture follows closely that of the stripes. For each value of $t'$ the spin textures are found at around $\sim 2 \,(10^{-3}t)$ above the stripes.
Moreover, for the whole range of values of $t'$ considered, the incommensurability of the spin texture is comparable to that of the lowest energy stripe.
For smaller values of $|t'|$ the AFM state is found at a relatively high energy with respect to stripes and spin textures, while at larger values of $|t'|$ the energy differences become considerably lower, being comparable to those of the spin textures. 
\clearpage
%


\end{document}